\documentclass[letterpaper,11pt]{article}
\usepackage{jheppub}

\usepackage{xcolor}
\usepackage{braket}
\usepackage{graphicx}
\usepackage{soul}
\usepackage{amssymb,amsmath}

\newcommand{\pd}{\partial}
\newcommand{\nn}{\nonumber\\}
\DeclareMathOperator{\tr}{tr}
\usepackage{verbatim}
\usepackage{anyfontsize}
\usepackage{dsfont}
\usepackage{tikz}
\usepackage{sidecap}
\usepackage{subfig}
\sidecaptionvpos{figure}{t}
\usepackage[english]{babel}
\usepackage{enumitem}  

\usepackage{graphicx, xcolor, varwidth}
\usepackage{setspace}
\usepackage[permil]{overpic}
\usepackage{graphicx}
\usepackage{mathtools}
\usepackage{wasysym}
\usepackage{graphicx}% Include figure files
\usepackage{placeins}
\usepackage{bm}
\newcommand{\be}{\begin{equation}}
\newcommand{\ee}{\end{equation}}
\usepackage{amsmath}
\usepackage{amsfonts}
\usepackage{amssymb}
\usepackage{graphicx}
\usepackage{wrapfig}
\usepackage{braket}
\usepackage{hyperref}
\usepackage{amsthm,verbatim,cancel}
\usepackage{color}
\usepackage{caption}
\newcommand{\bea}{\begin{eqnarray}}
\newcommand{\eea}{\end{eqnarray}}

\DeclareMathOperator{\U}{\mathrm{U}}
\DeclareMathOperator{\Cob}{\mathrm{Cob}}

\usepackage{tikz}
\usepackage{hyperref}
\usepackage{amssymb}
\usepackage{amsmath}
\usepackage{color}

\newcommand \mathtikz[1] {\quad \vcenter{\hbox{\tikz{#1}}} \quad}

\newcommand\SA[2] { %Open twist
\begin{scope}[xshift=#1,yshift=#2]
\draw (-0.25,0) to [out=-90,in=90] (0.25,-1);
\draw[line width=2mm, white] (0.25,0) to [out=-90,in=90] (-0.25,-1);
\draw (-0.25,0) -- (0.25,0) to [out=-90,in=90] (-0.25,-1) -- (0.25,-1);
\end{scope}
}

\newcommand\idC[2] { %Closed identity
\begin{scope}[xshift=#1,yshift=#2]
\filldraw[left color=lightgray, right color=white] (-0.25,0) -- (0.25,0) -- (0.25,-1) to [in=-90,out=-90] (-0.25,-1) -- (-0.25,0);
\filldraw[left color=white,right color=lightgray] (0,0) ellipse (0.25 and 0.1);, 
\draw[dotted] (0.25,-1) arc (0:180:0.25 and 0.1);
\end{scope}
}

\newcommand\idA[2] { %Open identity
\begin{scope}[xshift=#1,yshift=#2]
\filldraw[fill=white,draw=black] (-0.25,0) rectangle (0.25,-1);
\end{scope}
}

\newcommand\muC[2]{ % Closed product
\begin{scope}[xshift=#1,yshift=#2]
\filldraw[left color=lightgray, right color=white] (-0.25,0) to [out=-90,in=180] (0,-0.33) to [in=-90,out=0] (0.25,0) to  (0.75,0) to [in=90,out=-90] (0.25,-1) to [out=-90,in=-90] (-0.25,-1) to [in=-90,out=90] (-0.75,0);
\filldraw[left color=white,right color=lightgray] (-0.5,0) ellipse (0.25 and 0.1);
\filldraw[left color=white,right color=lightgray] (0.5,0) ellipse (0.25 and 0.1);
\draw[dotted] (0.25,-1) arc (0:180:0.25 and 0.1);
\end{scope}
}

\newcommand\pairC[2]{ % Closed pairing
\begin{scope}[xshift=#1,yshift=#2]
\filldraw[left color=lightgray, right color=white] (-0.25,0) to [out=-90,in=180] (0,-0.33) to [in=-90,out=0] (0.25,0) to  (0.75,0) to [out=-90,in=0] (0, -0.83) to [out=180,in=-90]
(-0.75,0);
\filldraw[left color=white,right color=lightgray] (-0.5,0) ellipse (0.25 and 0.1);
\filldraw[left color=white,right color=lightgray] (0.5,0) ellipse (0.25 and 0.1);
\end{scope}
}

\newcommand\copairC[2]{ % Closed pairing
\begin{scope}[xshift=#1,yshift=#2]
\filldraw[left color=lightgray, right color=white] (-0.25,0) to [out=90,in=180] (0,0.33) to [in=90,out=0] (0.25,0) to [out=-90,in=-90] (0.75,0) to [out=90,in=0] (0, 0.83) to [out=180,in=90]
(-0.75,0) to [out=-90,in=-90] (-0.25,0);
\draw[dotted] (-0.25,0) arc (0:180:0.25 and 0.1);
\draw[dotted] (0.75,0) arc (0:180:0.25 and 0.1);
\end{scope}
}

\newcommand\deltaC[2]{% Closed comultiplication
\begin{scope}[xshift=#1,yshift=#2]
\filldraw[left color=lightgray, right color=white] (-0.25,-1) to [out=90,in=180] (0,-0.66) to [in=90,out=0] (0.25,-1) to [out=-90,in=-90] (0.75,-1) to [in=-90,out=90] (0.25,0) to (-0.25,0) to [in=90,out=-90] (-0.75,-1) to [out=-90,in=-90] (-0.25,-1);
\filldraw[left color=white,right color=lightgray] (0,0) ellipse (0.25 and 0.1);
\draw[dotted] (-0.25,-1) arc (0:180:0.25 and 0.1);
\draw[dotted] (0.75,-1) arc (0:180:0.25 and 0.1);
\end{scope}
}

\newcommand\muA[2]{ % Open multiplication
\begin{scope}[xshift=#1,yshift=#2]
\draw (-0.75,0) -- (-0.25,0) to [out=-90,in=180] (0,-0.33) to [in=-90,out=0] (0.25,0) -- (0.75,0) to [in=90,out=-90] (0.25,-1);
\draw (-0.25,-1) -- (0.25,-1);
\draw (-0.75,0) to [in=90,out=-90] (-0.25,-1);
\end{scope}
}

\newcommand\pairA[2]{ % Open pairing
\begin{scope}[xshift=#1,yshift=#2]
\draw (-0.75,0) -- (-0.25,0) to [out=-90,in=180] (0,-0.33) to [in=-90,out=0] (0.25,0) -- (0.75,0) to [out=-90,in=0] (0,-0.83) to [out=180,in=-90] (-0.75,0);
\end{scope}
}

\newcommand\copairA[2]{ % Open copairing
\begin{scope}[xshift=#1,yshift=#2]
\draw (-0.75,0) -- (-0.25,0) to [out=90,in=180] (0,0.33) to [in=90,out=0] (0.25,0) -- (0.75,0) to [out=90,in=0] (0,0.83) to [out=180,in=90] (-0.75,0);
\end{scope}
}

\newcommand\deltaA[2]{ % Open comultiplication
\begin{scope}[xshift=#1,yshift=#2]
\draw (-0.75,-1) -- (-0.25,-1) to [out=90,in=180] (0,-0.66) to [in=90,out=0] (0.25,-1) -- (0.75,-1) to [in=-90,out=90] (0.25,0) -- (-0.25,0) to [in=90,out=-90] (-0.75,-1);
\end{scope}
}

\newcommand\zipper[2]{ % Zipper
\begin{scope}[xshift=#1,yshift=#2]
\draw (-0.25,-1) -- (0.25,-1);
\filldraw[right color=white,left color=lightgray] (-0.25,0) to (-0.25,-1) to [out=90,in=225] (0,-0.5) to [out=-45,in=90] (0.25,-1) to (0.25,0);
\filldraw[left color=white,right color=lightgray] (0,0) ellipse (0.25 and 0.1);
\end{scope}
}

\newcommand\cozipper[2]{ % Cozipper
\begin{scope}[xshift=#1,yshift=#2]
\draw (-0.25,0) -- (0.25,-0);
\filldraw[right color=white,left color=lightgray] (-0.25,-1) to (-0.25,0) to [out=-90,in=135] (0,-0.5) to [out=45,in=-90] (0.25,0) to (0.25,-1) to [in=-90,out=-90] (-0.25,-1);
\draw[dotted] (0.25,-1) arc (0:180:0.25 and 0.1);
\end{scope}
}

\newcommand\epsilonC[2]{ % Closed cap
\begin{scope}[xshift=#1,yshift=#2]
\filldraw[right color=white,left color=lightgray] (-0.25,0) to [out=-90,in=180] (0,-0.33) to [in=-90,out=0] (0.25,0);
\filldraw[left color=white,right color=lightgray] (0,0) ellipse (0.25 and 0.1);
\end{scope}
}

\newcommand\etaC[2] { % Closed cap
\begin{scope}[xshift=#1,yshift=#2]
\filldraw[right color=white,left color=lightgray] (-0.25,0) to [out=90,in=180] (0,0.33) to [in=90,out=0] (0.25,0) to [in=-90,out=-90] (-0.25,0);
\draw[dotted] (0.25,0) arc (0:180:0.25 and 0.1);
\end{scope}
}

\newcommand\epsilonA[2] {% Open cap
\begin{scope}[xshift=#1,yshift=#2]
\draw (-0.25,0) -- (0.25,0);
\draw (-0.25,0) to [out=-90,in=180] (0,-0.33) to [in=-90,out=0] (0.25,0);
\end{scope}
}

\newcommand\etaA[2] {% Open cap
\begin{scope}[xshift=#1,yshift=#2]
\draw (-0.25,0) -- (0.25,0);
\draw (-0.25,0) to [out=90,in=180] (0,0.33) to [in=90,out=0] (0.25,0);
\end{scope}
}

\newcommand\tauA[2] { % Open braiding
\begin{scope}[xshift=#1,yshift=#2]
\draw (-0.75,0) -- (-0.25,0) to [out=-90,in=90,looseness=0.5] (0.75,-1) -- (0.25,-1) to [out=90,in=-90,looseness=0.5] (-0.75,0);
\filldraw[fill=white,draw=black] (0.75,0) -- (0.25,0) to [out=-90,in=90,looseness=0.5] (-0.75,-1) -- (-0.25,-1) to [out=90,in=-90,looseness=0.5] (0.75,0);
\end{scope}
}

\begin{document}

\title{Entanglement entropy and edge modes in topological string theory: I}
\author[a]{William Donnelly,}
\author[b]{Yikun Jiang,}
\author[b]{Manki Kim,}
\author[c]{and Gabriel Wong}

\affiliation[a]{Perimeter Institute for Theoretical Physics, 31 Caroline St. N, N2L 2Y5, Waterloo ON, Canada}
\affiliation[b]{Department of Physics, Cornell University, Ithaca, New York, USA}
\affiliation[c]{Department of Physics, Fudan University, Shanghai, China}

\emailAdd{wdonnelly@perimeterinstitute.ca}
\emailAdd{phys.yk.jiang@gmail.com}
\emailAdd{mk2427@cornell.edu}
\emailAdd{gabrielwon@gmail.com }
\abstract{  Progress in identifying the bulk microstate interpretation of the Ryu-Takayanagi formula requires understanding how to define entanglement entropy in the bulk closed string theory.  Unfortunately, entanglement and Hilbert space factorization remains poorly understood in string theory.   As a toy model for AdS/CFT, we study the entanglement entropy of closed strings in the topological A-model in the context of Gopakumar-Vafa duality.  We will present our results in two separate papers.  In this work, we consider the bulk closed string theory on the resolved conifold and give a self-consistent factorization of the closed string Hilbert space using extended TQFT methods.  We incorporate our factorization map into a Frobenius algebra describing the fusion and splitting of Calabi-Yau manifolds, and find string edge modes transforming under a $q$-deformed surface symmetry group.  We define a string theory analogue of the Hartle-Hawking state and give a canonical calculation of its entanglement entropy from the reduced density matrix. Our result matches with the geometrical replica trick calculation on the resolved conifold, as well as a dual Chern-Simons theory calculation which will appear in our next paper \cite{secondpaper}. We find a realization of the Susskind-Uglum proposal identifying the entanglement entropy of closed strings with the thermal entropy of open strings ending on entanglement branes.  We also comment on the BPS microstate counting of the entanglement entropy. Finally we relate the nonlocal aspects of our factorization map to analogous phenomenon recently found in JT gravity.
}
\maketitle

\section{Introduction}
The holographic principle states that the number of degrees of freedom in a spacetime region scales with the area of its boundary, and is exemplified by the Bekenstein-Hawking (BH) entropy formula. In the context of the AdS/CFT correspondence \cite{1999IJTP...38.1113M, 1998PhLB..428..105G, 1998AdTMP...2..253W}, the Ryu-Takayanagi (RT) formula \cite{2006PhRvL..96r1602R} generalizes BH entropy to extremal surfaces in AdS which are anchored to the asymptotic boundary, and identifies the leading area term with the leading $\mathcal{O}(N^{2})$ contribution to the entanglement entropy of the boundary CFT \cite{2003JHEP...04..021M, Faulkner:2013ana}.  Given a factorization of the CFT Hilbert space\footnote{ Even though it is quite plausible,  such a factorization has never been carefully worked out.  However for rational CFT's, the question of  Hilbert space factorization and edge modes was recently addressed in \cite{2019arXiv191211201H}.}, this implies that the bulk extremal area is capturing the degrees of freedom for quantum states of a boundary subregion. However, the bulk micro-state interpretation of the entropy remains mysterious. One aspect of this puzzle is that the bulk supergravity only contains $\mathcal{O}(1)$ number of fields, while the classical area term is of $\mathcal{O}(N^2)$ \cite{2017arXiv170407763L, 2020JHEP...02..167T}. Where does this large number of degrees of freedom come from?

\begin{figure}
\centering 
\includegraphics[scale=1]{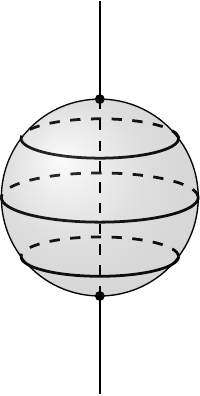}
\qquad \qquad
\includegraphics[scale=1]{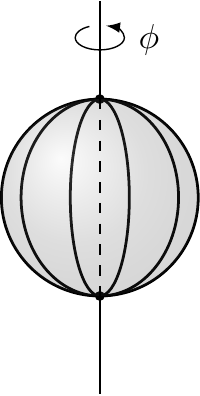}
\caption{
The partition function of the $A$-model on a line bundle over $S^2$ has two interpretations.
In the closed string channel (left), it represents the overlap $\braket{HH^*|HH}$ between the Hartle-Hawking state and its orientation reversal.
In the open string channel (right), it represents a trace in the Hilbert space of open strings.
Figure borrowed from Ref.~\cite{Donnelly:2016jet}.
} \label{fig:slicing}
\end{figure}

We want to understand this question directly in the bulk from the microscopic string theory.
In the case of BH entropy, Susskind and Uglum  \cite{Susskind:1994sm} proposed that the horizon area measures the entanglement entropy of closed strings across the horizon.  In the tree level replica trick calculation, the entanglement entropy is due to a sphere diagram which intersects the entangling surface, representing closed strings which are cut into open strings as depicted in figure \ref{fig:slicing}.  What distinguishes string theory from quantum field theory (QFT) is that this tree level closed string diagram has a one-loop open string interpretation, suggesting a trace over a quantum Hilbert space.  This led Susskind and Uglum to conclude that the BH entropy counts microstates of open string endpoints anchored on the horizon. In the language of Ref.~\cite{Donnelly:2016jet} the horizon is wrapped by \emph{entanglement branes}, which gives rise to entanglement edge modes responsible for the large $\mathcal{O}(1/{g_{\text{string}}^2}) = \mathcal{O}(1/G_\text{Newton})$ entropy.
Given the analogy between RT formula and BH entropy, it is tempting to apply this proposal to give a canonical interpretation of the RT entropy from the bulk string theory.

While the seminal work\cite{Strominger:1996sh} succeeded in reproducing the BH entropy for five dimensional extremal Reissner-Nordstrom black holes via counting BPS microstates in string theory, little is known about how to compute entanglement entropy and the associated Hilbert space factorization in string theory.  In field theory, the replica trick as computed by the Euclidean path integral offers a shortcut that circumvents the  factorization problem.  However a naive application to string theory requires putting an $n$-sheeted cover in the target space, which requires an off-shell formulation of  string theory that is not well understood.\footnote{ For the closed bosonic string, such an off-shell formulation was proposed by Tseytlin \cite{Tseytlin:1988tv} and applied by Susskind and Uglum in their proposal. }  Since string theory is well-defined in the presence of conical deficits, the references  \cite{1995NuPhB.439..650D, 2015JHEP...05..106H, 2019JHEP...01..126W} attempted worldsheet calculations of  entanglement entropy using an ``orbifold" replica trick.  However these calculations do not capture the sphere contribution to the entanglement entropy and the associated edge modes.\footnote{The sphere contribution vanishes at the orbifold point, and hence remains zero when analytically continued in the replica number. }  An attempt at an off-shell calculation was made in \cite{2018PhRvD..97f6025B} via Witten's open string field theory.   They showed that the symplectic structure of the string field theory in a subregion implies  that pure gauge (BRST) modes become dynamical edge modes at the entanglement cut, but did not go beyond this classical analysis. 

Edge modes are boundary degrees of freedom introduced to give a self consistent description of a subsystem. In string theory they appear due to the need to cut strings at the point where the string worldsheet intersects the entangling surface, leaving configurations where the strings end at the entangling surface.
A similar phenomenon occurs in Maxwell theory, where the edge modes can be thought of as configurations of ``electric strings'' with their endpoints anchored to the entangling surface \cite{Buividovich:2008gq}. As in string field theory, the presence of edge modes can be deduced from the symplectic structure of a subregion \cite{Donnelly:2016auv}.
These edge modes give an important contribution to the entanglement entropy \cite{Donnelly:2014fua,Donnelly:2015hxa}, in particular reproducing the contact interaction of \cite{Kabat:1995eq} which may be viewed as a field theory limit of a string worldsheet intersecting the entangling surface.
However, these field theory calculations can only capture the one-loop correction to the entropy, corresponding to toroidal worldsheets. 

In this work, we initiate a program to realize the Susskind-Uglum proposal for the topological A-model string in the context of Gopakumar-Vafa duality (GV duality) \cite{Gopakumar:1998ii,Gopakumar:1998jq,Gopakumar:1998ki,Gopakumar:1998vy,Ooguri:1999bv,Aganagic:2002qg,Dedushenko:2014nya}, which can be viewed as a topological version of AdS/CFT.  
The role of the bulk string theory is played by the topological A-model closed string on a resolved conifold geometry \cite{Candelas:1989js}.  This is a six-dimensional Calabi-Yau manifold which is a rank-2 bundle over a sphere of  complexified area $t$. The boundary CFT is replaced by the large-$N$ limit of $\U(N)$ Chern-Simons theory, with gauge coupling $g_{cs} =\frac{2 \pi }{k+N}$ and `t Hooft parameter $i g_{cs}N$.
The closed string coupling $g_{s}$ and the target space modulus $t$ in the bulk are related to the parameters of the CS theory by
\begin{align}
    g_{s}&=g_{cs}= \frac{2 \pi }{k+N},\nn
    t&=\frac{ 2 \pi i N }{N+k}.
\end{align}

The advantage of studying the A-model string is that it provides a setting similar to AdS/CFT where we can give precise accounting of edge modes and their entanglement entropy on both sides of the duality.  In this paper, we will focus on the closed string theory and define its Hilbert space via the categorical description of the A-model as a topological quantum field theory (TQFT) \cite{Bryan:2003aw,Aganagic:2004js,Bryan:2004iq}.   This allows us to apply the framework developed in \cite{Donnelly:2016jet}  to define the factorization of the string theory Hilbert space purely in terms of the categorical data of an open-closed TQFT.  For the A-model, the relevant TQFT can be viewed as coming from a large $N$, chiral limit of $q$-deformed 2D Yang-Mills theory (2DYM)\cite{2005NuPhB.715..304A}.\footnote{The $q$-deformed 2D Yang-Mills theory has been proposed as a non-chiral UV completion for the closed topological string theory, and in the discussion section we will discuss the implications of this completion and its connection to wormholes and baby universes.}  Using the TQFT description, we propose a factorization of the closed string Hilbert space that is consistent with the entanglement entropy as computed by the replica trick. 
For the Hartle-Hawking state of the closed string theory, we find that the entanglement entropy can be interpreted as the thermal entropy of open strings, with the aforementioned $\mathcal{O}(1/g_{s}^{2})$ scaling arising from the counting of Chan-Paton factors. 
 
 As in \cite{Donnelly:2016jet}, these Chan-Paton factors are the entanglement edge modes of the closed string theory, which we will interpret as coming from the stacks of entanglement branes at the entangling surface.   Interestingly, the coupling of these branes to the string endpoints endows them with nontrivial braiding statistics.  The corresponding edge modes thus behave like anyons and transform under a quantum group symmetry, which  plays the role of the surface symmetry group (c.f. \cite{Donnelly:2016auv}) for the topological string theory. 

In the follow-up paper \cite{secondpaper}, we will give a dual Chern-Simons gauge theory description of the entanglement entropy and the corresponding edge modes, thus giving an independent check of our closed string calculations.  In the closed string theory, we will define a Hartle-Hawking state obtained by summing over worldsheets ending on a stack of D-branes.  By applying GV duality, we will show that there is a \emph{local} mapping between these worldsheets and unknotted Wilson loops in the Chern-Simons theory, so that cutting the worldsheets correspond to cutting the Wilson loops. In gauge theory, the entanglement entropy $\delta S$ due to cutting a Wilson loop $W_{R}$ in a representation $R$ is
\cite{Lewkowycz:2013laa}
\begin{align}\label{W} 
\delta S= (1- n\pd_{n} ) \log \braket{W_{R}}= \braket{H_{\text{mod}}}_{W_{R}} + \log  \braket{W_{R}}.
\end{align}
This is the entanglement entropy relative to the vacuum state, also referred to as the defect entropy \cite{2013arXiv1307.1132J, 2013PhRvD..88j6006J}.  Here $\braket{H_{\text{mod}}}_{W_{R}}$ is the expectation value of the modular Hamiltonian in the presence of the Wilson loop, which vanishes in Chern-Simons theory.
Thus, the defect entanglement entropy  associated with the Wilson loop is just $\log \braket{W_{R}} $. For the unknot, $\braket{W_{R}}$ is precisely the quantum dimension which captures the topological degeneracy associated with the fusion Hilbert space of an anyon.  By superposing such Wilson loops in all possible representations, we will reproduce the string theory Hartle-Hawking state as well as its entanglement entropy in an appropriate large-$N$ limit of the quantum dimensions.   This limit gives a precise relation between the closed string edge modes and the anyons of Chern-Simons theory.

Our description of the relation between string worldsheets and Wilson loops has a direct analogue in AdS/CFT \cite{Ooguri:1999bv,Gomis:2006mv,Dymarsky:2006ve}. 
A Wilson loop in the CFT in the fundamental representation is dual to a probe string worldsheet in the bulk geometry, and equation \eqref{W} was used in \cite{Lewkowycz:2013laa} to compute its entanglement entropy.
In this calculation the entanglement entropy is $\mathcal{O}(\log(1/g_s))$, which is large at weak coupling but still much smaller than the $\mathcal{O}(1/g_s^2)$ RT entropy.  

A similar phenomenon was noted in Refs.~\cite{Donnelly:2016jet} for the string dual to 2DYM --- any state with $\mathcal{O}(1)$ number of strings has an entanglement entropy of $\mathcal{O}(\log(1/g_s))$.  However, for the Hartle-Hawking state, competition between the Chan-Paton factors and the string action leads to a saddle point with $\mathcal{O}(1/g_s^2)$ strings.  The counting of Chan-Paton factors at this saddle point leads to an $\mathcal{O}(1/g_s^2)$ entanglement entropy that is reminiscent of the scaling of ``spacetime" entropy \cite{Donnelly:2019zde}.

Similarly, in AdS/CFT, in the presence of Wilson loops corresponding to  ``large representations'' with order  $\mathcal{O}(N^{2})=\mathcal{O}(1/g_s^2)$ number of boxes, the dual branes backreact on the geometry.  In this case the defect entropy can be computed using the RT formula in the new background, and the $\mathcal{O}(N^{2})=\mathcal{O}(1/g_s^2)$ entropy is recovered \cite{Aguilera-Damia:2017znn} \cite{Gentle_2014}. In our computation, the resolved conifold itself is an emergent geometry arising from the  superposition of a large number of fundamental strings, dual to a large number of Wilson loops in the dual gauge theory.  By accounting for the contributions from the entire superposition of states, our entropy calculation captures the entanglement which  ``makes up the spacetime" itself.

Finally, we comment on how our work differs from the recent work \cite{Hubeny:2019bje} which also studied entanglement in the A-model string theory. The essential difference is twofold: first our choice of state cuts through the base manifold $S^2$ where the closed string worldsheets wrap, whereas the state defined in \cite{Hubeny:2019bje} does not.  As a result, our entanglement cut will directly probe the string edge modes that were not revealed by their computation. Second, rather than relying solely on the dual Chern-Simons field theory, we give a self-consistent Hilbert space factorization and entropy calculation on the closed string side.

\section{Summary, overview of GV duality and the Hartle-Hawking state }
\label{section:out} 
Here we give a summary of our paper, starting with an overview of the GV duality and a description of the closed string state whose factorization and entanglement entropy we will be studying. 

\subsection{Summary of the GV duality} 
\begin{figure}[h] 
\centering
\includegraphics[width=16cm,height=14cm,keepaspectratio]{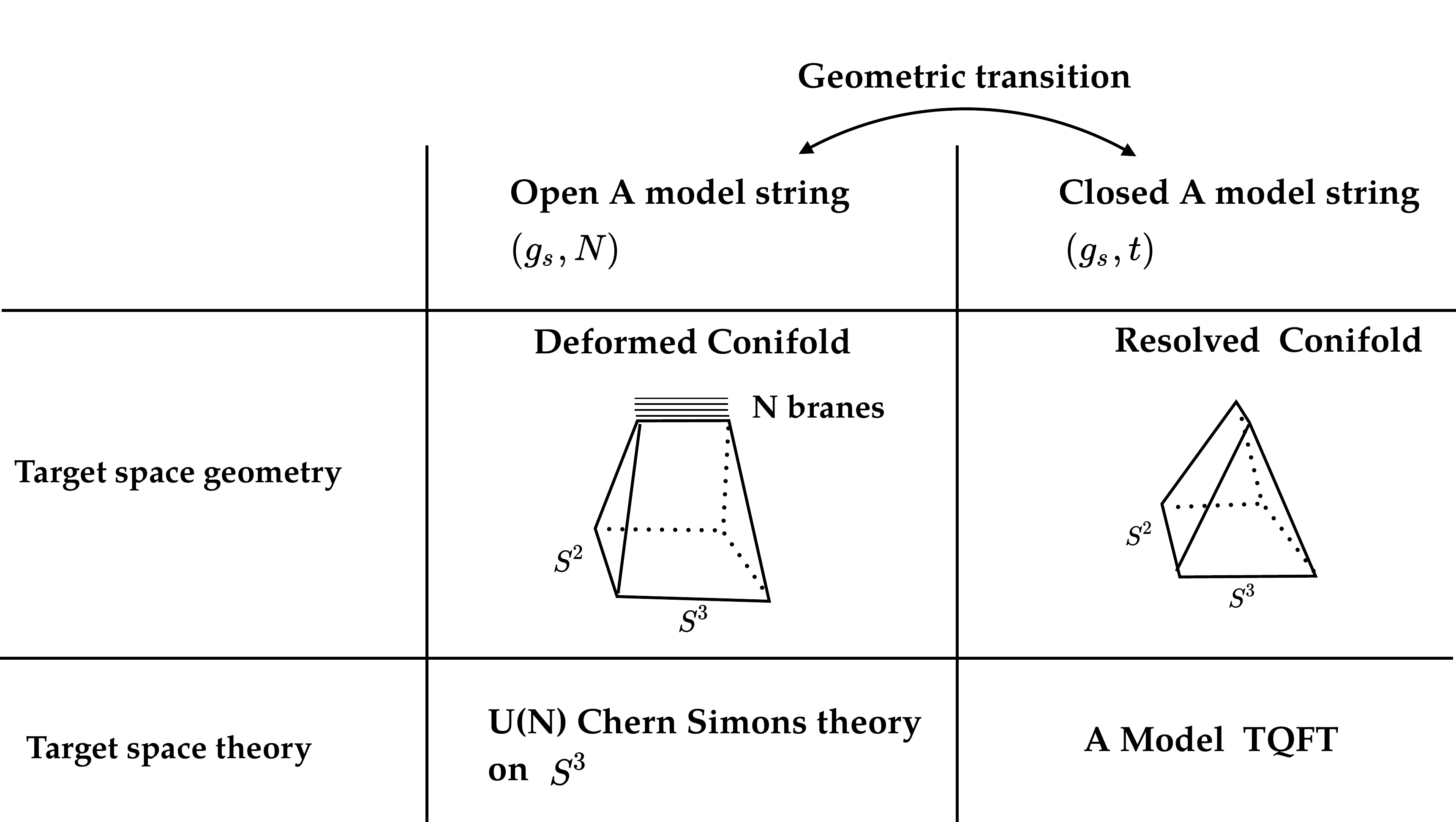}
\caption{Gopakumar duality relates closed A-model string on the resolved conifold to the open A-model string on the deformed conifold }\label{chart}
\end{figure}

Like AdS/CFT, the Gopakumar-Vafa duality is an open-closed string duality.   Figure \eqref{chart} shows the 6D target space geometries for the closed and open strings.  These  can be conveniently presented as two different ways to resolve the singularity of the conifold geometry, which is a cone over $S^3 \times S^2$.   The closed strings live on the resolved conifold where the conical tip is resolved into an $S^2$, whereas the open strings live on the deformed conifold, where the tip is deformed into an $S^{3}$. The defining equations and details of the geometries are summarized in appendix B.

An intuitive way to understand the GV duality is via `t Hooft's argument for the emergence of string theory from gauge theory \cite{tHooft:1973alw}. `t Hooft observed that in the large $N$ limit, the Feynman diagrams of a $\U(N)$ gauge theory can be represented by ribbon graphs which should be viewed as Riemann surfaces with holes.    These are open string worldsheets, corresponding to a free energy expansion in which  the gauge coupling $g_{\text{YM}}^{2}$ plays the role of the string coupling $g_{s}$:
\begin{align}
F= \sum_{g=0}^{\infty} \sum_{h=1}^{\infty} (g_{\text{YM}}^{2})^{2g-2+h} N^{h} F_{g,h}.
\end{align} 
Here $g$ is the genus of the worldsheet, $h$ is the number of holes,  and $N$ accounts for Chan-Paton factors of $\U(N)$.  The dual closed string theory is obtained by summing over holes, giving
\begin{align}
F &=\sum_{g=0}^{\infty}( g_{\text{YM}}^{2})^{2g-2} F_{g}(t),\nn
F_{g}(t) &=  \sum_{h=1}^{\infty} F_{g,h} t^{h}, \quad  \quad t= g_{\text{YM}}^{2} N.
\end{align} 
Here $t$ is the `t Hooft coupling of the gauge theory, which plays the role of a target space modulus for the closed string.   

In the `t Hooft paradigm, the gauge theory which is relevant to GV duality is $\U(N)$  Chern-Simons theory on $S^3$. A direct $1/N$ expansion of the Chern-Simons partition function leads to the connected amplitudes $F_{g,h}$  of the open topological string on the deformed conifold, which is the same as the cotangent bundle $T^{*}S^3$ \cite{Gopakumar:1998ki}. In this geometry the open string degenerates into a pointlike object and is restricted to live on the base $S^3$.  These degenerate Riemann surfaces of zero area correspond precisely to the ribbon graphs of the gauge theory.    Chern-Simons theory is thus the string field theory of these open strings \cite{Witten:1992fb}.    
Closed topological strings wrap minimal volume representative among homologous 2-cycles \cite{Witten:1988xj,Witten:1991zz};  in  the  resolved  conifold,  the  only  such  2-cycle  is  the $S^2$ at the tip. Open topological strings end on Lagrangian 3-cycles, and in the GV duality, they end on the base $S^3$ of the deformed conifold. Similar to AdS/CFT,  the open string theory with a large $N$ number of branes on $S^3$ is dual to a closed string theory where the branes have been dissolved and replaced by a nontrivial flux  $t= i g_{S} N$  of the $B$ field through the $S^2$. 

The dual closed string theory was derived from the worldsheet by directly summing over the holes in \cite{Ooguri:1999bv}. The resulting theory is the  A-model closed string on the resolved conifold, which should be viewed as the gravitational dual of Chern-Simons theory on $S^3$.   While the resolved conifold is locally a direct product, globally it has a nontrivial fiber bundle structure over the base $S^2$.   Denote by $\mathcal{O}(n)$  the complex line bundle over $S^2$ with Chern class $n.$
The resolved conifold can then be identified with the rank-2 vector bundle: 
\begin{align} 
    \mathcal{O}(-1)  \oplus \mathcal{O}(-1)  \rightarrow S^{2}.
\end{align}
More generally we can consider A-model closed strings on geometries of the form
\begin{align}
 X&= L_{1} \oplus L_{2} \rightarrow \mathcal{S} ,
\end{align}
where $\mathcal{S}$ is a general Riemann surface, and $L_{1},L_{2}$ are line bundles with Chern classes $(k_{1},k_{2})$.\footnote{Strictly speaking, when $\mathcal{S}$ is contractible, the Chern class $c_1\in H^2(\mathcal{S})$ is always trivial. Hence, the Chern class cannot keep track of the bundle data required for the gluing. For a manifold with boundary, such as a disk, we instead use the euler class $e(L)\in H_2(\mathcal{S},\partial\mathcal{S})$ which equals to the Chern class upon gluing.}
It was shown in \cite{Bryan:2004iq} that the all genus amplitudes on such vector bundles satisfy the gluing rules of a TQFT, which can be viewed as the string field theory for the A-model.   
Formally, the A-model TQFT is a functor from the category  $2\Cob^{L_{1},L_{2}}$ of 2-dimensional cobordisms with line bundles to the category of vector spaces. Physically, it can be interpreted as an appropriate large $N$ limit of $q$-deformed 2D Yang Mills on the base manifold $\mathcal{S}$. Figure \ref{chart} gives a summary of the geometries and target space theories on both sides of the duality. 
\subsection{The Hartle Hawking state in string theory} 
In QFT, quantum states live on a codimension-1 time slice $\Sigma$.  
Geometric states are defined by the Euclidean path integral on a manifold $\mathcal{M}$ with $\partial\mathcal{M}=\Sigma$. In particular, the Hartle-Hawking state $\ket{HH}$ is defined cutting the spacetime geometry at a moment of time reflection symmetry\cite{PhysRevD.28.2960}. Thus the QFT partition function can be expressed as the overlap\footnote{ Following conventions in  TQFT,  the geometric dual $\bra{M}$  denotes the amplitude on the manifold $M$ with orientation reversed.   The braket is then just a gluing of manifolds, viewed as a natural pairing. In particular a TQFT does not assume a Hermitian inner product.} 
\begin{align} \label{Hartle} 
    Z_{QFT}= \braket{HH|HH}.
\end{align}

However, in closed string theories the fundamental degrees of freedoms are closed loops, so the field theory construction of geometric states doesn't strictly apply.  In the first quantized theory, a single string state is a wavefunctional on the space of loops 
\begin{align}
    \Psi[X^{\mu}(\sigma)],\quad X^{\mu}(\sigma) \in \mathcal{F }.
\end{align}
Here, elements of $\mathcal{F}$ are closed string configurations specified by the embedding map $X^{\mu}(\sigma)$, with $\sigma \in S^{1}$.   By direct analogy with QFT, the operators of the second quantized theory are obtained by promoting $\Psi$ to a string field
\begin{align}\label{sfield}
\hat{\Psi} = \hat{\Psi}[X^{\mu} (\sigma)] ,
\end{align}
which  is an operator-valued function on the loop space $\mathcal{F}$ of the target manifold.  This is in contrast to QFT where the second-quantized field operators are functions of spacetime points $X^{\mu}$.   Thus the degrees of freedom in string theory are labelled by elements of the loop space $\mathcal{F}$, and the specification of time slices refers to subsets of $\mathcal{F}$.

Similarly, the second quantized string Hilbert space is defined on a time slice of $\mathcal{F}$ rather than on a time slice $\Sigma$ of spacetime \cite{Witten:1985cc, witten1986interacting, 1993NuPhB.390...33Z}. Nevertheless, we can associate a string Hilbert space $\mathcal{H}_{\Sigma}$ with $\Sigma$ by a choice of mapping between time slices 
\begin{align}
\Sigma \to \mathcal{F}_{\Sigma} \subset \mathcal{F}.
\end{align} 
For example, if $\Sigma$ is given by $X^{0}=0$, we could define a time slice in the loop space by
\begin{align}
    \mathcal{F}_{\Sigma} = \{ X^{\mu}(\sigma): X^{0}(\sigma)=0 , \sigma \in S^{1} \}.
\end{align}
However, as noted in \cite{2018PhRvD..97f6025B}, the mapping between $\Sigma$ and $\mathcal{F}_{\Sigma}$ is not unique;  for example, we can   restrict only the center of mass of the string to live in $\Sigma$.
\begin{figure}[h] 
\centering
\includegraphics[scale=.4]{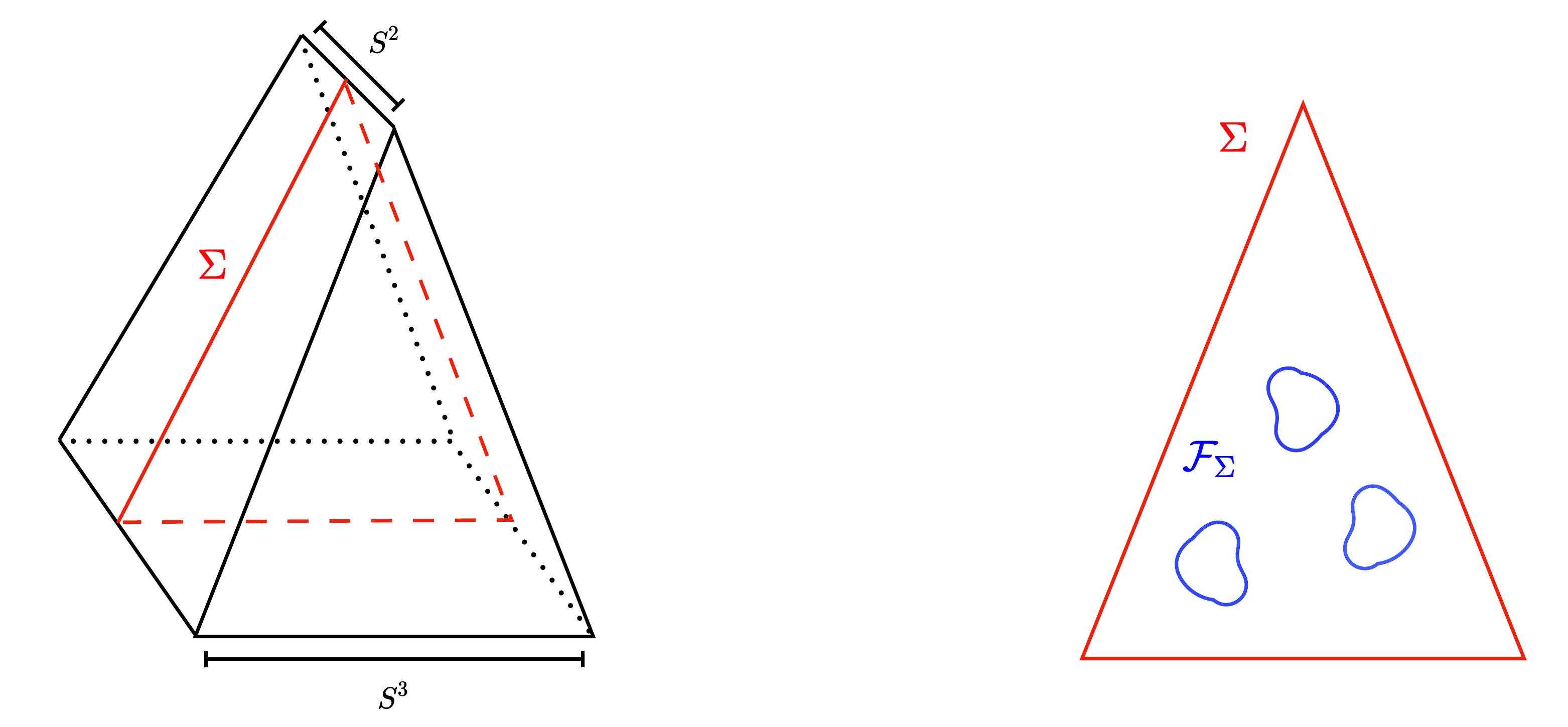}
\caption{The left figure shows the codimension-1 slice $\Sigma$ of the resolved conifold where a QFT state would be defined.  In the closed string theory, the analogue of a time slice is a set  $\mathcal{F}_{\Sigma}$ of loops configurations associated with $\Sigma$. For the A-model string, we will restrict these loop configurations to lie in a Lagrangian submanifold $\mathcal{L} \subset \Sigma$. The string wavefunctional assigns an amplitude to each configuration of such loops. } 
\label{sigma2}
\end{figure}  

To define the A-model Hilbert space, we choose $\Sigma$ to be the 5-dimensional region of the resolved conifold which intersects the base $S^{2}$ along the equator $C.$ This represents a symmetric cut through the target space geometry, and we would like to define the string theory analog of the Hartle-Hawking state on $\Sigma$. We  choose $\mathcal{F}_{\Sigma}$ to consist of noncontractible string loops living on a Lagrangian submanifold  $\mathcal{L} \subset \Sigma $.  A shown in figure \ref{Sigma}, this is a three-dimensional manifold with topology  $\mathbb{C} \times S^1$ and its defining equation is given in \eqref{lag} of appendix \ref{appB}. 
The topological vertex formalism \cite{Aganagic:2003db, 2008arXiv0809.3976M} can then be applied to define a string wavefunctional on $\mathcal{F}_{\Sigma}$ which gives a string theory analog of the Hartle-Hawking state. 

 The topological vertex encodes the A-model amplitude on $\mathbb{C}^3$ with three stacks of D-branes. It is a basic building block for partition functions on toric Calabi-Yau manifolds, such as the resolved conifold.   To compute the partition functions of more complicated geometries, one can glue the topological vertices by brane-antibrane annihilation. This gluing procedure allows us to cut and sew target space geometries as in Euclidean path integrals. In particular, we will define the Hartle-Hawking state $\ket{HH}$ using the topological vertex with a single stack of nontrivial D-branes on $\mathcal{L}$.   
 Denoting by $\bra{HH^*}$ the opposite vertex with antibranes inserted and opposite orientation, it can be shown that
\begin{align} \label{overlap} 
     Z = \braket{HH^*|HH},
 \end{align}
 where $Z$ is the partition function on the resolved conifold.  Note that that $Z$ is not a real norm of a state as in the QFT definition \eqref{Hartle}.   The is due to the holomorphic nature of the A model partition function, which is analogous to a chiral half of a conformal block.  From the point of view of $2\Cob^{L_{1},L_{2}} $, the HH state is given by a hemisphere with $(0,-1)$ Chern class, while $\bra{HH^*}$ is the oppositely oriented  hemisphere with $(-1,0)$ Chern class:
\begin{align}
\ket{HH}=\mathtikz{\etaC{0}{0};
\node at (0,3/5){(0,-1)}},    \quad \quad
\bra{HH^*}=\mathtikz{\epsilonC{0}{0};\node at (0,2/5){(-1,0)}}.
\end{align}
In our next paper we will derive the Chern-Simons dual of the HH state, which lives on the surface of a torus containing a specific superposition of Wilson loops inside. 
\subsection{Outline of the paper}
Our paper is organized as follows. We start with closed topological A model in section 3 and give a chiral boson description of the Hilbert space.  Using the topological vertex formalism, we obtain the Hartle-Hawking state of topological A model on the resolved conifold and compute its entanglement entropy using the geometric replica trick preserving the Calabi-Yau condition. We will also introduce the entanglement brane boundary state as a coherent state of chiral bosons.

In sections 4-5, we define a factorization of the closed string Hilbert space following the framework introduced in \cite{Donnelly:2018ppr}.  We first review the the relation between extended Hilbert space factorization and extensions of closed TQFT. We then present the A-model closed TQFT \cite{Bryan:2004iq} and propose an extension compatible with the entanglement brane axiom introduced in \cite{Donnelly:2018ppr}. The essential new ingredient in this factorization is the presence of an emergent quantum group symmetry which acts on the string edge modes. Compatibility with this symmetry leads us to modify the usual definition of Von Neumann entropy to: 
\begin{align}
S=- \tr_{q}( \rho \log \rho) =- \tr (D \rho \log \rho ),
\end{align} 
where the quantum trace $\tr_{q}$ is defined with the insertion of the  operator $D$, the Drinfeld element of a quantum group. We find that this definition of the entropy in the factorized Hilbert space agrees with the replica trick calculation in section 3.  This $q$-deformed notion of entropy has been studied previously in the context of quantum group invariant spin chains and non-unitary quantum systems \cite{Couvreur:2016mbr, Quella:2020aa}. $D$ is also the direct analogue of the defect operator introduced in \cite{Jafferis:2019wkd}\footnote{$\rho$ is equivalent to $\tilde{\rho}$ in \cite{Jafferis:2019wkd}. } to factorize the Hilbert space of JT gravity and in our case it is completely determined by the surface symmetry group.
In the end of section 5 we will revisit the geometric calculation of the replica trick and show how the preservation of the Calabi-Yau condition is enforced by the quantum trace. We will also show that the entanglement entropy has a natural interpretation in terms of the BPS microstate counting.

Finally, in the discussion section, we will comment on the relation between our work and factorization in JT gravity, particularly as it relates to the quantum group symmetry.

 \section{The closed string Hilbert space and entanglement entropy from the replica trick  } 
 \label{section:cboson}
  Topological string theory is a broad subject, so we will not try to give an extensive review in this paper. Nevertheless we give a short review on topological sigma model in Appendix A. In a similar spirit, we give a very short review on geometric transition between the deformed conifold and the resolved conifold in Appendix B. Curious readers may refer to \cite{hori2003mirror,Neitzke:2004ni,Marino:2004uf,Huang:2018tuh, Marino:2005sj, 2005hep.th....4147V}.
 \subsection{The Hartle-Hawking state from the all-genus amplitude} 
 Worldsheet topological string theory comes from applying topological twists to the $N=(2,2)$ supersymmetric sigma models, and the two inequivalent twisting procedures give the topological A-model and the topological B-model  \cite{Witten:1988xj}.
 In this paper, we will consider the A-model, whose target space is a six real dimensional Calabi-Yau manifold $X$.  The theory only depends on the Kahler modulus of the target space and is invariant under area preserving diffeomorphisms in the target space. 
 The free energy for the A model is a sum over all worldsheet instanton sectors corresponding to holomorphic worldsheets.   Let $[S_i]$ be a basis of $H_2(X,\mathbb{Z})$, so that a generic element $\beta \in H_2(X,\mathbb{Z})$ can be expressed as $\beta=\sum_{i} n_i [S_i]$.
Let $t_i=\int_{S_i} \omega$ be the complexified Kahler parameters and denote $Q^\beta = \prod_i e^{-n_i t_i}$.
The free energy of the A-model on $X$ then takes the form of a sum over all worldsheet instanton sectors:
\be\label{Free}
 F=\sum_{g} g_s^{2g-2} F_{g}(t_{i})=\sum_{g,\beta} g_s^{2g-2} N_{g,\beta} Q^{\beta}.
 \ee
$N_{g,\beta}$ is the genus-$g$ Gromov-Witten invariant that ``counts'' the number of holomorphic curves of genus $g$ in the two-homology class $\beta$ in an appropriate sense. 

A remarkable fact about the closed A-model string is that we can compute the all-genus amplitude using localization, connection to M-theory, mirror symmetry, and many other techniques \cite{Klemm:1999gm,Aganagic:2003db,Witten:1992fb,Gopakumar:1998ki,Dijkgraaf:2002fc,Aganagic:2003qj,Vafa:2004qa, 2008arXiv0809.3976M}. The free energy of the A-model can be resummed to be expressed in terms of the BPS index, Gopakumar-Vafa invariants $n_\beta^g$ for a curve $\beta,$ 
\begin{equation}
    F=\sum_{g,\beta,k} n_{\beta}^g \frac{1}{k}\left(2\sin \frac{kg_s}{2}\right)^{2g-2} Q^{k\beta}.
\end{equation}
In particular the partition function $Z=e^{F}$  on the resolved conifold is
 \begin{align} \label{ZA}
Z( \mathcal{O}(-1) \oplus \mathcal{O}(-1) \rightarrow S^{2})=\exp\left( \sum_{n=1}^{\infty}  \frac{1}{ n (2\sin (\frac{ng_{s}}{2} ))^{2}} e^{-n t} \right),
\end{align}
because $n_{S^2}^g=0$ for all $g>0$ but $n_{S^2}^0=1.$ 
In \eqref{ZA}, we have already summed over all genera. Note that although the partition function on the resolved conifold is well-defined for both weak and strong coupling, we presented an asymptotic form \eqref{ZA} which is valid for large values of the string coupling $g_{s}$. By expanding \eqref{ZA} in $g_{s}$, one can recover the free energy expression \eqref{Free} in terms of Gromov-Witten invariants, which is valid at weak string coupling. The $e^{-nt}$ factors correspond to holomorphic worldsheet instantons that wrap $n>0$ times  the base manifold $S^2$.  

\begin{figure}[h]
\centering
\includegraphics[width=14cm,height=10cm,keepaspectratio]{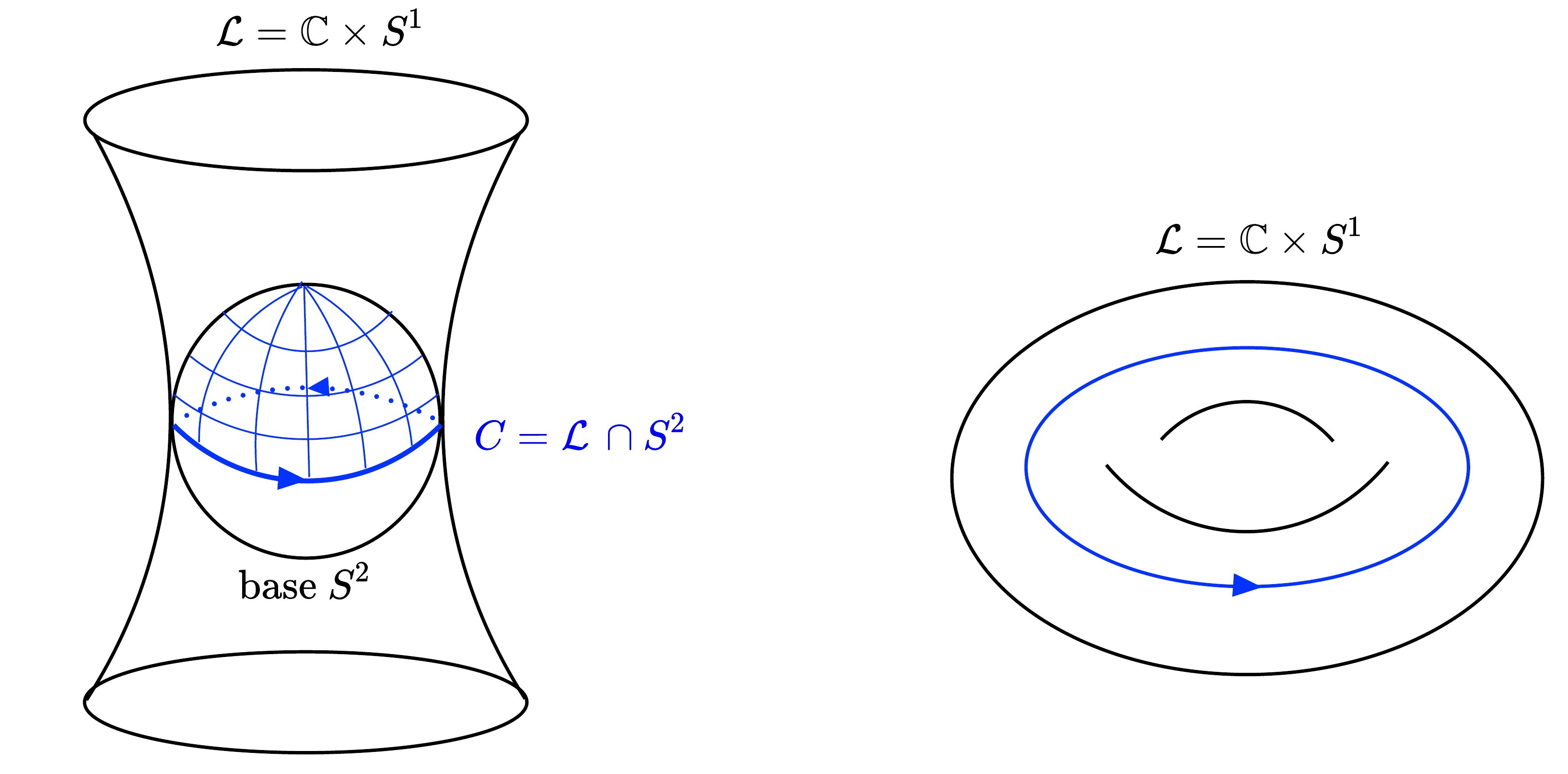}
\caption{The left figure shows a D-brane on $\mathcal{L} \subset \Sigma $ which intersects the base $S^2$ along the equator and extends in to the fiber directions along a hyperbola. In the right figure, we show the string loops in the time slice $\mathcal{F}_{\Sigma} $  which lives in $\mathcal{L}$.  The state $\ket{HH}$ state is defined via worldsheets which end on these loops and wrap the upper hemisphere, as shown in the left figure.  Similarly $\bra{HH^*}$ describes worldsheets on the southern hemisphere which end on anti-branes.} \label{Sigma}  
\end{figure}

As discussed in section \ref{section:out}, we want to define a Hartle-Hawking state for the resolved conifold as a wavefunctional of string loops inside the Lagrangian manifold $\mathcal{L}$.   To do this we apply the topological vertex formalism \cite{Vafa:2004qa} which express the string partition function \eqref{ZA} in terms of the overlap in  \eqref{overlap} by inserting branes and antibranes on $\mathcal{L}$. These branes cut through  the worldsheets along the equator while extending into the fiber directions as shown in figure  \ref{Sigma}.

Due to the coupling between the string endpoints and the branes, the A-model amplitude depends on the holonomy $U$ of the world volume gauge field. For the worldsheets ending on the branes and wrapping the upper hemisphere $D_{+}$, the amplitude is 
\begin{align}\label{TV}
Z_{+}( \mathcal{O}(0) \oplus \mathcal{O}(-1)\rightarrow D_{+}^{2}, U), \quad 
U= P\exp \oint_{C} A.
\end{align} 
 For $N$ branes the worldvolume gauge theory is $\U(N)$ Chern-Simons theory, and in the large $N$ limit the amplitude \eqref{TV} corresponds to the topological vertex with a single nontrivial stack of branes.  Similarly we can define another vertex via the amplitude $Z_{-}$ for the oppositely oriented worldsheets which  wrap the lower hemisphere and end on antibranes.  By  annihilating the brane-antibrane pairs, these vertices can be glued together to recover the partition function of the resolved conifold: 
\begin{align}
    Z( \mathcal{O}(-1) \oplus \mathcal{O}(-1) \rightarrow S^{2})= \int dU  \, Z_{+}(U) Z_{-}(U^{-1}).
\end{align}
Here the gluing is implemented by integration over the gauge group $\U(N)$ using the Haar Measure.

The vertex $Z_{+}$ can be interpreted as a closed string wavefunctional for the Hartle-Hawking state:
\begin{align}
\braket{U|HH} =  Z_{+}(U).
\end{align} 
$\ket{HH}$ is a state in the second-quantized string theory, and the path integral $Z_+$ includes all disconnected worldsheet configurations winding an arbitrary number of times in one orientation.  Each closed string configuration is defined by the occupation numbers $k_{j}$ which enumerate how many closed strings wind $j$ times around $C$, which should be identified with the non-contractible cycle of $\mathcal{L}$.

In terms of $k_{j}$, the wavefunctional is of the form  
\begin{align}
\braket{U|HH}\label{eqn:HH1}
&= \sum_{\vec{k}  } C_{00\vec{k}} (g_{s},t )  \prod_{i=1} \frac{ (\tr U^{i})^{k_{i}} }{z_{\vec{k}}} ,\quad k_{i}>0\nn 
z_{\vec{k}} &=  \prod_{j=1}^{\infty}  j^{k_{j}} k_{j} !,
\end{align}
where $C_{00\vec{k}} (g_{s},t ) $ are the vertex coefficients derived in \cite{Aganagic:2003db, 2008arXiv0809.3976M}, and the normalization $z_{\vec{k}}$ is a combinatorial factor associated with redundancy in labelling by $\vec{k}$.
\eqref{eqn:HH1} can be also derived by calculating the open Gromov-Witten invariants from the worldsheet theory \cite{2001math......3074K, 2008arXiv0809.3976M}. In the large $N$ limit the multi-trace factors form a linearly independent set called the winding basis $\ket{\vec{k}}$ 
\begin{align} \label{winding}  
  \lim_{N\rightarrow \infty} \prod_{i=1} \tr (U^{i})^{k_{i}}= \braket{U|\vec{k} } ,\quad k_{j} >0.
\end{align}
The overlap between the states $\ket{\vec k}$ is defined via the Haar measure $dU$:
\begin{align} \label{ip}
\braket{\vec{k}|\vec{k}'}=\int d U  \tr_{\vec{k}}(U)\tr_{\vec{k'}}(U^{-1}) = \delta_{\vec{k},\vec{k}'} z_{\vec{k}} .
\end{align}
This basis defines the chiral closed string Hilbert space $\mathcal{H}_{\Sigma}$ associated with strings winding around the equator.\footnote{
This Hilbert space can be viewed as a ``chiral half'' of the space of functions on $\U(N)$ in the large $N$ limit in the following sense.
In the large $N$ limit, the Hilbert space of functions on $\U(N)$ factorizes into two sectors consisting of positively oriented strings represented by wavefunctions $\tr(U^k)$ and negatively oriented strings represented by wavefunctions $\tr(U^{\dag^k})$ \cite{Gross:1993hu}.
The Hilbert space we consider consists only of the positively-oriented strings.} 

So far we have expressed the closed strings states as functions of the holonomies $U$.  Let us interpret these explicitly as  wavefunctionals of loops in $\mathcal{F}_{\Sigma}$.  Due to the topological invariance of the A-model, elements of $\mathcal{F}_{\Sigma}$  fall into equivalence classes labelled by their winding numbers around $C$.  If $X_{n}(\sigma)$ is a string loop winding $n$ times, then each single trace factor in \eqref{winding} should be treated as a single string functional:
\begin{align}
\Psi[X_{n}(\sigma)]  =\tr(U^{n}) = \tr P \left(\exp \oint X_{n}^{*}A \right).
\end{align}
Similarly, the Hartle-Hawking state is a multi-string functional obtained by treating multi loop configuration in $\mathcal{F}_{\Sigma}$ as boundary conditions for the string path integral.  
\subsection{The chiral boson description of $\mathcal{H}_{\Sigma}$ and D branes}  The Hilbert space  $\mathcal{H}_{\Sigma}$ has a second quantized description in terms of a chiral boson which can be viewed as a string field theory for the A-model.\footnote{This is the Hilbert space associated with the representations of $U(\infty)$. Strictly speaking, this is the string field theory for the topological B-model on the mirror manifold \cite{Aganagic:2003db, Bershadsky:1993cx,  Aganagic:2003qj}.} This is obtained by defining string creation/annihilation operators $\alpha_{\mp n} , n>0$ which create and annihilate closed strings winding $n$ times with positive orientation \cite{Bershadsky:1993cx,Aganagic:2003qj,Aganagic:2003db}: 
\begin{align}
\bra{U} \prod_{n} \alpha^{k_{n}}_{- n} \ket{0} = \prod_{n} \tr (U^{n})^{k_{n}}.
\end{align}  
In terms of these oscillators, the D-branes $ \ket{U}$ are coherent states
\begin{align}
\ket{U}= \exp \left( \sum_{n=1}^{\infty} \frac{ \text{tr}(U^n)}{n} \alpha_{-n} \right) \ket{0}.
\end{align}
This gives a more precise definition of $\ket{U}$ in the large $N$ limit, as we can apply the mapping 
 \begin{align}\label{U}
\ket{U} \rightarrow \ket{t}&=\exp \left( \sum_{n=1}^{\infty} \frac{t_{n}}{n} \alpha_{-n} \right) \ket{0}, \quad t_{n}= \tr U^n.
\end{align} 

In the large $N$ limit, $t_n$ can be viewed as formal variables without reference to the matrix $U$.
In particular, the HH state is given by evolution of such a coherent state \cite{Aganagic:2003db} 

\begin{align}\label{HHO}
\ket{HH} &=e^{-t\hat{H}/2} \exp \left( \sum_{n=1} \frac{1}{n (q^{n/2}-q^{-n/2} )} \alpha_{-n} \right) \ket{0}, \\
\hat{H}&=\sum_{n=1}^{\infty} \alpha_{-n} \alpha_{n},\quad q= \exp(i g_{s}) ,
\end{align} 
where $\hat{H}$ is the Hamiltonian of the closed string field theory and $e^{-\frac{t}{2} \hat{H}}$ is a string field propagator which evolves the geometry from an infinitesimal disk to a finite hemisphere of area $t/2$.  The dual state defined by the amplitude $Z_{-}$ with antibranes inserted is given by\footnote{This is a nontrivial adjoint operation which corresponds to changing the Chern class in addition to changing the orientation of the hemisphere\cite{Aganagic:2003db, Bershadsky:1993cx, 2001hep.th....1218V}. When $t$ is real, this is equal to the complex conjugation. When $t$ is complex, due to the holomophicity of the A-model, we shall not use the complex conjugation and our formula for the dual is correct for a generical complex $t$. }
\begin{align} \label{HH*}
\bra{HH^*} =\bra{0}\exp \left( \sum_{n=1} \alpha_{n} \frac{-1}{n (q^{n/2}-q^{-n/2} )} \right) e^{-t\hat{H}/2}.
\end{align} 
It can be verified directly from \eqref{HHO},\eqref{HH*},and \eqref{ZA}, that 
\begin{align}
Z=\braket{HH^*|HH} .
\end{align} 
\paragraph{The entanglement brane boundary state}
It is useful to identify the holonomy $D \in U(\infty)$ corresponding to the state on the infinitesimal disk :
\begin{align}
\ket{U=D} =\exp \left( \sum_{n=1} \frac{1}{n (q^{n/2}-q^{-n/2} )} \alpha_{-n} \right) \ket{0} .\label{eqn:e brane boundary state}
\end{align} 
From \eqref{U}, we know $D$ must satisfy 
\begin{align}
\tr D^n = \frac{1}{ (q^{n/2}-q^{-n/2} )}= \frac{1}{[n]_{q}},
\end{align} 
where we have introduced the $q$-deformed integer $[n]_{q}$.
 A diagonal matrix that satisfies this equation in the $N \rightarrow \infty$ limit  
has components: 
\begin{align}\label{D}
D_{jj}=  q^{-j+\frac{1}{2}},\quad j=1,\cdots N .
\end{align}
Deriving this holonomy requires a regularization of the trace. Note that
\begin{align} \tr D^n = \sum_{j=1}^{N} q^{n(-j+\frac{1}{2} )} = q^{-n/2} \sum_{j=0} (q^{-n})^{ j}=\frac{1-q^{-n(N+1) }}{q^{n/2}-q^{-n/2}},
\end{align} 
so we need to give $g_{s}=-i \log q$ a small imaginery part for the sum to converge.   This analytic continuation is possible because in topological string theory $g_{s}$ is a formal expansion variable rather than a physical coupling. 
In terms of $\ket{D}$ we can write $Z$ as a propagation amplitude
\begin{align} \label{amp}
Z&= \bra{D^*} e^{ - \hat{H}  t }  \ket{D}.
\end{align}

We will show in section \ref{section:open} that the state $\ket{D}$ is the analogue of the ``entanglement brane" boundary state described in \cite{Donnelly:2016jet}, and the holonomy $D$ determines the corresponding entanglement boundary condition.  
We can compute the amplitude \eqref{amp}  in the winding basis using the overlaps:
\begin{align} \label{kD}
\braket{\vec{k}|D} &= \prod_{n=1}  (\tr D^n)^{k_{n}} =\prod_{n=1}\left(\frac{1}{e^{i g_{s} n/2}-e^{-i g_{s}n/2}} \right)^{k_{n}},
 \end{align}
which gives another expression for the partition function:
 \begin{align}
     Z&=\sum_{\vec{k}}  (d_{q}(\vec{k}, g_{s}))^{2} e^{ - l(k)  t },\quad l(\vec{k}) = \sum_{j} j k_{j}, \nn
     (d_{q}(\vec{k}, g_{s}))^{2} &=  \frac{1}{z_{\vec{k}}} |\braket{k|D}|^{2}=\prod_{n=1} \frac{1}{z_{\vec{k}}}  (|[n]^{-1}_{q}|)^{2 k_{n}}.
 \end{align}
If we interpret $Z$ as a statistical partition function with Boltzmann factor $e^{ - l(k)  t }$, this expression suggests that $ (d_{q}(\vec{k}, g_{s}))^{2}$ is a degeneracy factor.  A small $g_{s}$ expansion of \eqref{kD} then shows that 
\begin{align}
 (d_{q}(\vec{k}, g_{s}))^{2}  \sim \prod_{n} \left( \frac{1}{g_{s}} \right)^{2 k_{n}}.
\end{align} 
We will see that this factor leads to a large $ O(\frac{1}{g_{s}})$ number of microstates per open string endpoint, as alluded to in the introduction.
The appearance of the quantum integers $[n]_{q}$ indicates an emergent quantum group symmetry in the target space. In the next subsection we will see addition evidence of this symmetry in the structure of the entanglement entropy as computed by the replica trick.
  
\paragraph{Boson representation of the topological vertex} As a final remark, we note that in the chiral boson language, the topological vertex can be viewed as a highly nontrivial choice of the ``pair of pants" amplitude. This is  a state $ \ket{\mathcal{V}} \in \mathcal{H}_{\Sigma}\otimes \mathcal{H}_{\Sigma}\otimes \mathcal{H}_{\Sigma} $. It's
wavefunction in the coherent state basis is defined by
\begin{align}
\braket{U_{1},U_{2},U_{3}|\mathcal{V}} =Z_{\mathbb{C}^3}, 
\end{align} 
where $Z_{\mathbb{C}^3}$ is the A-model  amplitude on $\mathbb{C}^3$ with  3 stacks of D-branes with holonomies $U_{i}, i=1,2,3$.  It is in this sense that the states $\ket{U}$ corresponds to degrees of freedom on A-branes.  In terms of the vertex, the Hartle-Hawking state in \eqref{eqn:HH1} is 
\begin{align}
   \braket{U|HH}=\bra{U}\otimes \bra{0}\otimes \bra{0}\ket{\mathcal{V}},
\end{align} where $\ket{0}$ corresponds to the state with no strings. 
\subsection{Entanglement entropy from the replica trick}
\label{ssection:replica}
In string field theories, an entanglement partition corresponds to a cut in the space $\mathcal{F}_{\Sigma}$ of field configurations. Given a spatial partition $\Sigma= \Sigma_{A}\cup \Sigma_{ B}$, one can consider string configurations $\mathcal{F}_{\Sigma_{A}}$ and $ \mathcal{F}_{\Sigma_{B}}
$, define the respective string Hilbert spaces $\mathcal{H}_{\Sigma_{A}} $ ,$\mathcal{H}_{\Sigma_{B}} $, and define the factorization map 
\begin{align}
\mathcal{H}_{\Sigma} \rightarrow \mathcal{H}_{\Sigma_{A}} \otimes \mathcal{H}_{\Sigma_{B}}.
\end{align} 
However, here we will bypass this procedure and apply the replica trick as suggested by Susskind and Uglum \cite{Susskind:1994sm}. We choose $\Sigma_{A}$ to be the subregion fibered over an arc $A \subset C$ of the equator, and $\Sigma_{B}$ to be region over the complementary arc $B$. The entangling surface is a codimension-2 surface fibered over two points on $C$ and separates the Lagrangian manifold $\mathcal{L}$ into two pieces,  cutting the closed strings winding around the equator into two open strings.

To apply the replica trick we have to compute the A-model partition function $Z(\alpha)$ on the $\alpha$-fold replicated geometry with opening angle $2\pi \alpha $ around $\pd \Sigma_{A}$. As we will show later, the replication can be applied in a way that preserves the bundle structure and the Calabi-Yau condition. As the topological A-model is invariant under area preserving diffeomorphisms, the replicated manifold thus remains $\mathcal{O}(-1) \oplus \mathcal{O}(-1) \rightarrow S^{2}$  with the volume rescaled by a factor of $\alpha$. The replica partition function is thus: 
\begin{align} \label{Zalpha}
Z(\alpha)= \sum_{k} (d_{q}(\vec{k},g_s))^{2} e^{-\alpha l(k)t} ,
\end{align} 
which gives the entanglement entropy:
\begin{align}\label{KEE}
    S_{\text{replica}}&=(1-\alpha \pd_{\alpha})|_{\alpha = 1} \log Z(\alpha)= \sum_k p(k)(-\ln{p(k)}+2 \ln d_{q}(\vec{k},g_s)),\nn
     p(k) &= \frac{(d_{q}(\vec{k},g_s))^2 e^{- t l(k) }}{Z}.
\end{align}

This formula is reminiscent of entanglement entropy in gauge theories.  To make this analogy more precise, we compute the amplitude expression \eqref{amp} for $Z(\alpha=1)$ in the representation basis.  At finite $N$, these basis elements $\ket{R}$ are defined by characters of $U(N)$.
\begin{align}
\braket{U|R} &=\tr_{R}(U), \nn
\braket{R'|R}&=\int dU \tr_{R'}(U^{-1}) \tr_{R}(U) =\delta_{RR'},
\end{align}
where $R$ labels irreducible representations(irreps) of $U(N)$.  They are related to the winding basis by the Frobenius relation
\begin{align} \label{frob}
    \ket{R}= \sum_{\vec{k} \subset S_{n}} \frac{\chi_{R}(\vec{k})}{z_{\vec{k}}}   \ket{\vec{k}}.
\end{align}
Here each $R$ is identified with a Young diagram with $n$ boxes, and $\chi_{R}(\vec{k})$ is the character of the symmetric group $S_{n}$ associated with the diagram.
In the $N \to \infty$ limit we take the expression on the RHS (which is independent of $N$) as a definition of $\ket{R}$.  
This limit captures states $\ket{R}$ whose diagrams have columns of arbitrary length.\footnote{This only captures a chiral half of the Hilbert space because it misses the representations obtained by tensoring anti-fundamental representations of $\U(N)$. }

In the representation basis we have 
\begin{align}
    \braket{R|D} = \tr_{R}(D)=(-i)^{l(R)}d_{q}(R) q^{\kappa_{R}/4}.
\end{align}
where $l(R)$ is the number of boxes in the Young diagram. The quantity $d_{q}(R)$ is the quantum dimensions of the symmetric group representation $R$. In term of the Young diagram, $d_q(R)$ is given by
\begin{align}
d_{q}(R) &= \prod_{\Box \in R}  \frac{i}{q^{ h(\Box)/2}-q^{- h(\Box)/2}}=\prod_{\Box  \in R} \frac{1}{2 \sin(\frac{h(\Box) g_s}{2})},
\end{align}
with $h(\Box)$ being the hook length, and the phase $q^{ \kappa_{R}/4}$ is given by
\begin{align}
    \kappa_{R}& =2 \sum_{\Box \in R} (i (\Box ) -j(\Box ) ),
\end{align}
here $ i (\Box ), j(\Box ) $  are the row and column numbers of the box.

It will be useful to view these quantities as arising from a particular large $N$ limit of the quantum dimensions $\dim_{q}(R)$ for $U(N)_{q}$: 
 \begin{align} \label{ldim} 
     \lim_{N\rightarrow \infty}  q^{-Nl(R)/2} \dim_{q}(R) =  (-i)^{l(R)}d_{q}(R) q^{\kappa_{R}/4},
 \end{align}
 where the prefactor $ q^{-Nl(R)/2} $ renormalizes the quantum dimension for $U(N)_{q}$, rendering it finite in the large $N$ limit. As we will show later, this is the same regularization used to determine the matrix $D$ in \eqref{D}. 
 
In the representation basis, the Hartle-Hawking state \eqref{HHO} can be written as    \begin{align}\label{JT1}
        \ket{HH}&=  \sum_{R}  e^{- t\hat{H}/2} \ket{R}\braket{R|D}= \sum_{R}d_{q}(R) (-i)^{l(R)}q^{\kappa_{R} /4}e^{-t l(R)/2 } \ket{R}, 
    \end{align}
and the partition function on the resolved conifold is
\begin{align}\label{Z1}
    Z= \sum_{R}  (d_{q}(R))^{2} e^{-t l(R)}.
\end{align}
Equations \eqref{JT1} and \eqref{Z1} are direct analogues of formulas for the Hartle-Hawking state in two dimensional gauge theories as well as in JT gravity.  Together with \eqref{ldim},  they suggest that  $(d_{q}(R))^2$ is a degeneracy factor due to a quantum group symmetry associated with the large $N$ limit of $\U(N)_{q}$. 

Applying the replica trick to \eqref{Z1} gives another expression for the entropy \eqref {KEE}:
   \begin{align}\label{RRep} 
       S_{\text{replica}}&=(1-\alpha \pd_{\alpha})|_{\alpha = 1} \log Z(\alpha),\nn
       &= \sum_R p(R)(-\ln{p(R)}+2 \ln d_{q} (R)),\quad
     p(R) = \frac{(d_{q}(R))^2 e^{- t l(R) }}{Z}.
   \end{align}
This is a direct analogue of the entropy in 2D nonabelian gauge theories \cite{Donnelly:2011hn,Gromov:2014kia,Donnelly:2014gva} with $R$ playing the role of representation labels for a surface symmetry,  $p(R)$ a probability factor, and $d_{q}(R)$ the dimension of each representation.  Indeed it can be shown \cite{secondpaper} that the Hartle-Hawking state and its entropy is a large $N$ limit of
\begin{align}
  \ket{HH}&=  \sum_{R}\dim_{q}(R) e^{-t l(R) } \ket{R},\nn
     S_{\text{replica}}(N) &= \sum_R p(R)(-\ln{p(R)}+2 \ln \dim_{q} (R)) ,\quad 
     p(R) = \frac{(\dim_{q}(R))^2 e^{- t l(R) }}{Z},
\end{align}
which are the Hartle-Hawking state and entropy for $q$-deformed 2DYM.  In the context of the $q$-deformed 2d Yang-Mills, the limit \eqref{ldim} has a very natural explanation.  Rather than removing the $N$ dependence of $\dim_{q}(R)$ by hand, we should view this as a renormalization procedure in which the divergent term $q^{Nl(R)/2}$ is absorbed into the Boltzmann factor $e^{-t l(R)}$.   The divergence arises due to the analytic continuation of $q$, and has precisely the right form so that it can be absorbed into a redefinition of the ``coupling" $t$.  In out next paper \cite{secondpaper}, we will explain this renormalization from the point of view of the geometric transition.

Given this limit, we expect that $2 \log d_{q}(R)$  has a state counting interpretation in terms of edge modes transforming in an irrep of a surface symmetry group.  
This symmetry group has been $q$-deformed, leading to quantum dimensions which do not have to be integers.

\section{The A-model closed TQFT and representation category of quantum groups } 
\label{section:closed} 
\subsection{General comments about factorization, E-brane axiom, and cobordisms  }
\begin{figure}[h]
\centering
\includegraphics[width=14cm,height=10cm,keepaspectratio]{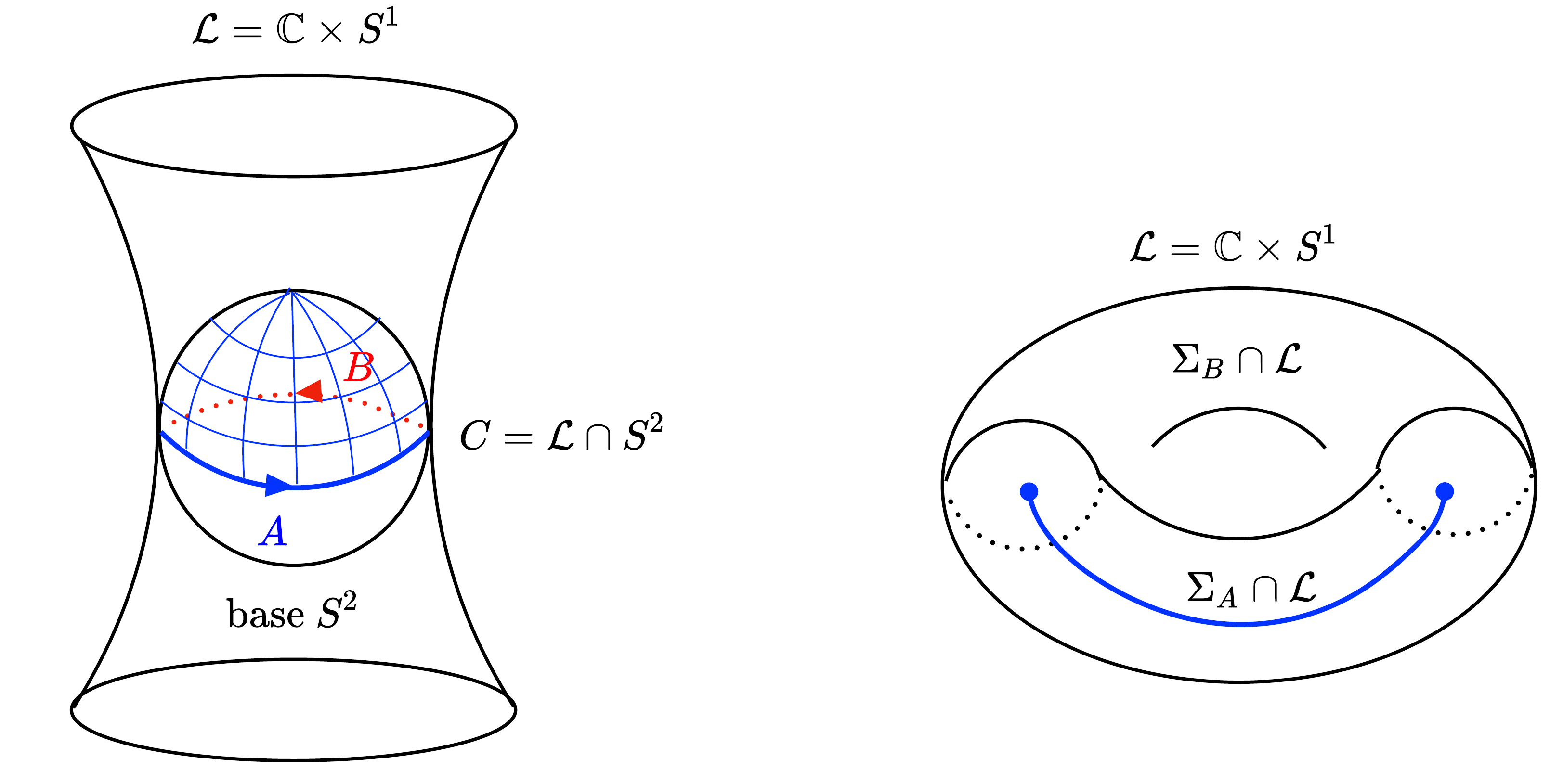}
\caption{On the left, we show the splitting of the worldsheet boundary into $A$  and $B$.   On the right, the brane $\mathcal{L}$ on which the closed string configurations $X(\sigma)$ live is split into subregions by the entanglement branes.  We show an open string configuration $X_{ij}(\sigma) \in \mathcal{F}_{\Sigma_{A}} $. These end on the entanglement branes intersecting $\mathcal{L}$ along two open disks.} \label{HHWS}
\end{figure}

In the following two sections, we give a  canonical interpretation of the replica trick entropy in \eqref{RRep} by defining a factorization of the closed string Hilbert space  $\mathcal{H}_{\Sigma}$ associated with the decomposition $\Sigma=\Sigma_{A} \cup \Sigma_{B}$ into the subregions. The intersection of these subregions with $\mathcal{L}$ are shown in the right of figure \ref{HHWS}. We start by defining the spaces  $\mathcal{F}_{\Sigma_{A}} $, $\mathcal{F}_{\Sigma_{B} }$ of string configurations associated  to these subregions. These spaces contain open string configurations $X_{ij}(\sigma)$ inside $\mathcal{L} \cap \Sigma_{A}$ which are stretched between entanglement branes (E-branes) which cut $\mathcal{L}$ in two disconnected slices. The E-branes wrap a submanifold \footnote{We expect $\mathcal{L}'$ to be a Lagrangian submanifold. see comments in the discussion section . } $\mathcal{L'}$ that intersects the base $S^2$ along a circle orthogonal to $C$.  
The indices $i,j$ are Chan-Paton factors labelling the $N \gg 1$ E-branes, which can be identified with the entanglement edge modes of the closed string.   We will give an explicit description of the open string Hilbert space $\mathcal{H}_{ \Sigma_{A}},\mathcal{H}_{ \Sigma_{B}}$ and the factorization map
\begin{align}\label{factor} 
\mathcal{H}_{\Sigma} \rightarrow \mathcal{H}_{\Sigma_{A}} \otimes \mathcal{H}_{\Sigma_{B}}.
\end{align}   This mapping embeds the Hilbert space of closed strings into an extended Hilbert space of open strings.      

Just as in QFT, the factorization problem is strongly ambiguous in the absence of locality constraints. For example, as noted in \cite{Jafferis:2019wkd}, we can always map the physical states into a maximally entangled state of some arbitrary extended Hilbert space, leading to an arbitrarily large entanglement entropy.  When the locality constraints are available, the strongest form of such constraints come from using the Euclidean path integral to split a time slice into subregions.  In 2 dimensions, such a factorization of a circle or an interval is obtained from the Euclidean evolution (read from top to bottom)
\begin{align} \label{splitting}
\mathtikz{ \zipper{0cm}{0cm} }: \mathcal{H}_{\text{circle}} \rightarrow \mathcal{H}_{\text{interval}}.\quad 
\mathtikz{ \deltaA{0cm}{0cm} }:
\mathcal{H}_{\text{interval} } \rightarrow \mathcal{H}_{\text{interval}} \otimes \mathcal{H}_{\text{interval}}
\end{align}   
with some appropriate choice of boundary conditions at the entangling surface.  In the previous work \cite{Donnelly:2018ppr}, we introduced a constraint on these factorization maps called the entanglement-brane (E-brane) axiom \eqref{holes}, which ensures that the factorized state preserves all the correlations of the original state.  This requires that \emph{all} holes traced out by the entangling surface can be closed up.  For example, we require that the cobordisms\footnote{The right diagram of \eqref{holes} was was refered to as isometry condition and employed to study factorization in JT gravity in \cite{Jafferis:2019wkd}.  It is one of the axioms of a ``special"  Frobenius algebra.} in \eqref{splitting} satisfy
\begin{equation}\label{holes}
\mathtikz{ \cozipper{0cm}{0cm} \zipper{0cm}{1cm} } 
= \mathtikz{ \idC{0cm}{0cm} },
\qquad
\mathtikz{ \muA{0cm}{0cm} \deltaA{0cm}{1cm} } = \mathtikz{ \idA{0cm}{0cm} } .
\end{equation}
This ensures that splitting the state does not change its correlations, since we can fuse it back and obtain the identity map by allowing the hole to contract.   

The E-brane axiom, generally requires that the factorization involves a sum over edge modes at the entangling surface. It axiomatizes the state counting interpretation of the replica trick entropy.   The replica trick, in both gravity and QFT, involves a path integral $Z(\alpha)$ on a background with a contractible circle around the entangling surface.   However a thermal interpretation 
\begin{align}
Z(\alpha)=\tr_{V} e^{-\alpha H}
\end{align}
requires a path integral in a background with a \emph{non-contractible} circle.  The E-brane axiom enforces the non-trivial requirement that these two are equal:
\begin{align}
\mathtikz{\epsilonC{0}{0}
\etaC{0}{0}}  = \mathtikz{\pairA{0}{0}\copairA{0}{0}}.
\end{align}
Previous works in gauge theory have shown that this can be satisfied provided we introduce appropriate edge modes into the Hilbert space of the subregion $V$ \cite{Donnelly:2018ppr}.

Unfortunately, demanding a path integral formulation of the target space physics is an overly restrictive requirement; in particular it is not generally  a useful assumption in CFT's or in string theory. However there is a categorical reformulation of the path integral in terms of cobordisms which does not presume a notion of path integration over local fields. From the categorical point of view, a path integral for a $D$-dimensional Euclidean theory is a rule which assigns a number (the partition function)  to a $D$-dimensional manifold, a Hilbert space to $D-1$ manifolds,  and linear maps to cobordisms, which are $D$-dimensional manifolds with ``initial" and ``final" boundaries.  Gluing of cobordisms along initial and final boundaries corresponds to composition of linear maps. The standard example of such a cobordism theory is a \emph{closed} 2D TQFT in which a Hilbert space $V^{\otimes{n}}$ is assigned to a disjoint union of $n$ circles,  and linear maps are assigned to cobordisms interpolating between collections of circles.  The theory on an arbitrary closed Riemann surface can then be constructed by gluing the basic cobordisms \cite{kock2004frobenius,atiyah1988topological}:
\begin{align}\label{data}
\mathtikz{\muC{0}{0}},  \quad \mathtikz{\deltaC{0}{0}}, \quad \mathtikz{\etaC{0}{0}}, \quad \mathtikz{\epsilonC{0}{0}}.
\end{align}
Consistency of different gluings for the same manifold is enforced by a set of sewing axioms which provide strong constraints on the cobordism data \eqref{data}. For a 2D TQFT, the resulting structure is a Frobenius algebra with multiplication defined by the pair of pants cobordism.  A similar formulation can be applied to 2D gauge theories and 2D conformal field theories \cite{2019arXiv191211201H}.

In the categorical framework \cite{Donnelly:2018ppr}, the path integral factorization maps \eqref{splitting} are viewed as additional cobordism data that defines an open  \emph{extension} of the closed TQFT.  This extension introduces  Hilbert spaces associated with codimension-one manifolds with boundaries (i.e. intervals) and additional set of cobordisms 
\begin{align}
    \mathtikz{\muA{0}{0}}, \quad \mathtikz{\deltaA{0}{0}}, \quad \mathtikz{ \etaA{0}{0} }, \quad \mathtikz{ \epsilonA{0}{0} },   \quad \mathtikz{\zipper{0}{0}}, \quad \mathtikz{\cozipper{0}{0}} .
\end{align}
which must be compatible with \eqref{data} according to the sewing relations 
\begin{align}\label{msegal}
\mathtikz{ \zipper{0cm}{1cm} \etaC{0cm}{1cm} } = 
\mathtikz{ \etaA{0cm}{0cm} } , \mathtikz{ \zipper{0cm}{0cm} \muC{0cm}{1cm} } = 
\mathtikz{ \muA{0cm}{0cm} \zipper{-0.5cm}{1cm} \zipper{0.5cm}{1cm} } ,
\quad \mathtikz {
\pairA{0cm}{0cm}
\zipper{-0.5cm}{1cm} \idA{0.5cm}{1cm} }
 = 
\mathtikz {
\pairC{3cm}{0cm}
\idC{2.5cm}{1cm} \cozipper{3.5cm}{1cm} },
\end{align}
\begin{align}
\mathtikz{\muA{0cm}{0cm} \zipper{-0.5cm}{1cm} \idA{0.5cm}{1cm} }
\quad = \mathtikz{  \muA{3cm}{-0.5cm} \tauA{3cm}{0.5cm} \zipper{2.5cm}{1.5cm} \idA{3.5cm}{1.5cm} }, \quad \mathtikz{ \deltaA{0cm}{0cm} \idA{0.5cm}{-3cm} \idA{0.5cm}{-2cm}\SA{-0.5cm}{-3cm} \SA{-0.5cm}{-2cm} \muA{0cm}{-4cm} \tauA{0cm}{-1cm} }=\mathtikz{ \zipper{0cm}{0cm} \cozipper{0cm}{1cm} },
\end{align}
that defines an open-closed TQFT \cite{Lazaroiu:2000rk,Moore:2006dw}. 

It was shown in \cite{Donnelly:2018ppr} that the sewing axioms for an open-closed TQFT can be consistently combined with the E-brane axiom: 
\begin{align} \label{Ebrane}
     \mathtikz{\etaC{0cm}{0cm} } = \mathtikz{ \etaA{0cm}{0cm} \cozipper{0cm}{0cm}
\draw (0cm,0.5cm) node {\footnotesize $e$};
\draw (0cm,-0.25cm) node {\footnotesize $e$};
},
\end{align}
to give a complete set of locality constraints that a consistent factorization should satisfy.  As explained in \cite{Donnelly:2018ppr}, when combined with \eqref{msegal} equation \eqref{Ebrane} is powerful enough to ensure that \emph{all} holes traced out by the entangling surface can be closed.  A solution to all of these constraints was given for 2DYM and its string theory dual, and led to a factorization consistent with the replica trick entropy.   

In the next 2 sections, we will apply the approach described above to define the factorization of the A-model string theory.   It was shown in \cite{Aganagic:2004js,Bryan:2004iq} that the closed string amplitudes on direct sums of line bundles 
\begin{align}
X&= L_{1} \oplus L_{2} \rightarrow \mathcal{S},
\end{align}
over a Riemann surface  $\mathcal{S}$  can be determined by a closed TQFT on $ 2\Cob^{L_{1},L_{2}} $.   This means that the A-model amplitudes on $X$ can be broken up into open string amplitudes by inserting brane-antibrane pairs as in our construction of the Hartle-Hawking state, and the gluing of these open string amplitudes satisfies the same rules as the category $2\Cob^{L_{1},L_{2}} $ of 2-cobordisms with line bundles. The A-model TQFT \cite{Bryan:2004iq} is a generalization of 2D TQFT, with the information about the higher-dimensional geometry captured by Chern classes $(k_{1},k_{2})$ of the line bundles $L_{1}, L_{2}$. It is generated by cobordisms in \eqref{data} with $(0,0)$ Chern class, together with the following four cobordisms
\begin{align}
\mathtikz{\etaC{0}{0};
\node at (0,1/2){(-1,0)};}, \quad \mathtikz{\etaC{0}{0};
\node at (0,1/2){(0,-1)};}, \quad \mathtikz{\etaC{0}{0};
\node at (0,1/2){(1,0)};}, \quad \mathtikz{\etaC{0}{0};
\node at (0,1/2){(0,1)};} .
\end{align}
Note that this generates a much larger category than the set $2\Cob$ of two-dimensional cobordisms, and the A-model TQFT has a more complicated set of sewing relations than an ordinary Frobenius algebra. However, in formulating the factorization of the A-model Hilbert space, we will restrict to target spaces which are Calabi-Yau manifolds.  This requires the Chern classes to satisfy
\begin{align}\label{CY}
k_{1} + k_{2}&= -\chi(\mathcal{S}),
\end{align}
where $\chi(\mathcal{S})$ is the Euler characteristic of the base manifold.  This is an important restriction that determines the form of the factorization map which we will propose.

\subsection{A model TQFT on Calabi Yau manifolds} 
The subcategory of $2\Cob^{L_{1},L_{2}} $ corresponding to Calabi-Yau manifolds defines a symmetric Frobenius algebra just like a 2D TQFT. The basic building blocks for this category are the same as the generators  in \eqref{data}, except they are now decorated by Chern class labellings satisfying \eqref{CY}.      Since both the Chern classes and the Euler characteristic of the base manifolds are additive under gluing, the Calabi-Yau condition \eqref{CY} is preserved under gluing.  The A-model TQFT is a functor which assigns a linear map to each generators \cite{Aganagic:2004js} :\begin{align}
 \mathtikz{\etaC{0}{0};
\node at (0,1/2){(0,-1)};} &= \sum_{R}(-i)^{l(R)} d_{q}(R) q^{\kappa_{R}/4} e^{-t l(R)} \ket{R}  \label{unit} \\ 
 \mathtikz{\epsilonC{0}{0};
\node at (0,1/2){(-1,0)};} &= \sum_{R} i^{l(R)} d_{q}(R) q^{-\kappa_{R}/4} e^{-t l(R)} \bra{R} 
\label{counit} \\
\mathtikz{\muC{0}{0};
\node at (0,1/2){(0,1)};} &=\sum_{R} \frac{i^{l(R)}}{d_{q}(R)} q^{-\kappa_{R}/4} e^{-t l(R)} \ket{R} \bra{R} \bra{R}
 \label{pants} \\ \mathtikz{\deltaC{0}{0};
\node at (0,1/2){(1,0)};} &=\sum_{R} \frac{(-i)^{l(R)}}{d_{q}(R)} q^{\kappa_{R}/4} e^{-t l(R)} \ket{R} \ket{R} \bra{R}
 \label{cop} \\
\mathtikz{\idC{0}{0} ;\node at (0,1/2){(0,0)}}&= e^{-t \hat{H}} =\sum_{R} e^{-t l(R)} \ket{R}  \bra{R}  \label{cylinder}
\end{align}

Note that each of the cobordisms describes a Riemann surface $\mathcal{S}$ with boundaries in the \emph{target} space.  Each (oriented) circle intersects a stack of Lagrangian branes on which worldsheets wrapping $\mathcal{S}$ ends.  Due to the area-dependent Boltzmann factors $e^{-t l(R)}$, the A-model TQFT is not exactly a Frobenius algbera. 
However the Frobenius algebra gluing rules are satisfied provided we keep track of the Kahler modulus $t$, which just adds upon gluing \cite{runkel2018areadependent}. 
\paragraph{Gluing rules}
To see the effect of introducing the Chern classes, we present some of the gluing rules here in detail. The pair of pants \footnote{Note that the  pair of pants amplitude here differs from the one defined by the topological vertex, because the location of the branes is different in the two cases.}  \eqref{pants} defines a multiplication on closed string states and the Hartle Hawking state \eqref{unit}, also known as the ``Calabi Yau cap"  is the unit element.  These satisfy 
\begin{align}
\mathtikz{ \muC{.5cm}{1cm}
   \node at (-.7,1){(0,-1)};  \node at (-.5,0){(0,1)} ;\etaC{0cm}{ 1.cm}} = \mathtikz{\idC{0cm}{0cm}  \node at (.7,0){(0,0)}}
\end{align} 
with the $(0,0)$ cylinder treated as the identity of the algebra.
Gluing the counit \eqref{counit} to the product \eqref{pants} we obtain a bilinear form we call the closed pairing:
\begin{align}\label{cpair}
    \mathtikz{  \node at (0,.5){(-1,1)} ;\pairC{0cm}{0cm} } := \mathtikz{  \node at (0,-.5){(-1,0)}; \node at (0,1.3){(0,1)} ;\epsilonC{0cm}{0cm} \muC{0cm}{1cm} } = \sum_{R} (-1)^{l(R)} q^{-2\kappa_{R}/4}    \bra{R} \bra{R}.
\end{align}
Note that the closed pairing has a different Chern class than the cylinder even though both have the same Euler characteristic.  This is required by the Chern class assignments of the counit and and unit, together with the fact that they are adjoint with respect to each other under the closed pairing. 

Applying the closed pairing to the unit gives the counit:
\begin{align} 
  \mathtikz{ \pairC{0cm}{0cm}\node at (-.5,.8){(0,-1)} ; \etaC{-.5cm}{0cm}}= \mathtikz{ \node at (0,-.6){(-1,0)} ;\epsilonC{0cm}{0cm} }.
\end{align}

This equation implements the mapping\footnote{In general the mapping $*$ which changes orientation while mapping branes to anti branes is given in the representation basis by $\ket{R} \to (-1)^{l(R)} \bra{R^{t}}$.  This agrees with the adjoint operation defined by \eqref{cpair} when acting on the unit \eqref{unit} and counit \eqref{counit}. }
\begin{align}
\ket{HH} \to \bra{HH^*}
\end{align} 
taking the Hartle-Hawking state to its adjoint as defined in section \ref{section:cboson}.

The pairing has an inverse, called the copairing, which is obtained by gluing the unit to the coproduct 
\begin{align}
    \mathtikz{  \node at (0,1){(1,-1)} ;\copairC{0cm}{0cm} } &:= \mathtikz{ \deltaC{0cm}{1cm} \node at (0,1.5){(0,-1)}; \node at (0,-.5){(1,0)} ;\etaC{0cm}{1cm}  }\nn
    \mathtikz{\node at (.8,1){(1,-1)};\node at (-.5,-1){(-1,1)}; \pairC{0cm}{0cm}; \idC{1.5cm}{0cm}; \copairC{1cm}{0cm}; \idC{-0.5cm}{1cm}; } &= \mathtikz{ \idC{0cm}{0cm}; \node at (0,0.5){(0,0)} }
\end{align}

The resolved conifold partition function is obtained by gluing the unit to the counit:
    \begin{align}
        Z= \mathtikz{
\node at (0,1/2){(0,-1)};
\epsilonC{0}{0};
\etaC{0}{0};
\node at (0,-1/2){(-1,0)}}=\sum_{R}  (d_{q}(R))^{2} e^{-t l(R)}.
    \end{align}
More generally, by gluing the generators, we can obtain the closed string partition function for a local Calabi-Yau manifold with base manifold $\mathcal{S}$ of genus $g$ and  Chern classes $(2g-2+p, p)$:
 \begin{align} 
 Z = \sum_{R} \left(\frac{1}{d_{q}(R)} \right)^{2g -2} q^{(g-1)\kappa_{R}/2} e^{-t l(R) }
 \end{align} 
 where $t$ is the complexified area of $\mathcal{S}$. 

\subsection{Quantum traces and q-deformation of the A model TQFT}
Following \cite{Aganagic:2004js}, we have expressed the linear maps \eqref{unit} to \eqref{cylinder} in an orthonormal basis $\ket{R}$ labelled by  representations of $U(\infty)$.  These linear maps should be viewed as string amplitudes. This becomes manifest when we express the basis $\ket{R}$ as wavefunctions on the group
\begin{align}\label{ctr}
    \braket{U|R}=\tr_{R}(U),
\end{align}
where $U= \exp \oint A $ gives the usual coupling of the worldsheet boundary  to the worldvolume gauge field.

This gives a consistent closed TQFT so long as we restrict to gluing of cobordisms along circles.   However, it was observed in \cite{deHaro:2006uvl} that for finite $N$ the use of the classical trace in \eqref{ctr} leads to inconsistencies when gluing along \emph{open} edges.  This is precisely the type of gluing which was needed to compute the replica trick entropy \eqref{KEE}, since this requires opening the base $S^2$ into a disk $D^2$ and then gluing a sequence of such disks along half of their boundary $\pd D^2$.  The same inconsistency appears if we apply the 2DYM factorization in \cite{Donnelly:2016jet} to the closed string wavefunction $\tr_{R}(U) $.  This was defined by splitting the Wilson loop $ U = U_{A} U_{B}$ into the product of Wilson lines in region $A$ and $B$, and then taking the classical trace: 
\begin{align}
   \tr_{R}(U) \rightarrow  \tr_{R}(U_{A}U_{B}) &=\sum_{i,j=1}^{\dim R}   R_{ij}(U_{A})R_{ji}(U_{B}),
\end{align}
where $R_{ij}(U_{A,B})$ are matrix elements in the $R$ representation, viewed as wavefunctions in the subregion $A,B$.   The indices  $i,j$ label entanglement edge modes transforming under the gauge group $U(\infty)$, and in the case of undeformed 2DYM, led to an entropy consistent with the replica trick.  However, for the A-model,  this naive counting of edge modes would lead to degeneracy factors of $\dim R$, which are incompatible with the quantum dimensions in the replica trick entropy \eqref{RRep}.  In terms of the sewing relations, the $U(\infty)$ edge modes fail to satisfy the E-brane axiom.

This problem arises because the A-model TQFT restricted to Calabi-Yau manifolds is really a functor which maps $2\Cob^{L_{1},L_{2}}$  to the representation category of a \emph{quantum} group.  This is suggested by the presence of the $q$-deformed dimension factor $d_{q}(R)$, which implies that the surface symmetry acting on the endpoints of the open strings, 
is $q$-deformed.  However, the classical trace employed in the wavefunction \eqref{ctr} is not invariant under this quantum group symmetry.  We will explain what this quantum group symmetry is in subsequent sections.  For now we note that \cite{deHaro:2006uvl} observed that gauge invariance under the quantum group symmetry can be achieved by replacing the classical trace with the quantum trace:
\begin{align}\label{Qtr}
    \braket{U|R}&=\tr_{q,R} (U) :=\tr_{R} (u U ),
\end{align}
where $u$ is the Drinfeld element of the quantum group.  This element is defined abstractly from quantum group data, and its classical trace gives the associated quantum dimension.

Thus for $\U(N)_{q}$ we  have
\begin{align}
    \tr_{q,R} (1) = \tr_{R} (u )  = \dim_{q}(R).
\end{align}
This equation remains valid for a general quantum group, with $\dim_{q}(R)$ the quantum dimension defined from its representation category data.  For the A-model string, the role of the Drinfeld element is played by the matrix $D$ defined in \eqref{D}, which may be viewed as a renormalized version of the Drinfeld element $u$ for $U(N)_{q}$:
\begin{align} \label{Du}
D &=q^{- N/2} u,  \nn
\lim_{N\rightarrow \infty} \tr_{R}(D) &=(-i)^{l(R)} d_{q}(R) q^{\kappa_{R}/4} .
\end{align}
This is the analogue of equation \eqref{ldim} and will be useful in relating the quantum group symmetry for the A-model string to $U(N)_{q}$. Finally note that the wavefunctions \eqref{Qtr} are orthonormal 
\begin{align}
    \int dU \tr_{q,R} (U) \tr_{q,R'} (U)= \delta_{R,R'},
\end{align}
and span the Hilbert space of class functions on the quantum group, which is isomorphic to $\mathcal{H}_{\Sigma}$ defined previously in section \ref{section:cboson}.  For this reason, the use of classical traces in \cite{Aganagic:2004js} was adequate for the purposes of computing A-model partition functions by sewing along circles. However, as we will see, the  quantum trace and the $q$-deformed nature of the holonomy $U$ becomes essential when we perform operations that effectively cut open the closed string loops. 
\subsection{String theory origin of the $q$-deformation}
\begin{figure}[h]
\centering 
\includegraphics[scale=.4]{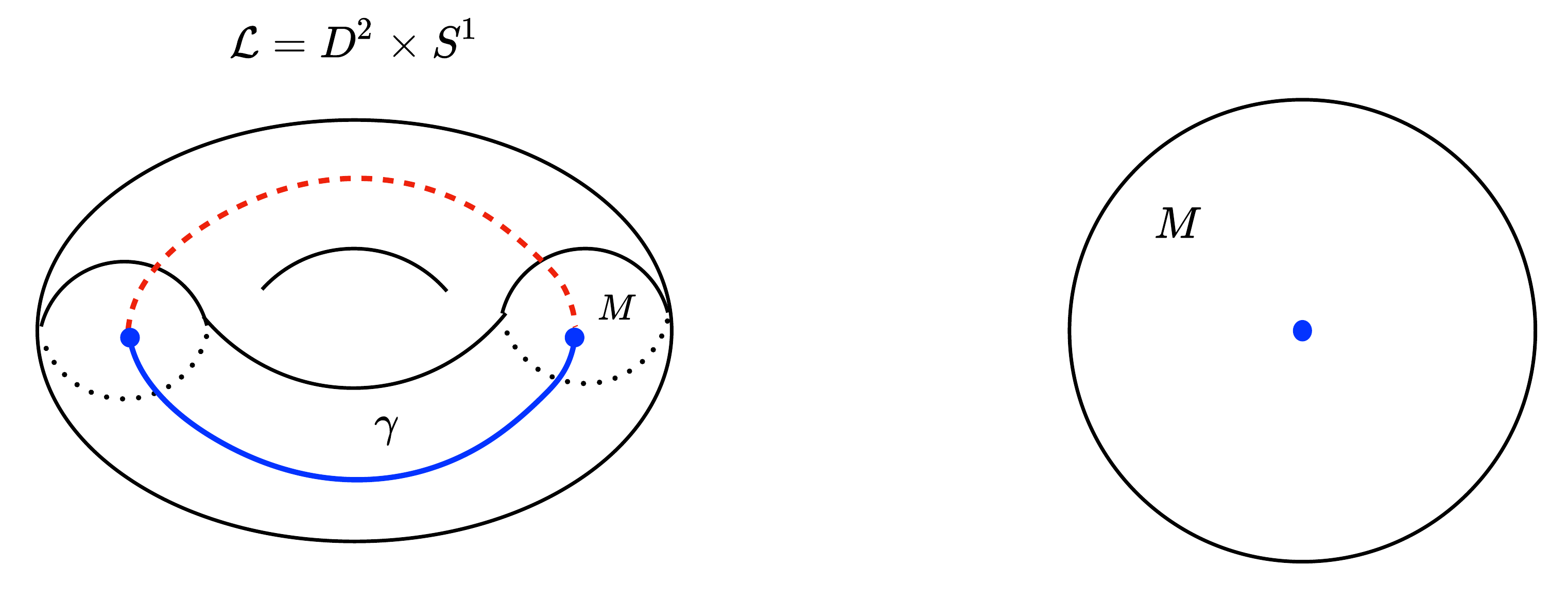} 
\caption{Quantizing the worldvolume gauge theory with time running around the non-contractible cycle of $\mathcal{L}$, we have to impose Gauss's law on $M$.  The puncture on $M$ corresponds to the anyon charge on the Wilson loop which sources Gauss Law.}
\label{QCS}
\end{figure}

In the previous discussion, we explained the necessity for quantum traces and the associated $q$-deformation of the closed string Hilbert space from consistency requirements of the TQFT.   Here we would like to explain how the quantum group symmetry emerges from the viewpoint of the worldvolume gauge theory on the D-branes.  
\paragraph{q-deformed connection in the worldvolume gauge theory} 
Replacing classical traces with quantum traces means that the coupling of the worldsheet boundary to the worldvolume gauge fields have been changed to 
\begin{align}
    \tr (u  P \exp \oint A).
\end{align}
This is because the \emph{classical} gauge field  should be viewed as a $q$-connection, whose components $A^{a}_{\mu} (X)$ ($a$ is a group index) are  noncommutative functions on the brane.  This $q$-deformation is a known property of the worldvolume $\U(N)$ Chern Simons theory on a stack of $N$ branes, and we will give a brief review here. 
Usually, the gauge fields components $A_{\mu}(X)$ are taken to be commutative functions of $X$.  However one can see how a $q$-deformed gauge field arises by considering the Gauss law constraint. This is a constraint applied in canonical quantization along a constant time slice $M$ (see figure \ref{QCS}), which we can take to be a surface at fixed angle along the non-contractible $S^1$  on the Lagrangian manifold $\mathcal{L}$ \eqref{lag}. 
In the presence of a Wilson loop around this cycle, corresponding to boundary $\gamma$ of the string worldsheet ending on the brane, Gauss law reads 
\begin{align}\label{Gauss}
     \frac{k}{8\pi } \epsilon^{ij}F_{ij}^{a} (X) =  \delta^{2}(X- P)T^{a},
\end{align}
where $i,j$ are spatial indices on $M$, $P$ is the location of the puncture where $M$ cuts the Wilson loop, $T^{a}, a=1 \cdots \dim \U(N) $ are generators  of $\U(N)$.  It was noted in \cite{Witten:1988hf} that this equation cannot be solved for an ordinary gauge field because $F_{ij}^{a}$ is a number while $T^{a}$ is a non commuting matrix. This mismatch occurs because the Wilson loop is a non-dynamical defect operator; there is no ``matter field" on the loop $\gamma$ that couples to $A$.  One solution is to ``integrate in"  dynamical degrees of freedom on the loop, which will couple to $A$ and render the objects on both sides of \eqref{Gauss} commutative\cite{Witten:1988hf, Elitzur:1989nr}. However to see the quantum group symmetry, we should apply the alternative prescription suggested in \cite{Witten:1988hf}, and $q$-deform the gauge field $A^{a}_{\mu}$ into a non-commutative object, i.e. a matrix in the lie algebra of $\U(N)$.  This idea was carried out in \cite{Guadagnini:1989tj}, where an explicit solution to \eqref{Gauss} was derived, giving a noncommutative connection that can be identified with the Knizhnik–Zamolodchikov connection in conformal field theory.   In appendix D, we give a string sigma model argument for noncommutative world volume gauge fields following \cite{1999JHEP...09..032S}

\section{Extension of the A-model closed TQFT} 
\label{section:open}
Having formulated the A-model closed TQFT in terms of representation categories of quantum groups, we now describe its extension to the open sector.   We begin by defining the open string Hilbert space associated to an interval on which the operators of the open sector act.
We give an explicit action of the  quantum group symmetry on this Hilbert space and the associated decomposition into irreducible representations.  Next, we derive the open-closed cobordisms which include diagrams describing the factorization of the closed string Hilbert space.   We then compute the $q$-deformed entropy from the reduced density matrix of the Hartle-Hawking state and show that it matches the geometric replica trick calculation in section \ref{ssection:replica}.  Finally we will revisit the geometric replica trick calculation and show that the preservation of the Calabi Yau condition requires the insertion of a ``defect" operator at the entangling surface, which plays the role of the (inverse) Drinfeld element of the quantum group.  
\subsection{The open string Hilbert space as the coordinate algebra $\mathcal{A}(\U(\infty)_{q})$ }

The q-deformation of the spacetime gauge field $A$ means that its holonomy $U= P \exp \oint A$ is an element of the quantum group $\U(N)_{q}$.  This can be defined by $q$-deforming the algebra $\mathcal{A}(\U(N))$  of functions on $\U(N)$, refered to as its  \emph{coordinate algebra}.  $\mathcal{A}(\U(N))$ is generated by matrix elements $U_{ij}$ satisfying the unitary constraint 
\begin{align} \label{uni} 
   \sum_{k} U_{ik} U_{jk}^{*}=  \sum_{k} U_{ki}^{*} U_{kj}  = \delta_{ij}.
\end{align}
As a vector space, $\mathcal{A}(\U(N))$ is defined over the complex numbers and spanned by the basis 
\begin{align}
     U_{i_1 j_1} U_{i_2 j_2} \cdots U_{i_n j_n}, 
     \quad  n=1,\cdots \infty.
\end{align}
In the undeformed algebra, the matrix elements themselves commute:
\begin{align}
    U_{ij}  U_{kl} =U_{kl}U_{ij}.
\end{align}
However, in the quantum group $\U(N)_{q}$ this multiplication law (distinct from the matrix multiplication rule) becomes noncommutative.  There exists a conjugate linear involution  $*$ of the coordinate algebra $\mathcal{A}(\U(N)_{q})$ for which the unitary constraint \eqref{uni} still holds. However, due to the noncommutativity, the placement of the $*$ is now crucial in \eqref{uni}.  In particular, it should be noted that for $U_{ij} \in  \mathcal{A}(\U(N)_{q}) $
\begin{align}
    \sum_{k} U_{ik}^{*} U_{jk}\neq \delta_{ij} .
\end{align}
It is customary to abuse language and refer to both the ``quantum space"  $\U(N)_{q}$ and the algebra of functions $\mathcal{A}(\U (N)_{q})$ as a quantum group.  This is done in the spirit of noncommutative geometry, where the geometry of a noncommutative space $X$ is defined by the algebra of noncommutative functions on $X$ \cite{connes1995noncommutative, 2001RvMP...73..977D}.

The precise nature of the noncommutative product in $\U(N)_{q}$ is determined by the R-matrix of the quantum group.   To express the product rule it is useful to consider an element $U\in \U(N)_{q}$ as a matrix acting in the fundamental representation. Thus it acts on a vector space $V$ according to
\begin{align} \label{V}
U:V &\rightarrow V,\nn
    v_{i}&\mapsto  \sum_{i} U_{ij} \otimes v_{j},
\end{align}
where the tensor product $\otimes$ symbol has been used to distinguish this product from the noncommutative product we wish to define. 
In the same fashion,  the R-matrix $\mathcal{R} \in \U(N)_{q} \otimes \U(N)_{q}$ can be regarded as an element $\mathcal{R} \in \text{End}(V\otimes V)$, i.e. a matrix operator acting on two copies of $V$.  If we define matrices
\begin{align}
    U_{1} &= U \otimes \mathbf{1},\nn
    U_{2} &= \mathbf{1} \otimes U.
\end{align}
Then the multiplication rule for the coordinate algebra on $U(N)_{q}$ is
\begin{align}\label{Ru}
    \mathcal{R} U_{1} U_{2}= U_{2}U_{1} \mathcal{R},
\end{align}
where the composition of the operators above is defined with ordinary matrix multiplication.   An explicit example of the  R-matrix ,  $*$ structure, and other quantum group properties of $\mathbf{SL}_{q}(2)$ is  presented in appendix \ref{section:QG}.

\paragraph{Definition of the open string Hilbert space} 
We now define  the open string Hilbert space $\mathcal{H}_{\Sigma_{A}}$ assigned to the subregion string configurations in $\mathcal{F}_{\Sigma_{A}}$  as the large $N$ limit of the coordinate algebra on $\U(N)_{q}$:
\begin{align}   
    \mathcal{H}_{\Sigma_{A}} &= \mathcal{A}(\U(\infty)_{q}), \nn
    q&= e^{i g_{s}}.
\end{align}
This a $q$-deformation of the open string Hilbert space defined in \cite{Donnelly:2016jet} for the string theory dual to 2DYM.  
In particular the subspace of $n$ open strings is spanned by the states $\ket{IJ} $ with wavefunctions 
\begin{align}\label{Uij} 
\braket{U| I,J} &= U_{i_1 j_1} U_{i_2 j_2} \cdots U_{i_n j_n},\nn
U_{i_a j_a} &= P\exp \int X_{i_aj_a}^{*} A,
\end{align} 
where in the second equation we have emphasized that these wavefunctions live on the space of subregion open string configurations $ X_{ij}(\sigma)\in \mathcal{F}_{\Sigma_{A}} $.  Due to the topological invariance, they are completely specified by the multi-index Chan-Paton factors $I,J$ labeling the entanglement branes.  In the undeformed case where $q=1$, the commutativity of the matrix elements $U_{ij}$  implies these open string are bosonic \cite{Donnelly:2016jet}, so that the $n$ string Hilbert space is 
\begin{align}\label{SnQ}
     \mathcal{H}_n =( V \otimes V^{*})^{\otimes{n}}/S_{n}.
\end{align}
Here $V^*$ denotes the dual of the fundamental representation, giving the strings an orientation.  Open string indistinguishability is enforced by the quotient of the permutations group $S_{n}$, which permutes the open strings by acting simultaneously on both endpoints $\ket{I,J} \to \ket{\sigma(I),\sigma(J)}$ for $\sigma \in S_n$.
\begin{figure}
\centering 
\includegraphics[scale=1]{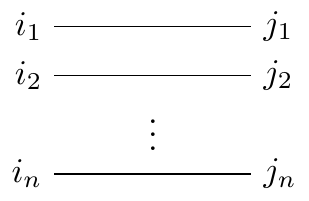} 
\caption{The state $\ket{IJ}$ represents a configuration of $n$ open strings with Chan-Paton factors $(i_1, j_1) \ldots (i_n, j_n)$. }
\label{openstrings}
\end{figure}
In the presence of nontrivial string interactions, $g_{s}>0, q\neq 1$, the open string endpoints become anyons \cite{Gomis:2006mv}.  This change in statistics is implemented by the equivalence relation \eqref{Ru}, which tells us that the exchange of open strings must be accompanied by an $R$ matrix transformation. The operation of permutating strings is therefore replaced by braiding, and the open string Hilbert space is
\begin{align}
\mathcal{H}_{\Sigma_{A} } &= \bigoplus_{n=1}^{\infty}  \mathcal{H}_n(q),\nn
    \mathcal{H}_n(q) &=( V^{*}\otimes V)^{\otimes{n}}/ (\mathcal{R} U_{1} U_{2}\sim  U_{2}U_{1} \mathcal{R}).
\end{align}
For $q\in \mathbb{R}$, the inner product on $\mathcal{H}_{\Sigma_{A} }$ is defined by the quantum group Haar measure and is given in terms of the representation basis in  \eqref{h}. 
\subsection{Quantum group symmetry on the open string Hilbert space}
Each open string in the state $\ket{I,J}$ transforms in the adjoint representation of the quantum group symmetry, which is the surface symmetry of the A-model string.

To describe the action of this symmetry and the associated decomposition of  $\mathcal{A}(\U(\infty)_{q})$, we need to introduce an operation called the antipode.  A more thorough presentation of the algebraic structure of a quantum group is given in appendix \ref{section:QG}.  Below we will work with $\mathcal{A}(\U(N)_{q})$ and then consider the $N \to \infty $ limit later. 
\begin{figure}
\centering 
\includegraphics[scale=.5]{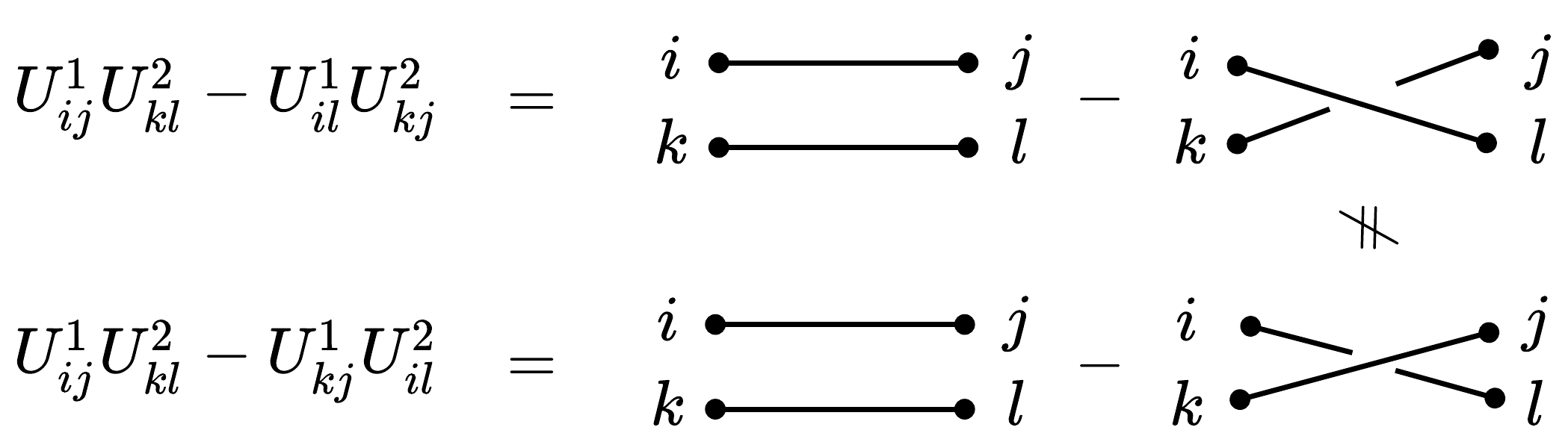} 
\caption{ The figure shows a state of two open strings.  Antisymmetrization of the right and left indices can be expressed by fixing the Chan-Paton factors while changing the pattern of connection between them.   In the figure we have put string number 1 on top of string number 2.  When the open strings have bosonic statistics as imposed by the $S_{n}$ quotient in \eqref{SnQ}, the antisymmetrization of the left or the right endpoints give the same state,  since the two operations are relating by commuting string number 1 and  2.  However, the A-model open string has anyonic statistics,  so the two orderings are not equal.   This corresponds a nontrivial braiding structure in the diagrams above.   }
\label{braid}
\end{figure}
\paragraph{The antipode and the conjugate representation}
Given a quantum group $\mathcal{A}$ the antipode is an anti-homomorphism
\begin{align}
S: \mathcal{A}\rightarrow \mathcal{A},  \\
S( U V) &= S(V)S(U) ,\quad U, V \in \mathcal{A} \label{anti} .
\end{align}
It acts on \emph{single} string elements  $f_{ij}(U)= U_{ij}  \in \mathcal{A}(\U(N)_{q})$ 
by giving the analogue of the matrix inverse:
\begin{align}
    \sum_{j} U_{ij} S(U)_{jk}= S(U)_{ij}U_{jk}= \delta_{ik}.
\end{align}
Note that due to the noncommutativty of  $U_{ij}$, $S(U)_{ij}$ is different from the usual inverse $U^{-1}_{ij}$, which is defined with respect to a commutative multiplication rule.
The definition of $S$ can be extended to the rest of the algebra recursively using the property \eqref{anti}.

Given a representation $R$ of $\mathcal{A}$, the antipode defines the conjugate (``anti-particle") representation $\bar{R}$ by
\begin{align}
    \bar{R}(U)= (R\circ S(U))^{t},
\end{align}
where $t$ denotes the transpose. 
\paragraph{The adjoint action, Drinfeld element, and the quantum trace }
We can now define how the quantum group acts on the open string Hilbert space via the adjoint action of the quantum group on itself. For an element $g \in \mathcal{A}(U(N)_{q})$, the adjoint action is defined using a combination of the coproduct and antipode:
\begin{align} \label{adjoint} 
    U_{ij}  &\rightarrow   (\text{Ad}_{g}(U))_{ij}  = \sum_{k,l}  U_{kl}\otimes  g_{ik}S(g)_{lj},
\end{align}
where we have used $\otimes$ in the same manner \eqref{V} to distinguish the objects $U_{ij}$ in the representation space $V^{*}\otimes V$ with the quantum group elements acting on that space.  It is important to note that  $U_{ij}$ commutes with $g_{ik}$ and $S(g)_{lj}$ but $g_{ik}$ and $S(g)_{lj}$ do not commute among themselves.   

As observed earlier, the ordinary trace $\tr_{R}(U)$ in any representation $R$ is not invariant under this transformation law. However, there exists an invariant ``quantum" trace function which can be defined purely in terms of quantum group data.   One first defines the ``Drinfeld" element $u$ in terms of the $\mathcal{R}$ matrix
\begin{align}
    \mathcal{R} = \sum_{i} a_{i} \otimes b_{i} \in \mathcal{A} \otimes \mathcal{A},
\end{align}
and the antipode $S$ according to
\begin{align}
 u&=\sum_{i} S(b_{i}) a_{i}.
\end{align} 

The quantum trace of an element $U \in \mathcal{A} $ in any representation $R$ can then be defined as: 
\begin{align}\label{qtrace}
    \tr_{q,R}(U)= \sum_{ij} u_{ij}^{R} R_{ji}(U) ,
\end{align}
where $R_{ij}(U)$ are the representation matrices for $U$ and we defined $u_{ij}^{R} = R_{ij}(u)$.
The properties
\begin{align} S^{2}(V) &=u V u^{-1}\nn
S( U V) &= S(V)S(U) ,\quad U, V \in \mathcal{A},
\end{align}  of $u$ and $S$ then imply that the quantum trace is invariant under the adjoint action \eqref{adjoint}  of the quantum group.  An explicit proof is given in eq.(5.8) of  \cite{deHaro:2006uvl}, and it gives a nice illustration of subtleties arising from the q deformed multipication rule.   For $\U(N)_{q}$ the Drinfeld element is a diagonal matrix of complex phases\footnote{The quantum group is an associative algebra over the complex numbers, so $u$ is a nongeneric element that consists of scalar elements of the algebra.} given explicitly by \cite{Ooguri:1999bv,deHaro:2006uvl}
\begin{align}  \label{u}
    u_{ii}= q^{\frac{N}{2}}  q^{-i+\frac{1}{2}}.
\end{align}
Finally, we note that the quantum trace is multiplicative under tensor products:
\begin{align} \label{mult} 
    \tr_{q}( A \otimes B) = \tr_{q}(A) \tr_{q}(B).
\end{align}

\paragraph{The representation basis and Schur-Weyl duality} Let us now consider how the open string Hilbert space is organized into irreducible representations of the quantum group symmetry.   Compact quantum  groups such as $\U(N)_{q}$ satisfy a Peter-Weyl theorem \cite{Woron}, which states that its space of functions is spanned by the matrix elements in all irreducible representations\footnote{A precise description of the representation theory for quantum groups is described in chapter 11 of \cite{Klimyk:1997eb}.} of the quantum group.   These (noncommutative)  matrix elements form an un-normalized basis of wavefunctions on $\mathcal{A}(U(N)_{q})$:
\begin{align}\label{Rij} 
\braket{U|R i j } &=R_{ij}(U) \quad i,j =1,\cdots \dim R,
\quad U\in U(N)_{q},
\end{align} 
where $\dim R$, distinct from $\dim_q(R)$, is the \emph{integer} dimension of the representation $R$.  Note that $\ket{Rij}$ labels a basis in $V_{R}\otimes V_{R}^{*}$, which is a vector space with an integer dimension.
There is also a $q$-analogue of the translation-invariant Haar measure, 
\begin{align} 
    h: \mathcal{A} \rightarrow \mathbb{C},
\end{align}
which can be used to define the inner product on $\mathcal{A}(U(N)_{q})$
\begin{align} \label{h}
     ( R_{ij}(U),  R'_{kl}(U) )  :=  h(R_{ij}^{*}(U),  R'_{kl}(U) ) = \delta_{R R'} \frac{(u^{R})^{-1}_{jk} \delta_{il}}{\dim_q R}.
\end{align}

We now relate the representation \eqref{Rij} and the open string basis \eqref{Uij} by applying  a $q$-deformed version of Schur-Weyl duality to the $n$-open string  states $\mathcal{H}_{n}(q) $.  This relation will be necessary to define the representation basis in the $N \to \infty$ limit.   
We first recall the undeformed Schur-Weyl duality.
The vector space $V^{\otimes n}$ carries a representation of $S_n$ which permutes the factors as well as a diagonal action of $\U(N)$. The Schur-Weyl duality states that $V^{\otimes n}$ decomposes into irreducible representations of these two groups as:
\begin{align}\label{SW}
V^{\otimes n} = \bigoplus_{R \in Y_n}  V_{R}^{U(N)} \otimes V_{R}^{S_{n}} ,
\end{align}
where $Y_{n}$ denotes the set of Young diagrams with $n$ boxes which label irreducible representations of both $\U(N)$ and $S_n$.  
Equation \eqref{SW} is the formal way of saying that irreducible representations of $\U(N)$ are obtained by symmetrizing/antisymmetrizing fundamental representations according to a Young diagram $R$.
To obtain the decomposition of the Hilbert space of $n$ strings, we apply the Schur-Weyl duality twice:
\begin{align}\label{SW2}
    \mathcal{H}_n &= ( V^{n}\otimes V^{*n})/S_{n}, \nn
    &= \left( \oplus_{R \in Y_n} V_{R}^{U(N)} \otimes V_{R}^{S_{n}}\right) \otimes \left( \oplus_{R' \in Y_n} V_{R'}^{U(N)} \otimes V_{R'}^{S_{n}}\right)^* / S_{n}, \nn
    &= \oplus_{R \in Y_{n} } V_{R}\otimes V_{R}^{*},  \\
    V_{R}&:= V_{R}^{U(N)} \otimes V_{R}^{S_{n}},
\end{align}
where the vector space $V_{R}\otimes V_{R}^{*} $ is spanned by the representation basis $\ket{R ij} \quad i,j=1,\cdots \dim R$. We can thus interpret  $\ket{R ij}$ as symmetrized/antisymmetrized linear combinations of $\ket{I,J}$. As a simple example, the projection on to the antisymmetric representation R for $n=2$ is given by:
\begin{align}
U^{1}_{ij}U^{2}_{kl} \rightarrow R_{ab}(U)= U^{1}_{ij}U^{2}_{kl}-U^{1}_{il}U^{2}_{kj} \in V_{R}\otimes V^*_{R},\nn
a,b=1,\cdots \dim R,
\end{align}
where the superscripts label the strings.
This decomposition \eqref{SW2} holds in the large $N$ limit, and leads to a dimension formula 
\begin{align}\label{dim}
    \dim \mathcal{H}_{n} =\sum_{R\in Y_{n}} (\dim R)^{2},
\end{align} which relates the counting of Chan-Paton factors to degeneracy factors of $\U(N)$.

In the $q$-deformed case, the  vector space  $V^{\otimes n}$ is a tensor product of $\U(N)_{q}$ fundamentals, so it can be organized into quantum group representations in a similar way.   
The operations which commute with the action of $U(N)_{q}$ belong to a $q$-deformed version of the symmetric group called the Hecke algebra $S^{q}_{n}$, which combines the permutation of the tensor factors with applications of the $R$ matrix.   Given a  transposition $\tau_{12} \in S_{n}$ which acts on a basis of $V_{1} \otimes V_{2} $ by 
\begin{align}
    \tau (\mathbf{e}_{1} \otimes  \mathbf{e}_{2}) = \mathbf{e}_{2} \otimes \mathbf{e}_{1}.
\end{align} 
  We define an element $h(\tau) \in  S^{q}_{n}$ in the Hecke algebra by 
\begin{align}
      h(\tau)= \tau \circ R.
\end{align}
  
The $q$-deformed Schur-Weyl duality states that the space $V^{\otimes n}$ decomposes under the commuting action of $(U(N)_{q})^{\otimes n}$ and $S^{q}_{n}$ as:\cite{deHaro:2006uvl}
\begin{align}
V^{\otimes n} = \oplus_{R \in Y_n}  V_{R}^{U(N)_{q}} \otimes V_{R}^{S^{q}_{n}}.
\end{align}
The $q$-deformed Hilbert space for $n$ strings decomposes into 
\begin{align}\label{Hn}
      \mathcal{H}_{n}(q)&= ( V^{n} \otimes V^{*n})/\sim,\nn
      &= \left( \oplus_{R \in Y_n} V_{R}^{U(N)_{q}} \otimes V_{R}^{S^{q}_{n}}\right) \otimes \left( \oplus_{R' \in Y_n} V_{R'}^{U(N)_{q}} \otimes V_{R'}^{S^{q}_{n}}\right)^* / \sim ,\nn
      &= \oplus_{R \in Y_{n} }   V^{q}_{R} \otimes V_{R}^{q*}, \\
    V^{q}_{R}&:= V_{R}^{U(N)_q} \otimes V_{R}^{S^{q}_{n}},
\end{align}
where $\sim$ refers to the equivalence relation 
\begin{align}
    \mathcal{R} U_{1}U_{2}=U_{2}U_{1}\mathcal{R},
\end{align}
which determines the braiding structure of the open strings.  
In direct analogy with the undeformed case, we should view $\ket{R{ij}}$ as a basis for the subspace $V^{q}_{R}\otimes V^{q*}_{R}$, obtained by symmetrizing/antisymmetrizing the Chan-Paton factors $\ket{IJ}$ using the Hecke algebra elements.  
The corresponding projectors labelled by Young diagrams $R\in Y_{n}$ were constructed in \cite{deHaro:2006uvl}.  
In contrast to the permutation group, the action of the Hecke algebra provides a representation of the braid group.  This is because the endpoints of the open strings behave as  anyons due to their coupling to the worldvolume Chern-Simons theory of the A-model branes \cite{Gomis:2006mv}.    
The quantum dimension of $\mathcal{H}_{n}(q)$ is the computed from the trace of the Drinfeld element in the representations given in the Hilbert space decompositions of Eq.~\eqref{Hn}:
\begin{align}
    \dim_{q} \mathcal{H}_{n}(q) := \tr_{\mathcal{H}_{n}}(u) =\sum_{R\in Y_{n}} (\dim_{q} R)^{2},
\end{align} 
which is the $q$-deformed version of equation \eqref{dim}.  In the large $N$ limit,  this formula will give a canonical interpretation to the total degeneracy factors in the resolved conifold partition function \eqref{Z1} and the replica trick entanglement entropy \eqref{RRep}.

\paragraph{The large $N$ limit of Schur-Weyl duality and the Drinfeld element} 
 Schur-Weyl duality continues to hold in the large $N$ limit of $\U(N)$.  As $N\to \infty$, we  continue to identify the representation basis $\ket{R{ij}}$ with Young diagrams describing the (anti)symmetrizations of Chan-Paton factors of the open string states $\ket{IJ}$. This basis spans the extended Hilbert space for the string theory dual to 2DYM with gauge group $\U(\infty)$ \cite{Donnelly:2016jet}.  At large $N$ the 2DYM partition function is determined by symmetric group data, which captures the wrapping of string worldsheets on the target space.  

In a similar fashion, the $q$-deformed Schur-Weyl duality also survives the large $N$ limit of $\U(N)_{q}$ and the corresponding basis $\ket{Rij}$ is once again determined by the symmetrization of the Chan-Paton factors by elements of the Hecke algebra \cite{deHaro:2006uvl}. This basis spans the extended Hilbert space of $q$-deformed 2DYM with gauge group  $\U(\infty)_{q}$.  
As in the undeformed case, we wish to identify these states with the extended Hilbert space of the A-model TQFT, which is also determined by $q$-deformed symmetric group data.  

Moreover in order for the counting of states in q2DYM to match with the A-model, we must identify the correct large $N$ limit of the Drinfeld element $u$ given in \eqref{u}.  Since $u$ determines the trace function \eqref{qtrace} on the extended Hilbert space, it can be viewed as determining the choice of measure on the open string states. 

We will define the large $N$ limit of $u$ according to \eqref{Du} in terms of the holonomy matrix $D$ of \eqref{D}.   As explained in the derivation of \eqref{D} this limit requires an analytic continuation of $q$ which regularizes the trace over the large $N$ Hilbert space. As a result, even though the dimension
\begin{align}
    \tr_{R}(u)=\tr_{R}(u^{-1}) = \dim_{q}R,
\end{align}
 is always a real quantity,  in the large $N$ limit $D$ has  a complex trace:
\begin{align}
    \tr_{R}(D)&=(-i)^{l(R)}d_{q}(R) q^{\kappa_{R}/4} \in \mathbf{C}, \nn
    \tr_{R}(D^{-1})&=\left(\tr_{R}(D)\right)^{*}=i^{l(R)}d_{q}(R) q^{-\kappa_{R}/4}.
\end{align}
Accordingly, we define the quantum trace for the large N limit
\begin{equation}\label{quantum trace}
    \tr_{q}(U)=\tr(DU).
\end{equation}
This feature is related to the holomorphic nature of the A model and essential to the emergence of the line bundle structure of the Calabi-Yau manifold.

With this definition of the large $N$ Drinfeld element, the quantum dimension of the $n$-string Hilbert space  becomes 
\begin{align}\label{deg}
    \dim_{q} \mathcal{H}_{n} := \tr_{\mathcal{H}_{n}}(D) &=\sum_{R\in Y_{n}}  \tr_{R \otimes \bar{R}} ( D)\nn
    &=\sum_{R\in Y_{n}} \tr_{R}(D)\left( \tr_{R}(D)\right)^{*}= \sum_{R\in Y_{n}} (d_{q}(R))^{2} ,
\end{align} 
where in the second to last equality we have used the multiplicative property of the quantum trace \eqref{mult} and the unitarity of the representations.

\subsection{A-model open-closed TQFT and factorization maps}
We have now assembled all the ingredients necessary to describe the extension of the A-model TQFT into a $q$-deformed open-closed theory which incorporates the factorization of the closed and open string states.  

We begin by defining the factorization maps in \eqref{splitting} and then  extend these into an interwoven set of open-closed cobordisms.

\paragraph{Factorization maps }
The factorization map which embedds closed string states into open string states in the extended Hilbert space as shown in the left of figure \eqref{splitting} is called the zipper $i_{*}$.  Our definition of the closed string wavefunction $\braket{U|R}$ as a quantum trace suggests that
\begin{align}\label{zip}
i_{*}=\mathtikz{ \zipper{0cm}{0cm} } &: \ket{R} \to \sum_{i,j} (D^{-1})^{R}_{ji} \ket{Rij}.
\end{align}

Compatibility with the E-brane axiom then requires the co-zipper to be
\begin{align}
i^{*}=\mathtikz{ \cozipper{0cm}{0cm} } &: \ket{Rij} \to (-i)^{l(R)}\frac{\delta_{ij}}{d_{q} (R) q^{-\kappa_{R}/4}} \ket{R},
\end{align}
so that 
\begin{align} \label{cwind}
      \mathtikz{ \zipper{0cm}{0cm} \cozipper{0cm}{-1cm}} &= \mathtikz{\idC{0cm}{0cm} }:\ket{R} \to \ket{R} ,
\end{align}
as can be shown by noting that $\sum_{i} (D^{-1})^{R}_{ii} =i^{l(R)} d_{q}( R)q^{-\kappa_{R}/4}$.

Next we consider the cobordism on the right of figure \eqref{splitting}, which embeds open string states of one subregion into the open string Hilbert space of two subregions.\footnote{The intervals in the cobordism diagrams really correspond to subregions of a time slice $\mathcal{F}_{\Sigma}$ in the space of string loops.} 
We identify this factorization map with the coproduct in the open sector of the A-model TQFT:
\begin{equation}\label{Delta}
\Delta=\mathtikz{\deltaA{0cm}{0cm} }: \ket{Rij} \to \sum_{k}   \ket{Rik} \ket {Rkj}.
\end{equation}
To see that this satisfies the E-brane axiom, we have to first define the open product
\begin{align}
  \mu_{O} = \mathtikz{\muA{0}{0}},
\end{align}
which fuses two subregions together.  This is the A-model version of the ``entangling product" \cite{Donnelly:2016auv}, and we propose that it is given by 
\begin{align}
    \mu_{O} = \mathtikz{ \muA{0cm}{0cm} } : \ket{Rij} \ket{R'kl} \to (i)^{l(R)} \frac{D^{R}_{jk}}{d_{q} (R) q^{\kappa_{R}/4}} \ket{Ril} . 
\end{align}
This satisfies the E-brane axiom which requires that splitting followed by fusion gives the identity map:
\begin{align}\label{spf}
     \mathtikz{ \muA{0cm}{0cm} \deltaA{0cm}{1cm} } = \mathtikz{ \idA{0cm}{0cm} }: \ket{Rij} \rightarrow \ket{Rij},
\end{align}
which follows from $\sum_{i}D^{R}_{ii} =(-i)^{l(R)} d_{q}( R)q^{\kappa_{R}/4}$. 
Finally, combining the zipper and coproduct gives the factorization map as promised in \eqref{factor}:
\begin{align} \label{cfact}
\mathtikz{ \zipper{0cm}{0cm}\deltaA{0cm}{-1cm} ; \node at (-.5,-2.3){$\Sigma_{A}$};\node at (.6,-2.3){$\Sigma_{B}$}; \node  at (0,.4){$\Sigma$}  }:\mathcal{H}_{\Sigma} &\to  \mathcal{H}_{\Sigma_{A}}\otimes \mathcal{H}_{\Sigma_{B} },\nn
\ket{R} &\rightarrow \sum_{i j} (D^{-1})^{R}_{ij} \ket{R ji}\to \sum_{ijk} (D^{-1})^{R}_{ij} \ket{R jk}\ket{R k i }.
\end{align}
We have seen from previous sections that open string Hilbert spaces $\mathcal{H}_{\Sigma_{A}},\mathcal{H}_{\Sigma_{B}}$ transform nontrivially under the quantum group symmetry $U(\infty)_{q}$.  However, by the invariance of the quantum trace, we know that the factorized state for $\ket{R}$ is invariant.  Thus the factorization map \eqref{cfact} into the extended Hilbert space respects the quantum group symmetry as promised.    

Notice that even though we have imposed the hole-closing conditions \eqref{spf}, \eqref{cwind}, this does not uniquely determine the factorization map.  In particular these conditions would have been satisfied with a factorization with respect to an un-deformed surface symmetry group\footnote{ In this case, the more natural undeformed surface symmetry group would be $S_{n}$ in each sector with $n$ strings, since the A model partition function depends on dimensions of the symmetry group } ,which does not involve the Drinfeld element.    As we show in section \ref{ssection:factor},  the necessity for the q-deformed edge mode symmetry and the insertion of the Drinfeld element can only be seen when we enforce the E-brane axiom with a choice of a \emph{geometric} state such as the closed unit \eqref{unit}.

\subsection{The open A-model TQFT and sewing relations}
As discussed in the beginning of section \ref{section:closed}, our choice of
factorization maps \eqref{zip}, \eqref{Delta} satisfies a set of sewing relations  in addition to the E-brane axiom.  Here we work out some of these relations explicitly in the open sector.  As in 2D extended TQFT, we find that the A-model open TQFT forms a Frobenius algebra under the product $\mu_{O}$.  We will taking the generating set for this algebra to be 
\begin{align}
      \mathtikz{\muA{0}{0}}, \quad \mathtikz{\deltaA{0}{0}}, \quad \mathtikz{ \etaA{0}{0} }, \quad \mathtikz{ \epsilonA{0}{0}}, \quad \mathtikz{ \tauA{0}{0} }  .
\end{align}
We have already defined the product and coproduct, which satisfy the Frobenius condition:
\begin{align}
    \mathtikz{ \deltaA{0}{-0.5cm} \muA{0}{0.5cm} }
&= \mathtikz{ \idA{4.5cm}{-0.5cm} \muA{3cm}{-0.5cm} \deltaA{4cm}{0.5cm} \idA{2.5cm}{0.5cm} }
= \mathtikz{ \idA{6.5cm}{-0.5cm}\muA{8cm}{-0.5cm}\deltaA{7cm}{0.5cm}\idA{8.5cm}{0.5cm} },
\end{align}
and are associative and co-associative.
Next we can determine the open unit $1_{O}$ and counit $\epsilon$ from the product and coproduct using the defining relations
\begin{align}
\mathtikz{ \etaA{-.5cm}{0 cm} \muA{0}{0}}=
\mathtikz{ \epsilonA{-.5cm}{-1 cm} \deltaA{0}{0}}&=\mathtikz{\idA{0}{0}}.
\end{align}
We find that 
\begin{align}
    1_{O}&=\mathtikz{ \etaA{0cm}{0cm} } = \sum_{R,i,j} (-i)^{l(R)} d_{q} (R) q^{\kappa_{R}/4} (D^{-1})^{R}_{ij} \ket{Rji},\nn
\epsilon&=\mathtikz{ \epsilonA{0cm}{0cm} } : \ket{Rij} \to  \delta_{ij}.
\end{align}
\paragraph{The open pairing, adjoint operation and the quantum trace}
Our open string Frobenius algebra also possesses a nondegenerate bilinear form (the Frobenius form) $\xi$, which defines an adjoint operation on the open string Hilbert space.  This is called the open pairing and can be obtained by gluing the counit $\epsilon$ to the product $\mu_{O}$ 
\begin{align}\label{xi}
\xi=\mathtikz{\pairA{0}{0}}=\mathtikz{ \muA{0cm}{0cm},\epsilonA{0cm }{-1cm}} : \ket{Rij} \ket{R'kl} \to (i)^{l(R)} \delta_{R R'} \frac{D^{R}_{jk} \delta_{il}}{d_{q} (R) q^{\kappa_{R}/4} }.
\end{align}
Notice that our definition of $\xi$ coincides precisely with the large $N$ limit of the bilinear form \eqref{h} and should therefore be related to the large $N$ limit of the $q$-deformed Haar measure.   Its inverse, called the copairing, can be obtained by gluing the unit to the product,
\begin{equation}
\xi^{-1} =\mathtikz{ \deltaA{0cm}{0cm}  \etaA{0cm}{0cm}}=\mathtikz{ \copairA{0cm}{0cm} }: 1 \to \sum_{R,i,j,k} (-i)^{l(R)}d_{q} (R) q^{\kappa_{R}/4} \, (D^{-1})^{R}_{ij} \ket{Rik} \ket{R kj}.
\end{equation}
and satisfies the zigzag identity 
\begin{align} \label{zigzag2}
    \mathtikz{ \copairA{0cm}{0cm} \pairA{1cm}{0cm}}  = \mathtikz{ \pairA{0cm}{0cm} \copairA{1cm}{0cm}} = \mathtikz{\idA{0cm}{0cm}  } : \ket{Rij} \to \ket{Rij} .
\end{align}
The pairing and copairing define an adjoint operation by turning the input Hilbert space to output Hilbert space and vice versa.   For example they relate the unit and product to the counit and coproduct:
\begin{align}
    \mathtikz{ \muA{0cm}{0cm} } = \mathtikz{ \deltaA{0cm}{0cm} \idA{1.5cm}{0cm} \pairA{1cm}{-1cm} \idA{-0.5cm}{-1cm} } ,\quad \mathtikz{\pairA{0cm}{0cm}  \etaA{.5cm}{0cm} } &= \mathtikz{\epsilonA{0cm}{0cm}} .
\end{align}
 They also define a canonical trace operation on open cobordisms by connecting the input Hilbert spaces to output Hilbert space:
\begin{align}\label{partial trace}
      \mathtikz{ \idA{-3.5cm}{0cm}; \pairA{-3cm}{-1cm}; \copairA{-3cm}{0cm}; 
\idA{-1.5cm}{1cm};
\idA{-1.5cm}{-1cm};
\draw[thick] (-3cm,-1cm) rectangle  (-1cm,0cm);} ,\quad \mathtikz{ \idA{-.5cm}{0cm}; \pairA{-1cm}{-1cm}; \copairA{-1cm}{0cm}; 
\idA{-2.5cm}{1cm};
\idA{-2.5cm}{-1cm};
\draw[thick] (-3cm,-1cm) rectangle  (-1cm,0cm);} .
\end{align}
Notice that we have drawn the partial trace to avoid braiding, so the trace on the left/right  side has to closed on the left/right side.  If we violate this rule, we would have to account for the nontrivial braiding that occurs when the strips cross.   Most importantly, this categorical definition of the trace coincides with the quantum trace $\mathcal{A}(\U(\infty)_{q})$ as defined in \eqref{quantum trace}, \eqref{deg}. 

Using the pairing and copairing we can calculate the annulus:
\begin{align} \label{QAtrace}
      \mathtikz{ \pairA{0cm}{0cm} \copairA{0cm}{0cm} }&= \sum_{R}\tr_{R}(D) \tr_{R}(D^{-1} ) ,\nn
      &=\sum_{R} \tr_{R\otimes \bar{R}} (D) = \tr_{q}(1),
\end{align}
where we used the multiplicative property of the trace and unitarity of the representations.  The final expression above is just the quantum trace of the identity operator on the total open string Hilbert space. 

The final generator of our open Frobenius algebra is the braiding operator 
\begin{align} \label{braiding} 
    \mathcal{B}= \tau \circ \mathcal{R}_{\text{string}}  = \mathtikz{\tauA{0}{0}}: \mathcal{A} \otimes \mathcal{A} \to\mathcal{A} \otimes \mathcal{A},
\end{align}
where $\tau $ is the operation that exchanges two copies of the open string Hilbert space.
The operation $\mathcal{R}_{\text{string}}$ refers to the $R$ matrix which describes the braiding of the open strings.  
This is nontrivial, in contrast with the usual 2D open-closed TQFT \cite{Moore:2006dw}, since the left/right string endpoints themselves have nontrivial braiding.    However, since we will not require the braiding operation in our calculation of entanglement entropy, we will leave this for future work. 

\subsection{The open closed sewing axioms and factorization of the Hartle-Hawking state}
\label{ssection:factor} 
We have seen that the factorization map $\Delta$ extends consistently to a Frobenius algebra describing the open sector of the A-model TQFT.   We now consider the open-closed sewing axioms \cite{Moore:2006dw} which enforce the compatibility of the open string algebra with the closed string algebra defined by  \eqref{unit}-\eqref{cylinder}.  We will pay particularly close attention to the consistency of the Chern class labelings, which places additional constraints on the factorization.  This is because one could obtain factorization maps that satisfy \eqref{cwind} and \eqref{spf}, and extend to a consistent open Frobenius algebra, but is nevertheless incompatible with the A-model TQFT restricted to Calabi-Yau manifolds\footnote{For example, since the $(0,0)$ sector of the A model TQFT is a closed algebra which is isomorphic to the Frobenius algebra of an ordinary 2D TQFT, we could simply use the factorization maps with respect to an undeformed $S_{n}$ surface symmetry.  The E-brane condition would then be defined with the $(0,0)$ cap, giving a factorization that is compatible with the $(0,0)$ sector of the A-model TQFT, but incompatible with the Calabi-Yau condition.}.  As an application of this machinery, we give a simple factorization of the Hartle-Hawking state.

The relation between the closed and open sector is given by the zipper and cozipper, which are algebra/coalgebra homomorphisms between the respective Frobenius algebras. Keeping track of the Chern class on the closed cobordisms, the homomorphism property is equivalent to the sewing relations
\begin{align}\label{ziph}
\mathtikz{ \zipper{0cm}{1cm} \etaC{0cm}{1cm} \node at (0,1.5){(0,-1)}; } = 
\mathtikz{ \etaA{0cm}{0cm} },\quad \quad 
\mathtikz{ \zipper{0cm}{0cm} \muC{0cm}{1cm} \node at (0,1.5){(0,1)}; } =  
\mathtikz{ \muA{0cm}{0cm} \zipper{-0.5cm}{1cm} \zipper{0.5cm}{1cm} },
\end{align}
\begin{align}\label{cziph}
\mathtikz{  \epsilonC{0cm}{0cm}\node at (0,-0.5){(0,-1)}; \cozipper{0cm}{1cm} } = 
\mathtikz{ \epsilonA{0cm}{0cm} },\quad \quad 
\mathtikz{  \deltaC{0cm}{-1cm} \node at (0,-2.5){(1,0)}; \cozipper{0cm}{0cm} } =  
\mathtikz{ \deltaA{0cm}{0cm} \cozipper{-0.5cm}{-1cm} \cozipper{0.5cm}{-1cm} },
\end{align}
which is satisfied by our open-closed cobordisms.   The left diagrams above express the fact that the unit/counit is preserved by the zipper/cozipper.  
\paragraph{Compatibility of E-brane and Calabi-Yau condition}
The E-brane axiom for the A model restricted to Calabi-Yau manifolds is
\begin{align} \label{EbraneCY}
     \mathtikz{\node at (0,.6){(0,-1)}; \etaC{0cm}{0cm} } = \mathtikz{  \etaA{0cm}{0cm} \cozipper{0cm}{0cm}
\draw (0cm,0.5cm) node {\footnotesize $e$};
}. 
\end{align}
The Chern class labelling on the left, as defined by the Calabi-Yau condition, places a strong and non local constraint on the edge modes and factorization map.   We have seen in section \ref{section:cboson} that the A-model amplitude on the ``Calabi-Yau cap" on the LHS of \eqref{EbraneCY} gives the entanglement boundary state 
\begin{align}\label{Bstate}
    \ket{D}= \sum_{R} (-i)^{l(R)}d_{q}(R) q^{\kappa_{R}/4} \ket{R} .
\end{align}
with a prescribed holonomy matrix $D$.  The nonlocality of the Calabi-Yau condition is expressed by the fact that $D$ is not the identity, so it cannot be equivalent to a local boundary condition at the entangling surface.   This nonlocality requires that the extension of the A-model closed TQFT be compatible with a q-deformed surface symmetry group $\U(\infty)_{q}$.  In particular, the quantum trace defined in \eqref{quantum trace} automatically incorporates the entanglement boundary condition by insertion of the Drinfeld element. 

When viewed from the open string channel, the boundary state \eqref{Bstate} inserts a large $N$ number of E-branes at the entangling surface, giving a geometric realization of the string edge modes.   The $\ket{D}$ is therefore the E-brane boundary state which realizes Susskind and Uglum's proposal in the A-model target space.  

\paragraph{Factorization of the HH state}
We now apply our factorization map to the Hartle-Hawking state and its dual at $t=0$:
\begin{align}\label{HHF}
    \ket{HH} &\to \mathtikz{ \node at (0,1.5){(0,-1)};\deltaA{0cm}{0cm} \zipper{0cm}{1cm} \etaC{0cm}{1cm} }
= \mathtikz{ \deltaA{0cm}{0cm} \etaA{0cm}{0cm} } 
= \mathtikz{ \copairA{0cm}{0cm} }
=  \sum_{R,i,j,k} (-i)^{l(R)} d_{q} (R) q^{\kappa_{R}/4} (D^{-1})^{R}_{ij} \ket{Rik}\ket{Rkj}, \nn
    \bra{HH^*} &\to \mathtikz{\node at (0,-2.7){(-1,0)}; \epsilonC{0cm}{-2cm} \muA{0cm}{0cm} \cozipper{0cm}{-1cm} }
= \mathtikz{  \epsilonA{0cm}{-1cm} \muA{0cm}{0cm} } 
= \mathtikz{ \pairA{0cm}{0cm} }: \ket{Rij} \ket{R'kl} \to (i)^{l(R)} \delta_{R R'} \frac{D^{R}_{jk} \delta_{il}}{d_{q} (R) q^{\kappa_{R}/4} }.
\end{align}

Using this factorization map the A-model partition function on the resolved conifold can be given a canonical open string interpretation:
\begin{align}\label{sph}
    Z= \braket{HH^{*}|HH} &\to  \mathtikz{\pairA{0}{0}\copairA{0}{0}} = \tr_{q}(e^{-t H_{\text{open}}}), \nn
    &= \sum_{R} ( d_{q}(R))^{2} e^{-tl(R)} =\mathtikz{
\node at (0,1/2){(0,-1)};
\epsilonC{0}{0};
\etaC{0}{0};
\node at (0,-1/2){(-1,0)}} .
\end{align}
where we have defined the open string modular Hamiltonian 
\begin{align}\label{mod}
     H_{\text{open}} \ket{Rij}= l(R) \ket{Rij}.
\end{align}
Even though we have drawn the same diagram as in the $t=0$ case, we have included an open string propagator $e^{-t H_{\text{open}}} $ which introduces an explicit $t$ dependence.     Equation \eqref{sph} gives an explicit realization of the Susskind-Uglum proposal to interpret the closed string amplitude as a trace over an open strings which end on the entangling surface.  

\paragraph{Compatibility of open and closed string pairings}
The closed pairing \eqref{cpair} defines the adjoint operation in our closed string TQFT which maps the Hartle-Hawking state to its dual.  Having defined an extension to the open TQFT with an adjoint operation given by the open pairing \eqref{xi}, we should check that these two adjoint operations are compatible. This is a consequence of the  E-brane axiom together with the right diagrams in \eqref{ziph} \eqref{cziph}, which states that zipper/cozipper respects the  multiplication/comultiplication.  
Explicitly, we can glue the counit to both sides of the right diagram in \eqref{ziph} :
\begin{align}
\mathtikz{ \zipper{0cm}{0cm} \muC{0cm}{1cm} \node at (0,1.5){(0,1)}; \epsilonA{0cm}{-1cm} } =  
\mathtikz{ \muA{0cm}{0cm} \zipper{-0.5cm}{1cm} \zipper{0.5cm}{1cm} \epsilonA{0cm}{-1cm} }.
\end{align}
On the left diagram, we apply the E-brane axiom in the form:
\begin{align} \label{czipe}
   \mathtikz{ \zipper{0cm}{0cm} \epsilonA{0cm}{-1cm}}= \mathtikz{\epsilonC{0}{0};
\node at (0,1/2){(-1,0)}},
\end{align} which then implies
\begin{align} \label{ocpair}
    \mathtikz{ \pairC{0}{0} \node at (0,.5){(-1,1)}  }  =  
\mathtikz{  \pairA{0cm}{0cm} \zipper{-0.5cm}{1cm} \zipper{0.5cm}{1cm} }.
\end{align}
This expresses the compatibility of the open and closed pairing, with analogous relations holding for the copairing. 
Note that in both \eqref{czipe}  and \eqref{ocpair}, the E-brane axiom is satisfied only for a specific Chern class labelling compatible with the Calabi-Yau constraint defining our closed string algebra.

Finally note that the zipper and cozipper are adjoint operations, which is implied by the third sewing axiom in \eqref{msegal}:
\begin{align}
\mathtikz {\pairA{0cm}{0cm}
\zipper{-0.5cm}{1cm} \idA{0.5cm}{1cm} }
\quad &= \quad\mathtikz{ \node at (0,-1.5){(-1,1)};
\pairC{0cm}{0cm}
\idC{-.5cm}{1cm} \cozipper{.5cm}{1cm} } .
\end{align}
By gluing the copairing to the right input of this relation and applying the zigzag identity, we find that the cozipper is the adjoint of the zipper
\begin{align}
    \mathtikz{\cozipper{0}{0}}= \mathtikz{\zipper{-.5cm}{1cm} \copairC{0cm}{1cm} \pairA{-1cm }{0cm}}.
\end{align}

\subsection{The reduced density matrix for the Hartle-Hawking state and a canonical calculation of entanglement entropy}
\label{entropy}
The reduced density matrix for the Hartle-Hawking state is easily derived from the factorization map \eqref{HHF}. First note that unnormalized density matrix\footnote{ $\tilde{\rho}$ is defined by $\tr (\tilde{\rho} O) = \braket{HH^{*}|O|HH}$ for any operator $O$ with the trace defined by the closed pairing/copairing.  This has the same structure as density matrices in non-Hermitian systems\cite{Couvreur:2016mbr}.} $\tilde{\rho}$ for the Hartle Hawking state factorizes as
\begin{align}
    \tilde{ \rho }=  \ket{HH}\bra{HH^*}= \mathtikz{ \node at (0,.7){(-1,0)}; \node at (0,-.75){(0,-1)};  \epsilonC{0cm}{.4cm}\etaC{0cm}{-.4cm}} \to \mathtikz{\pairA{0cm}{1cm} \copairA{0cm}{-1cm} } .
\end{align}

The corresponding reduced density matrix is given by the (quantum) partial trace, which is defined by \eqref{partial trace}, over the subregion $\Sigma_{B}$ :
\begin{align} \label{red}
    \tilde{\rho}_{A} =\tr_{B}\tilde{\rho}  = \mathtikz{\pairA{0cm}{1cm} \copairA{1cm}{1cm} \copairA{0cm}{-1cm} \pairA{1cm}{-1cm} \idA{1.5cm}{0cm}\idA{1.5cm}{1cm}} =\mathtikz{\idA{0cm}{0cm}\idA{0cm}{1cm}}=\sum_R e^{-t l(R)} \mathbf{1}_{R}=\sum_{R,i,j} e^{-t l(R)}  |Rij\rangle \bra{Rij}.
\end{align}
where we have applied the zigzag identity and absorbed the area dependence into the propagator represented by the strip.  Note that we have applied a quantum partial trace defined by the pairing and copairing.  This operation cancels the insertions of $D$ and $D^{-1} $ in the density matrix which would have led to non local boundary conditions for the modular Hamiltonian.  We will comment more on this in the next section. 

As in the case of undeformed gauge theory, the form of the reduced density matrix \eqref{red} is dictated by symmetry.  The action of $\U(\infty)_{q}$ must commute with $\tilde{\rho}_{A}$, since our factorization map \eqref{cfact} respects the quantum group symmetry.  Schur's lemma then requires the reduced density matrix to act as the identity in each irreducible representation $R$, leading to the block-diagonal form of \eqref{red}.  Note that while the degeneracy associated with each irreducible representation $R$ is generic (it holds for any gauge-invariant state in the theory), the modular Hamiltonian \eqref{mod} actually has a much larger degeneracy, since all representations with the same number of boxes has the same modular energy.

Tracing over $\Sigma_{A}$ gives the expected normalization 
\begin{align}
Z=\tr_{A}( \tilde{\rho}_{A} )= \sum_{R} (d_{q}(R))^{2} e^{-t l(R)}.
 \end{align}
It is useful to express the normalized reduced density density matrix  $\rho_{A}=\tilde{\rho}_{A}/Z $ as a direct sum over normalized\footnote{Normalized according to the quantum trace.} operators $ \frac{ \mathbf{1}_{R} }{(d_{q}R)^2} $ in each superselection sector labelled by $R$:
\begin{align} \label{rhoA}
    \rho_{A} &= \oplus_{R} p(R) \frac{ \mathbf{1}_{R} }{(d_{q}R)^2},\nn
    p(R)&= \frac{(d_{q}(R))^{2} e^{-t l(R)}}{Z}.
\end{align}
The $q$-deformed entanglement entropy can be directly evaluated from
\begin{align}
S= - \tr_{q} (\rho_{A}\log \rho_{A})=- \tr (D \rho_{A}\log \rho_{A}).
\end{align} 
This type of $q$-deformed entropy has been studied previously in the context of quantum group invariant spin chains \cite{Couvreur:2016mbr, Quella:2020aa}.  Here we have seen that the use of the quantum trace arises naturally from the requirement of quantum group symmetry as dictated by the cobordisms of the open-closed TQFT.  

Since the spectrum of the $\rho_{A}$ can be read off from \eqref{rhoA}, we can compute the entanglement entropy without appealing to the replica trick: 
\begin{align}
S&=-\sum_{R}\tr_{q} \left(  \frac{ p(R) \mathbf{1}_{R} }{(d{q}R)^2} \log   \frac{ p(R) \mathbf{1}_{R} }{(d_{q}R)^2}\right )=-\sum_{R}\tr_{q}(\mathbf{1}_{R} )   \frac{ p(R)  }{(d_{q}R)^2} \log   \frac{ p(R) }{(d_{q}R)^2}, \nn
&=\sum_{R} \left(- p(R) \log p(R) + 2 p(R) \log d_{q}R\right).
\end{align}
This gives the sought after canonical calculation of entanglement entropy which agrees with the replica trick answer in section \ref{ssection:replica}  

\subsection{Revisiting the replica trick on the resolved conifold}
As discussed previously, the resolved conifold is a nontrivial vector bundle  $\mathcal{O}(-1)\oplus \mathcal{O}(-1)\rightarrow S^{2}$. In section \ref{ssection:replica} we gave a prescription for the replication of this geometry in which the volume of the base manifold is replicated without affecting the bundle structure.   Here we explain this prescription, first in terms of the our categorical formulation of the reduced density matrix,  and then by appealing to a direct geometric construction of the replica manifold. 

\paragraph{Replica trick in terms of cobordisms}
Using the reduced density matrix \eqref{red} for $\ket{HH}$, we can apply  the replica trick in the form
\begin{align} \label{CYrep}
   S  =\pd_{n} \tr_{q} (\rho_{A}^n) |_{n=1}=\pd_{n} \tr (D \rho_{A}^n) |_{n=1}.
\end{align}
Note that we did not replicate $D$ because it is merely part of the definition of the quantum trace. In terms of cobordisms, the $n$th power of $\rho_{A}$ is simply a long strip, and $\tr_{q} (\rho_{A}^n)$ is a large annulus  with one insertion of $D$ and $D^{-1}$ as in the $n=1$ case.
\begin{align}
    \rho_{A}^n= \oplus_{R} \frac{d_{q}(R)^{2} e^{- n t H_{\text{open}}}}{Z_{1}^n}  \frac{\mathbf{1}_R}{d_{q}(R)^{2}}, \nn
    \tr_{q} \rho_{A}^n = \frac{\sum_{R} d_{q}(R)^{2} e^{- n t H_{\text{open}}} }{Z_{1}^n},
\end{align}
and the only effect of the replication is to rescale the area factor $t$ by a factor of $n$.
This replicated partition function agrees with the prescription given in \eqref{Zalpha} and is proportional to the resolved conifold partition function, indicating that Calabi-Yau condition is preserved. 

\begin{figure}[ht]
\centering
\includegraphics[scale=.6]{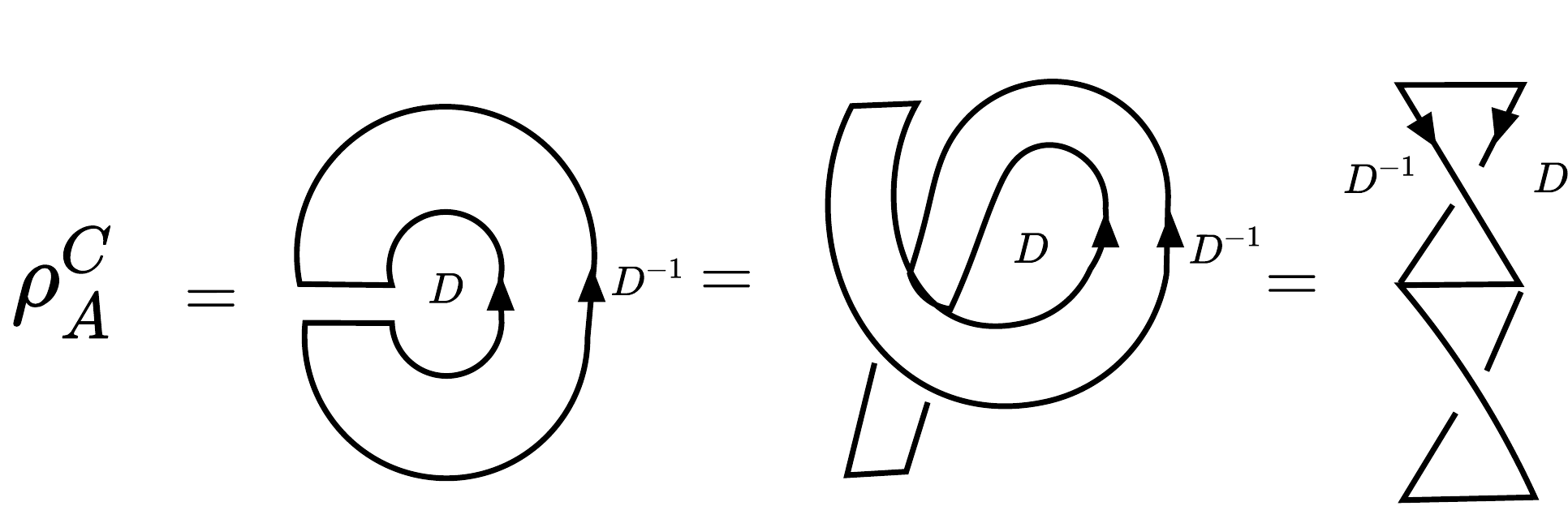}
\caption{ The reduced density matrix $\rho_{A}^C$ defined using a noncanonical trace operation fails to satisfy the E-brane axiom when it is replicated. It also does not commute with the edge mode symmetry group }
\label{Cfig}
\end{figure}

The main reason that the Calabi-Yau condition is preserved is the use of the quantum partial trace in \eqref{red}.  To see this, consider an alternative replication in which we use a naive trace, corresponding to simply gluing the Hartle-Hawking state and its dual along region $\Sigma_{B}$ without the application of state-channel map as in \eqref{red}. 
\begin{align}
    \rho^{C}_{A}= \mathtikz{\copairA{0cm}{.5cm}  \pairA{0cm}{-.5cm} \idA{.5cm}{.5cm} } : \ket{Rij} \to D^{R}_{lj} (D^{-1})^{R}_{ik} e^{-tl(R) } \ket{Rkl}. 
\end{align}
As shown in figure \ref{Cfig}, $\rho^{C}_{A}$ differs from $\rho_{A}$ because of the nontrivial braiding of the open string, so that when we ``straighten" the cobordism for $\rho_{A}^{C}$ we get a nontrivial double twist diagram instead of a strip.  $\rho_{A}^{C}$ also does not commute with the quantum group symmetry that permutes the edge modes.   When we replicate $\rho_{A}^{C}$, the Wilson lines $D$ and $D^{-1}$ do not cancel. As a result we find that
\begin{align}
     \tr (\rho^{Cn}_{A}) =\sum_{R} \tr_{R}(D^n) \tr_{R}(D^{-n}) ,
\end{align}
which does not satisfy the Calabi-Yau condition and gives an entropy inconsistent with the replica prescription of section \ref{ssection:replica}. The problem is that the entanglement boundary condition is violated each time we replicate this density matrix. 
A simple way to compensate for this is to insert a factor of $D^{-1}$ each time we replicate $\rho_{A}^{C}$.  The replica entropy \eqref{CYrep}  can then by expressed as 
\begin{align}
   S= \pd_{n} \tr (D \rho_{A}^n) |_{n=1}=  \pd_{n} \log \tr (D^{1-n}  (\rho^{C}_{A})^n) |_{n=1},
\end{align}
where we have introduced the $\log$ to account for normalizaton.   
These are the direct analogue of  the replica entropy in \cite{Jafferis:2019wkd}, with $D$ playing the role of the ``defect" operator.

\paragraph{Geometric replication of the resolved conifold geometry }
Rather than appealing to the categorical formulation of the density matrix, we can also try to replicate the resolved conifold geometry directly by appealing to the usual multi-sheeted construction of the replica manifold.  This will show more explicitly the geometrical role played by the defect operator as a topological twisting. 

In \cite{Aganagic:2004js,Bryan:2004iq}, it was shown that one can compute topological A-model partition function on $L_1\oplus L_2\rightarrow \Sigma_{g},$ for a 2 dimensional surface $\Sigma_{g}$ with genus g. It was then further shown that one can glue $L_1\oplus L_2\rightarrow \Sigma_1$ and $L_1'\oplus L_2'\rightarrow \Sigma_2,$ given a gluing map $i:\partial \Sigma_1\rightarrow \partial\Sigma_2,$ to compute the topological A-model partition function on $(L_1+L_1')\oplus(L_2+L_2')\rightarrow \Sigma_1\cup\Sigma_2.$ 

We define the Hartle-Hawking state to be the topological A-model partition function on $\mathcal{O}_1\oplus\mathcal{O}_2(-1)\rightarrow D^2_1$
\begin{equation}
    |HH\rangle=    \mathtikz{
        \draw[thick] (0,0) arc (-90:90:1);
        \draw[dashed] (0,2) arc (90:270:1);
        \node[black] at (-1.5,1){$A_1$};
        \node[black] at (1.5,1){$B_1$};
        \filldraw[black] (0,1) circle (2pt);
    },\label{disk HH}
\end{equation}
where the black dot in \eqref{disk HH} represents a pole of a local section in $\mathcal{O}_2(-1).$ For later use, we split $\partial D^2_1=A_1\cup B_1$. In the similar way, we define a dual of the Hartle-Hawking state to be topological A-model partition function on $\mathcal{O}_1(-1)\oplus\mathcal{O}_2\rightarrow D^2_2$
\begin{equation}
    \langle HH^*|=    \mathtikz{
        \draw[dashed] (0,0) arc (-90:90:1);
        \draw[thick] (0,2) arc (90:270:1);
        \node[black] at (-1.5,1){$B_2$};
        \node[black] at (1.5,1){$A_2$};
        \filldraw[blue] (0,1) circle (2pt);
    },\label{disk HH dual}
\end{equation}
where the blue dot in \eqref{disk HH dual} represents a pole of a local section in $\mathcal{O}_1(-1).$ 

To construct $|HH\rangle \langle HH^*|,$ we prepare  $\mathcal{O}_1\oplus \mathcal{O}(-1)\rightarrow D_1^2$ and $\mathcal{O}_1(-1)\oplus \mathcal{O}_2\rightarrow D_2^2.$ 
\begin{equation}
    |HH\rangle \langle HH^*|=  \mathtikz{
        \draw[thick] (0,0) arc (-90:90:1);
        \draw[dashed] (0,2) arc (90:270:1);
        \node[black] at (-1.5,1){$A_1$};
        \node[black] at (1.5,1){$B_1$};
        \filldraw[black] (0,1) circle (2pt);
    }\otimes   \mathtikz{
        \draw[dashed] (0,0) arc (-90:90:1);
        \draw[thick] (0,2) arc (90:270:1);
        \node[black] at (-1.5,1){$B_2$};
        \node[black] at (1.5,1){$A_2$};
        \filldraw[blue] (0,1) circle (2pt);
    }.
\end{equation}
Now we can obtain a reduced density matrix $\rho_{red}=\tr_{B_1\sim B_2}( |HH\rangle\langle HH^*|)$ by identifying $B_1\in \partial D_1^2$ and $B_2\in \partial D_2^2.$ We expect that $\rho_{red}$ is equivalent to $\rho_A,$ but we have not explicitly verified this claim.
\begin{equation}
    \rho_{red}=\mathtikz{
        \draw[thick] (0,0) arc (-90:90:1);
        \draw[dashed] (0,2) arc (90:270:1);
        \node[black] at (-1.5,1){$A_1$};
        \node[black] at (1.5,1){$A_2$};
        \filldraw[black] (-0.5,1) circle (2pt);
        \filldraw[blue] (0.5,1) circle (2pt);
    }.\label{reduced density HH}
\end{equation}
One can check, under the identification $A_1\sim A_2,$ $\tr_{A_1\sim A_2}(\rho_{red})$ computes topological A-model partition function of $\mathcal{O}_1(-1)\oplus\mathcal{O}_2(-1)\rightarrow S^2.$ 

Let us consider a replicated geometry of \eqref{reduced density HH}. First we prepare two copies of \eqref{reduced density HH}
\begin{equation}
    \rho_{red}\otimes\rho_{red}=\mathtikz{
        \draw[thick] (0,0) arc (-90:90:1);
        \draw[dashed] (0,2) arc (90:270:1);
        \node[black] at (-1.5,1){$A_1$};
        \node[black] at (1.5,1){$A_2$};
        \filldraw[black] (-0.5,1) circle (2pt);
        \filldraw[blue] (0.5,1) circle (2pt);
    }\otimes\mathtikz{
        \draw[dashed] (0,0) arc (-90:90:1);
        \draw[thick] (0,2) arc (90:270:1);
        \node[black] at (-1.5,1){$A_3$};
        \node[black] at (1.5,1){$A_4$};
        \filldraw[blue] (-0.5,1) circle (2pt);
        \filldraw[black] (0.5,1) circle (2pt);
    }.
\end{equation}
In order to compute $\rho_{red}^2,$ we then identify $A_2\sim A_3$
\begin{equation}
    \rho_{red}^2=\mathtikz{
        \draw[thick] (0,0) arc (-90:90:1);
        \draw[dashed] (0,2) arc (90:270:1);
        \node[black] at (-1.5,1){$A_1$};
        \node[black] at (1.5,1){$A_4$};
        \filldraw[black] (-0.6,1) circle (2pt);
        \filldraw[black] (-0.3,1) circle (2pt);
        \filldraw[blue] (0.3,1) circle (2pt);
        \filldraw[blue] (0.6,1) circle (2pt);
    }.\label{naive replica HH}
\end{equation}
Note that as a result of the replication, volume of the base manifold is doubled. This na\"ive replica trick \eqref{naive replica HH} has a problem. To illustrate the problem, let us compute $\tr_{A_1\sim A_4} (\rho_{red}^2).$ Because there are two poles for each section of line bundles $L_1$ and $L_2,$ one can deduce that $L_1=\mathcal{O}(-2)$ and $L_2=\mathcal{O}(-2)$ whereas topology of the base manifold is still of $S^2.$ Then the manifest problem occurs as $\mathcal{O}_1(-2)\oplus \mathcal{O}_2(-2)\rightarrow S^2$ is not a Calabi-Yau manifold. To ensure that the replicated geometry is Calabi-Yau, we apply a topological twisting by $\mathcal{O}_1(1)\oplus\mathcal{O}_2(1),$ which we will represent by $\mathcal{O}_{\text{twist}}\equiv D^{-1}$ 
\begin{equation}
    \mathcal{O}_{\text{twist}}\rho_{red}^2=\mathtikz{
        \draw[thick] (0,0) arc (-90:90:1);
        \draw[dashed] (0,2) arc (90:270:1);
        \node[black] at (-1.5,1){$A_1$};
        \node[black] at (1.5,1){$A_4$};
        \filldraw[black] (-0.5,1) circle (2pt);
        \filldraw[blue] (0.5,1) circle (2pt);
    }.
\end{equation}
As a result, $\tr_{A_1\sim A_4}(\mathcal{O}_{\text{twist}}\rho_{red}^2)$ computes the topological A-model partition function of $\mathcal{O}_1(-1)\oplus\mathcal{O}_2(-1)\rightarrow S^2,$ where the volume of the base manifold is doubled,
\begin{equation}
    Z_2=\tr_{A_1\sim A_4}(\mathcal{O}_{\text{twist}}\rho_{red}^2)=\sum_R (d_q R(g_s))^2 e^{-2 l(R)t}.\label{eqn:replicated partition function}
\end{equation}

One can easily generalize \eqref{eqn:replicated partition function} to
\begin{equation}
    Z_n=\tr(\mathcal{O}_{\text{twist}}^{n-1}\rho_{red}^n)=\sum_R (d_q R(g_s))^2 e^{-n l(R)t}.
\end{equation}
As a result, we obtain the entanglement entropy
\begin{align}
    S=&\frac{\partial}{\partial n}\left.\frac{Z_n}{Z_1^n}\right|_{n=1}=-\sum_R p(R)(\log(p(R))-2\log(d_qR)),
\end{align}
where
\begin{equation}
    p(R)=\frac{ (d_qR)^2 e^{-l(R)t}}{\sum_R (d_qR)^2 e^{-l(R)t}}.
\end{equation}

Finally, we want to express the entanglement entropy in terms of the BPS index. First, we rewrite the n-sheeted partition function in terms of the BPS index.
\begin{equation}
    Z_n=\exp\left(\sum_k n_{S^2}^0\frac{1}{k}\left(2\sin \frac{kg_s}{2}\right)^{-2} e^{-nkt}\right),
\end{equation}
where $n_{S^2}^0=1$ is the only non-vanishing GV invariant of the resolved conifold. As a result, the entanglement entropy is expressed as
\begin{equation}
S=\sum_{k} n_{S^2}^0 \left(\frac{1}{k}+t\right)\left(2\sin\frac{kg_s}{2}\right)^{-2} e^{-kt}. \label{EE bps resolved}  
\end{equation}
It is very interesting to observe that \eqref{EE bps resolved} is proportional to the number of BPS states, including the multi-particle states. In fact, the linear dependence of the EE on the BPS index is not a special feature of the resolved conifold. For a general non-compact Calabi-Yau of the form
\begin{equation}
    L_1\oplus L_2\rightarrow \mathcal{S},
\end{equation}
the linear dependence continues to hold
\begin{equation}
    S=\sum_{\beta,g,k} n_{\beta}^g\left(\frac{1}{k}+t_\beta\right)\left(2\sin\frac{k g_s}{2}\right)^{2g-2}Q^{k\beta},\label{EE BPS}
\end{equation}
if one replicates the geometry while fixing the topology of the replicated Calabi-Yau.

\section{Discussion}
In this work we have given a factorization of the A-model closed string Hilbert space and a canonical calculation of the entanglement entropy for the Hartle-Hawking state  on the resolved conifold.  
The factorization maps \eqref{spf}, \eqref{cwind} and associated string edge modes are determined by solving the sewing relations of the A-model extended TQFT. These sewing relations, particularly the E-brane axiom, were chosen to be compatible with the Calabi-Yau condition.   
This constraint imposes a nontrivial holonomy $D$ \eqref{D} along the entangling surface, which is captured by the entanglement boundary state $\ket{D}$.
This boundary condition is local in the sense that it can be introduced without affecting the state, but is nonlocal with respect to the ``modular time'' going around the entangling surface.
We then interpret this as an E-brane boundary state by showing that in the open string channel it corresponds to the insertion of a large $N$ number of E-branes at the entangling surface.  We view this as a realization of Susskind and Uglum's proposal \cite{Susskind:1994sm} in the target space of the A-model string theory.  Finally we found that the compatibility of the E-brane axiom with the Calabi-Yau condition requires edge modes to transform in a $q$-deformed edge mode symmetry group.
This $q$-deformation changes the statistics of the open strings: they are no longer bosonic strings but obey anyonic statistics.
Invariance under the quantum group symmetry requires the introduction of the Drinfeld element into the factorization map, and leads to the appearance of quantum dimensions in the entanglement entropy.
In a follow-up paper, we will relate this calculation to the dual Chern-Simons description of the A-model, where quantum dimensions also appear.

The use of extended TQFT techniques was crucial in making our closed string factorization maps self-consistent.  
However our proposed extension of the A-model TQFT is not yet complete, since we did not consider sewing relations which involve the braiding operator \eqref{braiding}. 
We also worked entirely in the target space theory, whereas D-branes are usually formulated in the first-quantized, worldsheet point of view and we do not know how to formulate the E-brane boundary condition on the worldsheet.
A direct check along this direction would be to quantize open strings stretched between intersecting D-branes on $\mathcal{L}$ and $\mathcal{L}'$ as shown in figure 10 and check whether this description agrees with the E-brane calculation we present in this paper.
We leave these problems to future work.

\begin{figure}[h]  \label{intersects}
\centering
\includegraphics[scale=.4]{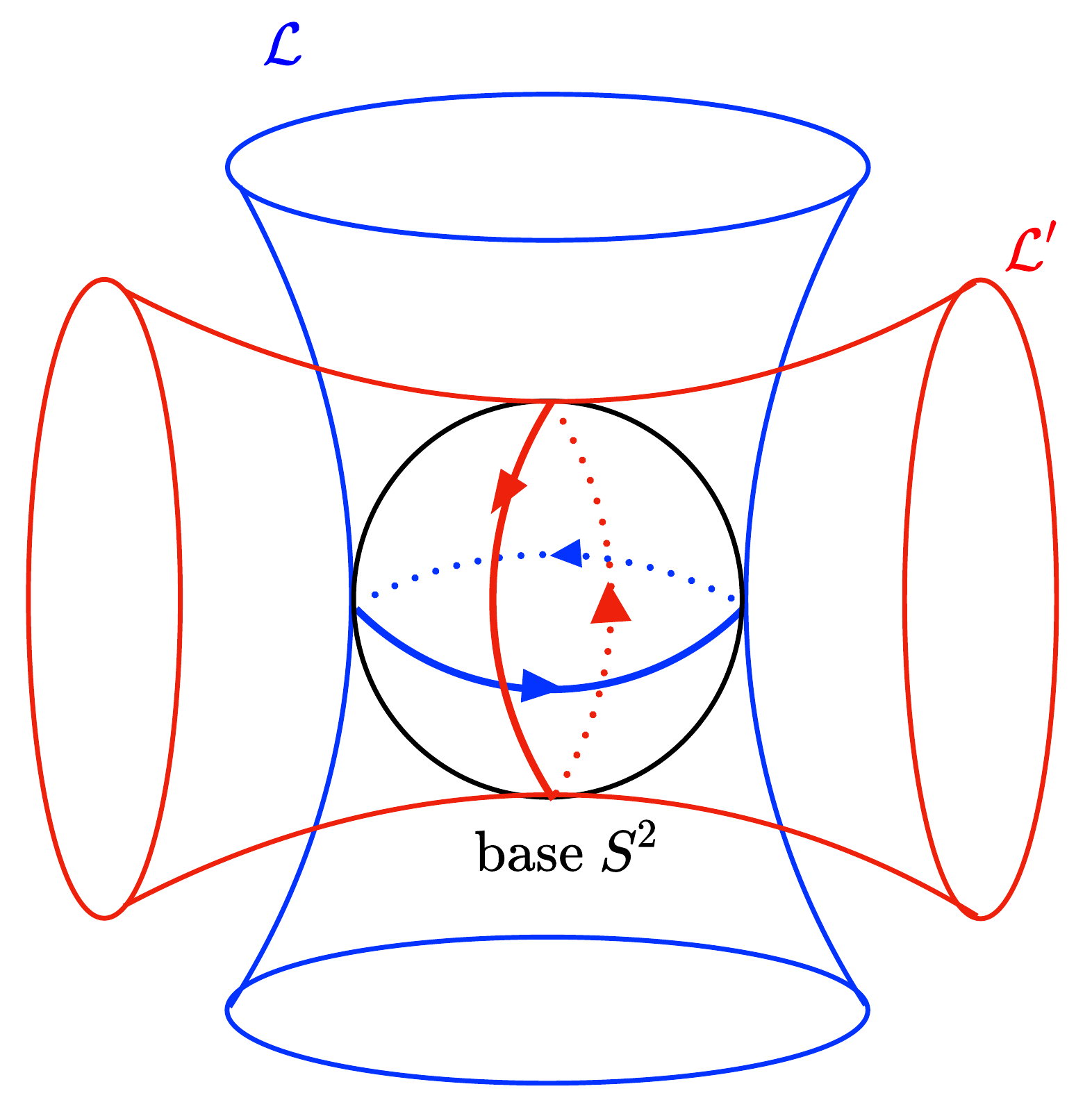}
\caption{D-branes on $\mathcal{L}'$ intersect with D-branes on $\mathcal{L}$.}
\end{figure} 
\paragraph{Analogy to JT gravity}
 The Drinfeld element $D$ can be viewed as an operator on the open string Hilbert space. It is incorporated into the definition of the quantum trace \eqref{Qtr}, which agrees with the categorical trace defined by elements of the open string Frobenius Algebra.  However, as shown in section \ref{entropy}, we can also interpret $D$ as a  ``defect" operator whose insertion at the entangling surface enforces a topological constraint, which is equivalent to filling in the hole with a Calabi-Yau cap. 

An analogous defect operator was found in the factorization of JT gravity \cite{Jafferis:2019wkd}\footnote{\cite{Kitaev_2019} considered a statistical mechanical model for JT gravity which also gave rise to the analogue of this defect operator when attempting to write the partition function on the disk as a trace over a Hilbert space.}.  In that work, the topological  constraint analogous to the Calabi Yau condition in the A-model is the gravitational constraint imposed on the BF gauge theory description of JT gravity.   This constraint is needed because while the variables of JT gravity can be mapped to the BF gauge theory, there are gauge theory configurations such as those with trivial Chern class which are not allowed in the JT gravity path integral.  In particular, the analogue of the E-brane condition for JT gravity requires that the hole can be filled in such a way to reproduce the Einstein-Hilbert term on a disk, which is a topological invariant that can only be captured with nontrivial holonomy of the BF gauge field around the hole.   The defect operator in JT gravity implements this nontrivial holonomy around the entangling surface, just like the Drinfeld element in the A-model.   These similarities suggest that the defect operator in JT gravity might also be viewed as a limit of the 
Drinfeld element of a quantum group surface symmetry. 

There are other indications that quantum groups play an important role in JT gravity as well.  In \cite{2019JHEP...09..066B},  it was proposed that the edge mode symmetry of JT gravity is given by the semi-group $SL^{+}(2,\mathbb{R})$ as a $q \to 1$ limit of $SL_q^{+}(2,\mathbb{R})$, based on the fact that JT gravity can be obtained from the extremal limit of the dimensional reduction of 3D gravity, whose dynamics is connected to the representation theory of the quantum group $SL_q^{+}(2,\mathbb{R})$\cite{Witten:1988hc, 2001CQGra..18R.153T, 2013JHEP...11..208M, 2020arXiv200610105D, 2019JHEP...09..066B, 2020arXiv200607072M}. In \cite{2013JHEP...11..208M}, it was also observed that the Bekenstein-Hawking entropy for 3d BTZ black holes can be reproduced in the large charge limit by the topological entanglement entropy related to the quantum dimensions in Liouville theory.  It will be very interesting to see if there is a canonical way to directly justify the origin of the above observation.

One way to see quantum group symmetry appearing in JT gravity is via the Sachdev-Ye-Kitaev model, for which JT gravity can be viewed as an infrared effective theory.  
Specifically, Ref.~\cite{Berkooz:2018jqr} studied correlation functions in a ``double-scaling'' limit of SYK and found evidence of quantum group symmetry such as $q$-deformed $6j$ symbols.
This suggests that the bulk dual of the double-scaled SYK model could be identified as a TQFT with quantum group symmetry like the one described here for the A-model string.
Such a TQFT would be a $q$-deformation of JT gravity which might elucidate the appearance of $q \to 1$ limits of quantum group structures in JT gravity.

In this work, we have calculated entanglement entropy of topological A-model on a fixed geometry: the resolved conifold. We have made use of a TQFT formalism in which the topology of spacetime is fixed rather than summed over. 
This is analogous to the entanglement entropy on the hyperbolic disk in JT gravity \cite{2018arXiv180706575L, Jafferis:2019wkd, 2019JHEP...09..066B}. However, JT gravity can be UV completed by a random matrix model by summing over all different topologies, which is interpreted as being dual to an ensemble average of theories \cite{2019arXiv190311115S}. Recently, it has also been shown in the context of models related to JT gravity that topology-changing processes play an important role in understanding the black hole information paradox when we calculate the entropy using the replica trick \cite{2019arXiv191111977P, 2020JHEP...05..013A}. There is an analogue UV completion of topological string theory by including topology-changing processes via nonperturbative effects in the context of $q$-deformed 2d Yang-Mills theory \cite{2004hep.th....6058V, 2005NuPhB.715..304A}. In \cite{2006PhRvD..73f6002D, 2007NuPhB.778...36A}, it was further shown that the inclusion of baby universes doesn't lead to naive loss of quantum coherence, in accordance with earlier arguments from \cite{Coleman:1988cy, Giddings:1988cx, Giddings:1988wv}. On the other hand, the ensemble average interpretation and the lack of factorization in JT gravity is clearly in tension with the standard AdS/CFT correspondence. In \cite{2017arXiv170301519J, 2020JHEP...08..044M}, another perspective is given, interpreting the ensemble average as coming from gravitational constraints and different superselection sectors of the baby universe Hilbert space. Based on this observation, it was further conjectured in \cite{2020arXiv200406738M} based on \cite{McNamara:2019rup} that the constraint is so strong in $d>3$ that the baby universe Hilbert space is always one-dimensional in a consistent theory of quantum gravity, thus resolving the contradiction. As we have a UV completion for a theory of quantum gravity involving topology changes \cite{2004hep.th....6058V, 2005NuPhB.715..304A, 2006PhRvD..73f6002D, 2007NuPhB.778...36A}, we find it appealing that we might be able to test all these ideas in this context, and may directly identify an ``information paradox'' in string theory where calculations of entropy without the inclusion of topology-changing procedures leads to violation of unitarity.

    \paragraph{Comment on the BPS formula for the EE} At strong string coupling, fundamental degrees of freedom are no longer string states rather D-brane particle states. Furthermore, the degeneracy of the BPS states is typically expected to be exponential in the number of the BPS states \cite{Vafa:1995zh}. This exponential scaling of the degeneracy equates well with \eqref{EE BPS} as is proportional to the BPS index $n_\beta^g.$ Hence, \eqref{EE BPS} implies that the entanglement entropy counts how many BPS states (including the multi particle states) there are across the entangling surface. Interestingly enough, in \cite{Strominger:1996sh} the Bekenstein-Hawking entropy computed via the BPS microstate counting is also polynomial in the BPS index due to the exponential scaling of the degeneracy. It will be therefore interesting to explictly show that the degeneracy of the Calabi-Yau manifold is exponential in the number of the M2-brane BPS states in M-theory on $\text{CY}_3\times S^1.$

\section*{Acknowledgements}
We thank Thomas Hartman for collaboration in the early stages of this work and providing countless suggestions and discussions. The work of M.K. was supported in part by NSF grant PHY-1719877. The
work of Y.J. is supported by the Simons Foundation through the Simons Collaboration on the Nonperturbative Bootstrap. 
The work of W.D. is supported by Perimeter Institute for Theoretical Physics.
Research at Perimeter Institute is supported in part by the Government of Canada through the Department of Innovation, Science and Economic Development Canada and by the Province of Ontario through the Ministry of Colleges and Universities.
G.W. thanks Leopoldo Pando Zayas for encouraging him to embark on a general research program to understand entanglement entropy in string theory. G.W. also thanks Ling Yan Hung and Ce Shen for discussions, and is supported by Fudan University and the Thousands Young Talents Program. Y.J. thanks Ling Yan Hung and Chen-Te Ma for extended discussions on this topic.  Finally G.W. also thanks his parents for hosting and feeding him in their home during the coronavirus epidemic.   
\appendix

\section{Topological twist and topological sigma model on the worldsheet}

Let us briefly review $N=2$ supersymmetric non-linear sigma model defined on a Riemann surface $\Sigma$ with a Kahler manifold $X$ as a target space. This theory consists of the following data: holomorphic map/coordinate function $\Phi:\Sigma \rightarrow TX,$ superpartners of $\Phi.$ Because of the complex structure of $X,$ the complexified tangent bundle $TX$ decomposes into holomorphic and anti-holomorphic tangent bundle
\begin{equation}
    TX=T^{1,0}X\oplus T^{0,1}X.
\end{equation}
Respective to the decomposition of the complexified tangent bundle, we denote the holomorphic components of $\Phi$ by $\phi^i\in T^{1,0}X$ and similarly for the anti-holomorphic components. With this holomorphic decomposition, we can think of $\phi^i$ as a holomorphic tangent vector, of the target space, valued scalar field on the worldsheet. A superpartner of such field then should live in holomorphic tangent vector valued $spin$ bundle, which reads
\begin{equation}
    \sqrt{K_\Sigma}\otimes(\mathcal{O}_\Sigma\oplus \Omega_\Sigma ^{0,1})\otimes\Phi^*(TX^{1,0}), 
\end{equation}
where $\sqrt{K_\Sigma}$ is an algebraic square root of canonical bundle of $\Sigma,$ $\mathcal{O}_\Sigma$ is structure sheaf of $\Sigma,$ and $\Omega^{0,1}_\Sigma\equiv\overline{K_\Sigma}$ is anti-holomorphic cotangent bundle of $\Sigma.$ As anti-holomorphic canonical bundle is dual of canonical bundle, the corresponding spinor bundle can be written as
\begin{equation}
    (K_\Sigma^{1/2}\oplus \overline{K_\Sigma}^{1/2})\otimes \Phi^*(TX^{1,0}).
\end{equation}
We will then denote the fermions living in $K_\Sigma^{1/2}\otimes \Phi^*(TX^{1,0})$ and $\overline{K}_\Sigma^{1/2}\otimes \Phi^*(TX^{1,0})$ by $\psi_+^i$ and $\psi_-^i,$ respectively. We will use the similar convention for $\psi_+^{\bar{i}}$ and $\psi_-^{\bar{i}}.$ Given the field contents, the worldsheet action is
\begin{equation}
    S=2t \int_\Sigma \left(\frac{1}{2} g_{IJ}\partial_z\phi^I\partial_{\bar{z}}\phi^J+ig_{i\bar{i}}\psi_-^{\bar{i}}D_z\psi_-^i +ig_{i\bar{i}}\psi_+^{\bar{i}}D_{\bar{z}}\psi_+^i+R_{i\bar{i}j\bar{j}}\psi_+^i\psi_+^{\bar{i}}\psi_-^{j}\psi_-^{\bar{j}}\right),
\end{equation}
where $g$ is the hermitian metric of the target space.

Topological string model is then obtained by a topological twist to the bundle \cite{Witten:1991zz}, in which fermionic fields live in, that preserves the form of kinetic terms of fermionic fields. 
The topological twist of A model can be understood as moving the non-trivial bundle $\sqrt{K_\Sigma}$ from $K_\Sigma^{1/2}\otimes\Phi^*(TX^{1,0})$ to $K_\Sigma^{1/2}\otimes \Phi^*(TX^{0,1})$ and similarly for $\overline{K}_{\Sigma}^{1/2}.$ 
As a result of this topological twist, $\psi_+^i$ and $ \psi_-^i$ becomes (anti)-holomorphic tangent vector valued scalar field on the worldsheet. 
Then we can focus on transformation that transforms $\phi^i$ into $\psi_+^i$ and $\phi^{\bar{i}}$ into $\psi_-^{\bar{i}},$ as those transformations can be represented by a globally well defined functions and others not in general.

Given the topological twist, let us rename the fermionic fields as $\chi^i=\psi_+^i$ and $\chi^{\bar{i}}=\psi_-^{\bar{i}}.$ Supersymmetry transformation is concisely repackaged as
\begin{align}
    \{Q,\Phi\}=&\chi,\nonumber\\
    \{Q,\chi\}=&0,\nonumber\\
    \{Q,\psi_{-}^I\}=&i\partial_{\bar{z}}\Phi^I-\chi^J\Gamma^{I}_{JK}\psi_{-}^K,\nonumber\\
    \{Q,\psi_{+}^{\bar{I}}\}=&i\partial_{\bar{z}}\Phi^{\bar{I}}-\chi^{\bar{J}}\Gamma^{\bar{I}}_{\bar{J}\bar{K}}\psi_{+}^{\bar{K}},
\end{align}
where $Q^2=0$ on-shell thus supersymmetry becomes BRST symmetry. The action is 
\begin{equation}
    S=2t \int_\Sigma \left(\frac{1}{2} g_{IJ}\partial_z\phi^I\partial_{\bar{z}}\phi^J+ig_{i\bar{i}}\psi_-^{i}D_z\chi_-^{\bar{i}} +ig_{i\bar{i}}\psi_+^{\bar{i}}D_{\bar{z}}\chi^i-R_{i\bar{i}j\bar{j}}\psi_-^i\psi_+^{\bar{i}}\chi^j\chi{\bar{j}}\right).
\end{equation}
A very important observation is that this action is a sum of a Q-exact term and a topological term
\begin{equation}
    S=it\int_\Sigma d^2z\{Q,V\}+t\int_\Sigma \Phi^*(J),
\end{equation}
where $V=g_{i\bar{j}} (\psi_+^{\bar{i}}\partial_z\phi^j+\partial_z\phi^{\bar{i}}\psi_-^j)$ and $\Phi^*(J)$ is pullback of the K\"ahler form defined on $X.$ One can add pullback of two-form tensor $B$ to the action to complexfy the K\"ahler form.

We have not specified yet if $\Sigma$ has boundaries or not. If $\Sigma$ does not attain a boundary, then the worldsheet theory is a closed string theory. Similarly, if $\Sigma$ has boundaries, then the worldsheet theory is an open string theory.

Topological strings wrap volume minimizer, which is energetically stable, among homologous 2 cycles in $X.$ This means that for closed string theory, worldsheet instanton is classfied by homology class 
\begin{equation}
    \Phi_*([\Sigma])\in H_2(X,\Bbb{Z}).
\end{equation}

This classification can be generalized to open string theory directly. Open string worlsheet can be regarded as a Riemann surface with $h$ holes due to the conformal invariance. As there are $h$ boundaries of the Riemann surface, one should impose boundary conditions. Let us denote $h$ boundaries of $\Sigma$ by $C_i,$ where $i=1,\dots, h.$ In \cite{Witten:1992fb}, Witten showed that the physical boundary condition is given by
\begin{equation}
    \Phi(C_i)\subset\mathcal{L}
\end{equation}
for some $\mathcal{L}$ which is a Lagrangian submanifold of $X.$ Note that a submanifold $\mathcal{L}$ is Lagrangian if $J|_\mathcal{L}=0.$ This condition implies that supersymmetric D-branes in topological A model wrap Lagrangian three-cycles in $X.$\footnote{In this work, we do not focus on torsion one or five cycles.} Therefore, open string worldsheet instanton is naturally classified by relative homology class
\begin{equation}
    \Phi_*(\Sigma)\in H_2(X,\mathcal{L}).
\end{equation}

One important class of observable in closed A model is a three points function which has various interpretations in physical string theory. Let us consider a non-trivial 2 form $[D_i]\in H^2(X).$ Then one can consider an operator
\begin{equation}
    \mathcal{O}_{D_i}=(D_i)_{i_1,i_2}\chi^{i_1}\chi^{i_2}.
\end{equation}
If we assume that $X$ is a Calabi-Yau threefolds, when computed on string worlsheet $\Bbb{P}^1,$ the three points function of $\mathcal{O}(D_i)$ is \cite{Candelas:1990rm}
\begin{equation}
    \langle\mathcal{O}_{D_1}\mathcal{O}_{D_2}\mathcal{O}_{D_3} \rangle = \mathcal{K}_{D_1D_2D_3}+\sum_\beta N_{0,\beta}(D_1,D_2,D_3)\prod_i \int_\beta [D_i]Q^{\beta},
\end{equation}
where $\mathcal{K}_{D_1D_2D_3}$ is an intersection number and $N_{0,\beta}(D_1,D_2,D_3)$ is a genus 0 Gromov-Witten invariant for an integral curve $\beta\in H_2(X),$ and $Q= e^{-\int_\beta J}.$ Note that this three points function can be obtained from the third derivative of the genus 0 prepotential, which is free energy of genus 0 worldsheet theory,
\begin{equation}
    \partial_{t_1}\partial_{t_2}\partial_{t_3}F_0(t)=\langle\mathcal{O}_{D_1}\mathcal{O}_{D_2}\mathcal{O}_{D_3} \rangle,\label{prepotential yukawa}
\end{equation}
where $t_i= \int_{D^i}J.$ Genus 0 prepotential receives classical and instanton contributions
\begin{equation}
    F_0=F_0^{cl}+F_0^{inst},
\end{equation}
where
(to add prepotential at LCS).
Coupling to gravity \cite{Witten:1992fb}, genus g free energy can be computed as well which reads
\begin{equation}
    F_g(t)=\sum_\beta N_{g,\beta}Q^\beta,
\end{equation}
where $N_{g,\beta}$ is a genus g Gromov-Witten invariant. Combining all genera prepotential, we get a generating functional the all genera free energy
\begin{equation}
    F(g_s,t)=\sum_g F_g(t)g_s^{2g-2}.
\end{equation}

\section{Topological String on Conifolds and Geometric Transition}
\label{appB}
In this appendix, we briefly the geometric transition of interest. Let us consider A-model open topological string theory on the deformed conifold $T^*S^3.$ We wrap N D-branes on $S^3,$ whose low energy effective theory is $U(N)$ Chern-Simons theory \cite{Witten:1992fb}. Wilson lines can be introduced, if M D-branes wrap on a lagrangian submanifold\footnote{In topological string theory. Unlike physical string theory, Lagrangian is good enough to ensure supersymmetry. Note that in the conifold, Lagrangian submanifolds we consider are in fact special Lagrangian.} $\mathcal{L}$ of $T^*S^3$ which intersects $S^3$ at $S^1.$ This corresponds to U(N) Chern-Simons theory on $S^3$ with M knots on $S^1.$ Under the geometric transition at large N, we obtain A-model topological string theory on the resolved conifold $\mathcal{O}(-1)\oplus\mathcal{O}(-1)\rightarrow \Bbb{P}^1,$ in which the N D-branes are desolved into B-flux and M D-branes are still wrapped on the same special lagrangian $\mathcal{L}$ \cite{Ooguri:1999bv}.

Let us first study the deformed conifold. Cotangent bundle of $S^3$ can be embedded into $\Bbb{C}^4$ by an equation
\begin{equation}
y_1^2+y_2^2+y_3^2+y_4^2=a^2,
\end{equation}
$y_i$'s$\in\Bbb{C}.$ We assume that $a$ is a real number. The bundle structure is more vivid when we write $y_i=x_i+i p_i,$ then the embedding equation is written as
\begin{equation}
\sum_i x_i^2=a^2+\sum_i p_i^2,~~~ \sum_i x_i p_i=0.
\end{equation}
It is then clear when $p_i=0,$ for all $i,$ then the equations are reduced to
\begin{equation}
\sum_i x_i^2=a^2.
\end{equation}
Thus $a$ describes radius of $S^3.$ When $a$ is sent to 0, the deformed conifold in the limit described by 
\begin{equation}
y_1^2+y_2^2+y_3^2+y_4^2=0.\label{eqn:sing con}
\end{equation}
As Jacobian of the defining equation vanishes at the origin $y_1=y_2=y_3=y_4=0,$ the conifold at the origin is singular. 
\subsection{Blow up of the resolved conifold} 
\label{ssection: blowup}
To fix the singularity at the origin, we blow up the origin such that $y_1=y_2=y_3=y_4=0$ is replaced with a smooth manifold. If we reparametrize the coordinates as
\begin{equation}
    z_{ij}=\sum_n \sigma^n_{ij}y_n,
\end{equation}
then \eqref{eqn:sing con} is written as
\begin{equation}
    \det z_{ij}=0.\label{eqn:sing con2}
\end{equation}
In this presentation, the singularity occurs when the matrix coordinates $z_{ij}$ are trivial. It is important to note that we can view \eqref{eqn:sing con2} as a condition for the following equation to have a non-trivial solution
\begin{equation}
    \left(\begin{array}{cc} z_{11} & z_{12} \\ z_{21} & z_{22} \end{array}\right)\left(\begin{array}{c} \lambda_1 \\ \lambda_2 \end{array}\right)=0,\label{eqn:res con}
\end{equation}
for some complex variable $\lambda_1$ and $\lambda_2$ which cannot be simultaneously zero, because $\lambda_1=\lambda_2=0$ results in no constraints on $z_{ij}$ matrix. Furthermore, \eqref{eqn:res con} provides a resolution of the singularity because when $z_{ij}$ is non trivial $\lambda_1$ and $\lambda_2$ are fixed up to rescaling and $z_{ij}=0$ is replaced with coordinates $(\lambda_1,\lambda_2).$ This implies that equation \eqref{eqn:res con} is an embedding of the resolved conifold into $\Bbb{C}^4\times \Bbb{P}^1$ in which $z_{ij}$ is a coordinate of $\Bbb{C}^4$ and $[\lambda_1,\lambda_2]$ is a homogeneous coordinate of $\Bbb{P}^1.$ Note that, when $\det(z_{ij})=0$ the non-homogeneous coordinate $z$ of $\Bbb{P}^1$ is related to the rest of the coordinates by
\begin{equation}
z:=\frac{\lambda_1}{\lambda_2}=\frac{-y_1+iy_2}{y_3+y_4}=\frac{y_3-y_4}{y_1+iy_2}.
\end{equation}
\subsection{Lagrangian Submanifolds}
\label{ssection:lag} 
Lagrangian submanifolds can be easily found by finding symmetric locus of an anti-holomorphic involution. We consider an anti-holomorphic involution
\begin{equation}
y_{1,2}=\overline{y}_{1,2},~~~y_{3,4}=-\overline{y}_{3,4}.\label{eqn:knot}
\end{equation}
In the deformed conifold, the invariant locus of \eqref{eqn:knot}, a lagrangian submanifold $\mathcal{L},$ is
\begin{equation}\label{lag}
p_{1,2}=0,~~~x_{3,4}=0.
\end{equation}
At the symmetric locus of \eqref{eqn:knot}, the embedding equation becomes
\begin{equation}
x_1^2+x_2^2=a^2+p_3^2+p_4^2.
\end{equation} 
Hence $\mathcal{L}$ intersects $S^3$ at
\begin{equation}
x_1^2+x_2^2=a^2,
\end{equation}
which is a $S^1.$

Similarly, in the resolved conifold, the lagrangian submanifold is defined by
\begin{equation}
    \left(
\begin{array}{cc}
 i p_3+i p_4 & x_1-i x_2\\
 x_1+i x_2 & i p_4-i p_3 \\
\end{array}
\right)\left(\begin{array}{c}
     \lambda_1  \\
      \lambda_2
\end{array}\right)=0.
\end{equation}

\section{Quantum groups and their representations}
\label{section:QG}
\subsection{Hopf algebra structure} The quantum group  $\mathcal{A}(U(N)_{q})$ is a quasi-triangular Hopf algebra.  To explain what this is, we start with the simpler structure of a bi-algebra $\mathcal{A}$, which is an algebra endowed with 4 operations
\begin{align}
     \text{product}& \quad\nabla: \mathcal{A}\otimes \mathcal{A} \rightarrow \mathcal{A}\nn
     \text{unit}& \quad\eta: \mathbb{C} \rightarrow \mathcal{A}\nn
     \text{coproduct}&\quad  \Delta : \mathcal{A}\rightarrow \mathcal{A} \otimes \mathcal{A}\nn
     \text{counit}& \quad   \epsilon: \mathcal{A} \rightarrow \mathbb{C}   
\end{align}
These operations satisfy various sewing relations \cite{Klimyk:1997eb}; in particular the product and coproduct are associative and co-associative respectively.  A basic example is the set of $\mathcal{A}(G)$ of $\mathbb{C}$ valued-functions on a group $G$, where $\nabla$ is pointwise multiplication, $\eta =1 $, and the coproduct and  counit are defined to act on $f \in \mathcal{A}(G)$  as
\begin{align}\label{DE}
    \Delta(f) (U,V) &= f(UV ),\quad U,V \in G \nn
    \epsilon (f) &= f(1_{G})
\end{align}
Here $U V$ denotes the group multiplication of $U$ and $V$, and $1_{G}$ is the identity element of $G$.   The formulas \eqref{DE} show that the coproduct and counit are dual to the product and unit on the group $G$.
For $G=U(N)$, this describes the algebraic structure of the Hilbert space for 2DYM and its string theory dual.

In the coordinate algebra $\mathcal{A}(U(N)_{q})$, $\nabla $ is  q-deformed into a  non-commutative product, while $UV$ remains the same as the ordinary matrix multiplication and $1_{G}$ is still the identity matrix. In particular, the actions of the coproduct and counit on single string wavefunctions  $f_{ij}(U)=U_{ij}$ are given by
\begin{align} \label{UF}
\Delta(U_{ij}) &= \sum_{k} U_{ik} \otimes U_{kj}\nn
\epsilon(U_{ij}) &= \delta_{ij}  
\end{align} 
Meanwhile the counit defines the trivial, or ``vacuum"  representation.  

This bi-algebra structure is upgraded into a Hopf algebra by the introduction of a mapping called the antipode
\begin{align}\label{S}
         \text{antipode}&\quad   S:\mathcal{A} \rightarrow\mathcal{A} 
\end{align}
which acts as an inverse on the quantum group:
\begin{align}
   \sum_{k} S(U)_{ik}U_{kj} &=\sum_{k} U_{ik}S(U)_{kj} = \delta_{ij} 
\end{align}

The final element that makes a Hopf Algebra into a quantum group is the $R$ matrix, which makes it a quasi-triangular Hopf algebra. This can be viewed as an element
\begin{align}
    \mathcal{R}& \in \mathcal{A} \otimes \mathcal{A} \nn
    \mathcal{R}&=\sum_{i} a_{i} \otimes b_{i} 
\end{align}
We can also interpret this as a linear operator on $V \otimes V$.  It satisfies the Yang-Baxter equation. 

\subsection{R matrix and antipode for $SL_{q}(2)$}
To illustrate this definition, consider the quantum group $SL_{q}(2)$. Its  coordinate algebra is generated by 4 elements (a,b,c,d) of a matrix 
\begin{align}
U= \begin{pmatrix} a && b\\c&&d\end{pmatrix}
\end{align} 
and the $R$ matrix is 
\begin{align}
    \mathcal{R} = \begin{pmatrix} 
    q^{1/2} && 0&&0&&0 \\
    0&&q^{-1/2}&&0&&0\\
    0&& q^{1/2}-q^{-3/2} &&1&&0\\
    0&&0&&0 &&q^{1/2} 
    \end{pmatrix} 
\end{align}
Then the multiplication rule  \eqref{Ru} is equivalent to
\begin{align}
ab&= q ba, \quad  ac= q ca,\quad  bd= q db,\quad cd= q dc\quad  \nn
cb&=bc,\quad ad-da = (q-q^{-1}) bc\nn
ad-q bc&= 1
\end{align}
The antipode is given by
\begin{align} \label{antipode}
S(a)=d, ,\quad S(c) = -q c ,\quad S(b) = -q^{-1} b ,\quad S( d) = a 
\end{align} 
\subsection{$*$ structure and unitary representations}
\paragraph{* on the coordinate algebra} 
we define an involution of the $SL(2)_{q}$ algebra  which plays the role of complex conjugation by
\begin{align}
a^{*} = d,\quad b^{*} = -q c ,\quad c^{*} = -q^{-1} b ,\quad d^{*} = a 
\end{align} 
From the antipode \eqref{antipode} we find the relation
\be
U^{t*}=S(U)
\ee
where $t$ stands for transpose.
$SU(2)_{q}$ refers to $SL(2)_{q}$ equipped with the above star structure.

 \section{Spacetime non-commutativity from B fields}
 \label{app:Bfield}
Here we show how s non commutative worldvolume gauge field in arises from the string sigma model due to the coupling to a nontrivial $B$ field flux in the base $S^2$.  For the physical string, it is known \cite{1999JHEP...09..032S} that the presence of the $B$ field alters the boundary conditions for the open string, and leads to an anti-symmetric part to the worldsheet propagator.  This in turn leads to nontrivial commutation relations of the open string endpoint,resulting in a non-commutative worldvolume gauge theory on the D branes. 

For the A model, we can see how this phenomenon arises from the bosonic part of the sigma model action in the presence of the $B$ field:
\begin{align}
 S= \int_{W}  \frac{1}{2} g_{IJ} \delta ^{ab}\pd_{a} X^I\partial_{b} X^J + i B_{IJ} \epsilon^{ab} \pd_{a} X^{I} \pd_{b} X^{j}  d^{2}\sigma  
\end{align} 
where $W$ denotes the 2 dimensional worldsheet.  For a constant B field $B_{IJ}=B \epsilon_{IJ}$, the second term  is a total derivative that can be written as a boundary term: 
\begin{align}
     S= \int_{W}  \frac{1}{2} g_{IJ} \delta ^{ab}\pd_{a} X^I\partial_{b} X^J +  i  \int_{\pd W} B \epsilon_{IJ} X^{I} \epsilon^{ab}  \pd_{s} X^{j}  ds
\end{align}
where $s$ is the "time" coordinate along the boundary. We can treat the boundary term as the integral of the canonical one-form $\int p d q $ for a quantum mechanical particle corresponding to the open string endpoint.   This implies that the $B \epsilon_{IJ} X^{I}$ is the canonical momentum conjugate to $X^J$, and therefore the equal time commutation relations in $g_{ij}\rightarrow 0 $ limit are
\begin{align}
    [X^{I},X^{J}] =  i \frac{\epsilon^{IJ}}{B}
\end{align}
for the open string endpoints.   
 
\newpage

\bibliographystyle{utphys}
\bibliography{ZEEtopostrings.bib}

\end{document}